\documentclass{jfm}
\usepackage{graphicx}
\usepackage{newtxtext}
\usepackage{newtxmath}
\usepackage{xfrac,bigints}
\usepackage{natbib}
\usepackage{amsmath}
\usepackage{hyperref}
\usepackage{mathrsfs}
\usepackage{cancel}
\usepackage{pdftricks}
\usepackage{mathtools}
\usepackage{dcolumn}
\usepackage{bm}
\usepackage{xcolor}
\usepackage{amsmath}
\usepackage{rotating}
\usepackage{epstopdf, epsfig}
\usepackage{booktabs}
\usepackage{dashrule}
\usepackage[justification=justified]{caption}
\usepackage{tikz}
\begin{psinputs}
    \usepackage{lmodern}
    \usepackage{pstricks}
    \usepackage{xcolor}
\end{psinputs}
\usepackage{mwe}
\usepackage{subcaption}
\usepackage{mathtools}
\usepackage{epstopdf,epsfig}
\hypersetup{
    colorlinks = true,
    urlcolor   = blue,
    citecolor  = black,
}

\newcommand{\RomanNumeralCaps}[1]

\title{The flow field due to a sphere moving in a viscous, density stratified fluid}

\author{Ramana Patibandla\aff{1},
  Anubhab Roy\aff{1}
 \and Ganesh Subramanian\aff{2}\corresp{\email{sganesh@jncasr.ac.in}}}

\affiliation{\aff{1} Department of Applied Mechanics, Indian Institute of Technology Madras, Chennai - 600036, India,
\aff{2}Engineering Mechanics Unit, Jawaharlal Nehru Center for Advanced Scientific Research, Bangalore - 560064, India.}

\begin{document}
\defcitealias{varanasi2022motion}{VS22}

\maketitle

\begin{abstract}
In this paper, we study the disturbance velocity and density fields induced by a sphere translating vertically in a viscous density-stratified ambient. Specifically, we consider the limit of a vanishingly small Reynolds number $(Re = \rho U a/\mu \ll 1)$, a small but finite viscous Richardson number $(Ri_v = \gamma a^3 g/\mu U\ll 1)$, and large Peclet number $(Pe = Ua/D\gg 1)$. Here, $a$ is the sphere radius, $U$ its translational velocity, $\rho$ an appropriate reference density within the framework of the Boussinesq approximation, $\mu$ the ambient viscosity, $\gamma$ the absolute value of the background density gradient, and $D$ the diffusivity of the stratifying agent. For the scenario where buoyancy forces first become comparable to viscous forces at large distances, corresponding to the Stokes-stratification regime defined by $Re \ll Ri_v^{1/3} \ll 1$ for $Pe \gg 1$, important flow features have been identified by \citet{varanasi2022motion} - these include a vertically oriented reverse jet, and a horizontal axisymmetric wake, on scales larger than the primary (stratification) screening length of $\mathcal{O}(aRi_v^{-1/3})$. Here, we study the reverse jet region in  more detail, and show that it is only the central portion of a columnar structure with multiple annular cells concentric about the rear stagnation streamline. In the absence of diffusion, corresponding to $Pe = \infty$ $( \beta_\infty = Ri_v^{1/3}Pe^{-1} = 0)$, this columnar structure extends to downstream infinity with the number of annular cells diverging in this limit. We provide expressions for the boundary of the structure, and the number of cells within, as a function of the downstream distance. For small but finite $\beta_\infty$, 
two length scales emerge in addition to the primary screening length - a secondary screening length of $\mathcal{O}(aRi_v^{-1/2}Pe^{1/2})$ where diffusion starts to smear out density variations across cells, leading to exponentially decaying density and velocity fields; and a tertiary screening length, $l_t \sim \mathcal{O}\left(aRi_v^{-1/2}Pe^{1/2}\left[\zeta + \dfrac{13}{4}\ln{\zeta} + \dfrac{13^2}{4^2}\dfrac{\ln\zeta}{\zeta}\right]\right)$ with $\zeta = \dfrac{1}{2}\ln\left(\dfrac{\sqrt{\pi}Ri_v^{-1}Pe^3}{2160}\right)$, beyond which the columnar structure ceases to exist. The latter causes a transition from a vertical to a predominantly horizontal flow, with the downstream disturbance fields reverting from an exponential to an eventual algebraic decay, analogous to that prevalent at large distances upstream.
\end{abstract}

\section{Introduction}\label{sec:Intro}
\begin{figure}
      \centering
      \includegraphics[width=0.9\textwidth]{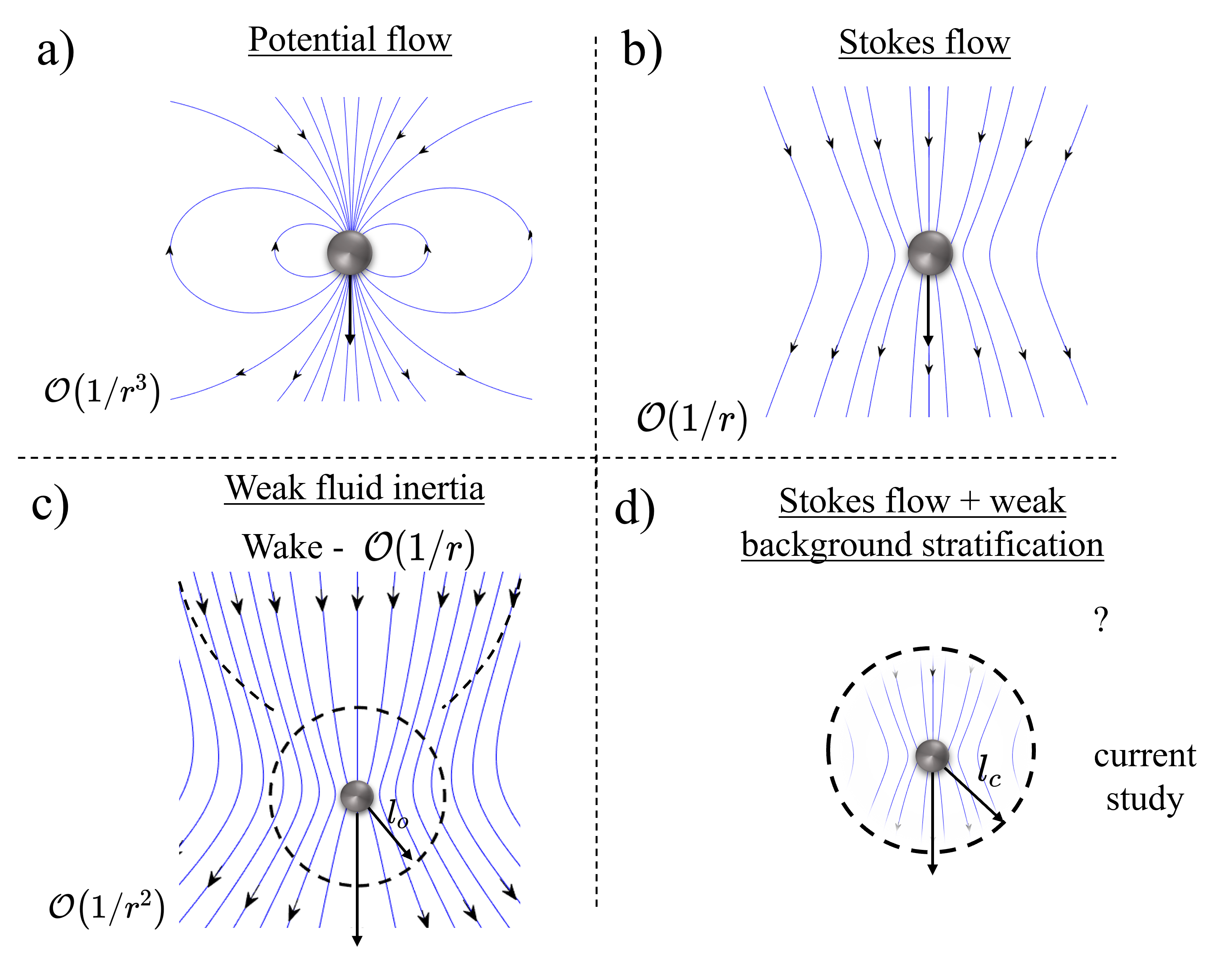}
  \caption{A schematic (not to scale) of the disturbance flow fields due to a translating sphere (of radius $a$) in different physical scenarios: (a) Potential flow, (b) Stokes flow, (c) in presence of weak fluid inertia, and (d) with a weak background stratification. Here, $l_o = \mathcal{O}(a Re^{-1})$, the Oseen length, is the scale at which inertial forces become comparable to viscous ones, and $l_c = \mathcal{O}(a Ri_v^{-1/3})$, the stratification screening length, is the scale at which buoyancy forces become comparable to viscous ones\,(for large $Pe$). The focus of the current work is the flow field beyond $l_c$ in (d).}
  \label{fig:5_1}
\end{figure}
It has been suggested that the diurnal vertical migration of marine organisms \citep[termed the largest migration on earth; see][]{martin2020oceans} could mix the oceans and is likely as important as the well-known energy sources due to winds and tides, towards the meridional overturning circulation \citep{dewar2006does, katija2009viscosity}. This ``biogenic mixing'' hypothesis concerns the motion of both passive particles and active swimmers at low (eg. bacteria) to order unity (eg. copepods) Reynolds numbers ($Re$). Such motion results in intricate flow patterns on scales comparable to and larger than the individual particles/swimmers, with the nature of the patterns being governed by various factors such as their swimming characteristics, fluid physical properties, interactions with other particles/swimmers, etc\,\citep{katija2012biogenic}. An important element in the aforesaid hypothesis is the amount of fluid displaced by a single entity, either active or passive. This so-called drift volume can, in principle, act as a source of available potential energy which can then couple to motion on scales much larger than the swimmer size. The drift volume may be calculated by considering an initial plane of material points with the particle infinitely far upstream to begin with. The plane gets deformed as the particle translates towards it, and eventually to downstream infinity. For the initial plane being infinite in extent, the volume enclosed between it and the final deformed one is defined as the `total drift volume' \citep{darwin1953note, lighthill1956drift}. For a sphere in potential flow, with a fore-aft symmetric disturbance velocity field of $O(1/r^3)$\,($r$ being the radial distance from the sphere center; see figure \ref{fig:5_1}a), the total drift volume is half the sphere volume, this being equal to the added mass divided by the fluid density\,\citep{lighthill1956drift}. In the opposite limit of a viscous fluid ambient ($Re = 0$), the slower decaying $\mathcal{O}(1/r)$ disturbance velocity field\,(see figure \ref{fig:5_1}b) causes the total drift volume to diverge over any finite time interval. Thus, it becomes necessary to define a `partial drift volume', as the volume enclosed between initial and final material planes of a finite spatial extent, and with the initial plane located at a finite distance away from the sphere \citep{eames1994drift, chisholm2018partial}. Accounting for the effects of weak fluid inertia (see figure \ref{fig:5_1}c) results in a faster $\mathcal{O}(1/r^2)$ source-flow-like decay of the velocity field, in almost all directions, at distances larger than the Oseen length\,($l_o = \mathcal{O}(aRe^{-1})$). However, the original $\mathcal{O}(1/r)$ decay persists within a parabolodial wake region behind the sphere, and as a result, despite there being no finite-time divergence, the drift volume does diverge linearly in the infinite time limit for any non-zero $Re$\,\citep{subramanian2010viscosity,chisholm2017drift}. 

\begin{figure}
    \hspace{-1.5cm}
    \begin{minipage}[t]{0.65\textwidth}
      \centering
      \includegraphics[width=\textwidth]{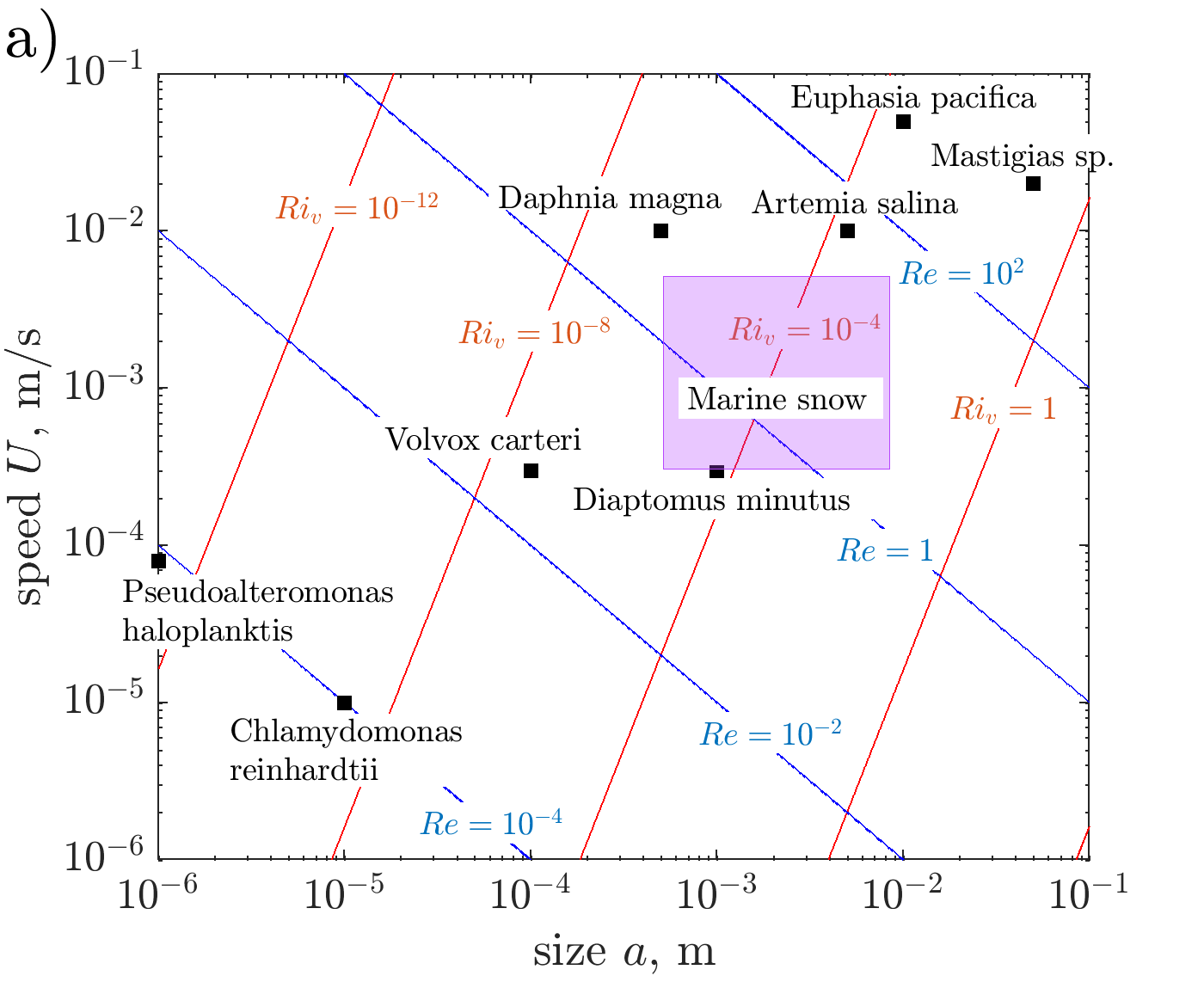}
    \end{minipage}
     \hspace{-0.7cm}
    \begin{minipage}[t]{0.65\textwidth}
      \centering
      \includegraphics[width=\textwidth]{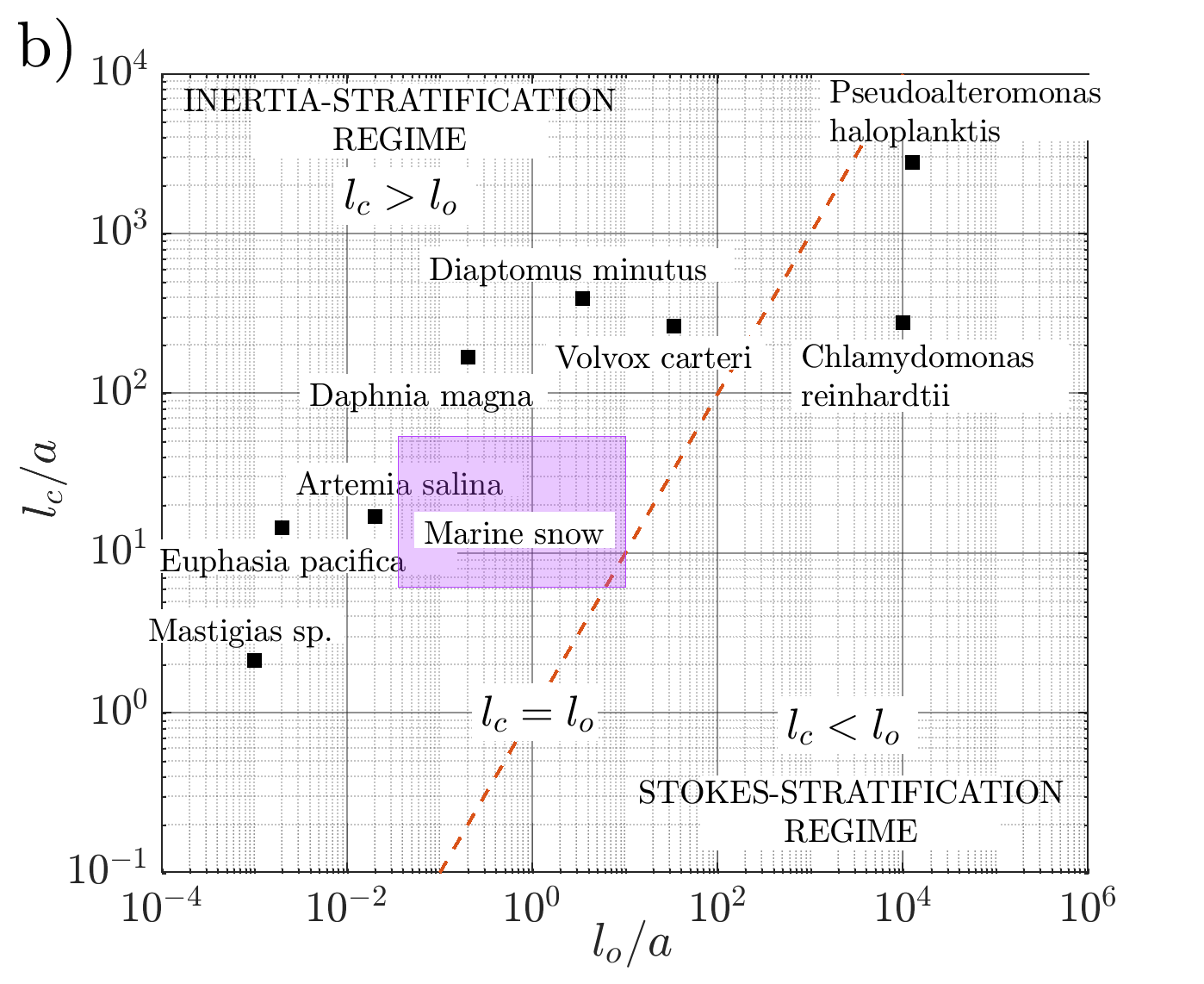}
    \end{minipage}
  \caption{(a) Typical sizes and speeds of different groups/classes of marine particulate matter: Pseudoalteromonas haloplanktis  \citep{wagner2014mixing}, Chlamydomonas reinhardtii \citep{more2020motion}, marine snow \citep[relevant range of values indicated by a purple box, see][]{turner2002zooplankton}, Volvox carteri \citep{drescher2009dancing}, Diaptomus minutus \citep{malkiel2003three}, Daphnia magna \citep{noss2014direct}, Artemia salina \citep{houghton2018vertically}, Euphasia pacifica \citep{kunze2006observations} and Mastigias sp. \citep{katija2009viscosity}. The blue and red lines correspond to constant Reynolds and viscous Richardson numbers, respectively. $U \propto a^{-1}$ and $U\propto a^3$ along the blue and red lines, respectively. (b) A classification of the marine particulate matter in (a), in dimensionless terms, on a parameter plane comprising the ratio of the stratification screening length to particle size ($l_c/a$) and the ratio of Oseen length scale to the particle size ($l_o/l_a$); refer to third paragraph of \S~1 for definitions of $l_c$ and $l_o$.}
  \label{fig:5_2}
\end{figure}

The divergence of the drift volume over a finite time interval, or for infinite time, led to the suggestion, by \citet{katija2009viscosity}, of the aforementioned diurnal migration being a biogenic source of oceanic mixing. Although this migration is known to affect the transport of nutrients and entrapment of CO$_2$ in the ocean interior \citep{falkowski1998biogeochemical}, its contribution to oceanic mixing has been proposed to be negligible based on various physical arguments\,\citep{visser2007biomixing, subramanian2010viscosity, wagner2014mixing}, with supporting evidence from numerical simulations\,\citep{ wang2015biogenic}. The arguments rely on two points: (a) consideration of an active instead of a passive particle in the observations of \cite{katija2009viscosity}, and (b) neglect of the stable stratification of the oceanic ambient by the said authors, both of which should lead to a faster decay of the disturbance velocity field, and thereby, a convergent drift volume. In being applicable to both passive and active particles, the ambient stratification is a more important factor than particle activity \citep{subramanian2010viscosity}. In addition to the above, mixing efficiencies in stratified fluids have been found to be very low for small Peclet numbers\,($Pe$), corresponding to a rapidly diffusing stratifying agent; in this limit, the energy injected by the particles/swimmers is primarily dissipated, rather than contributing to mixing or an increase in the potential energy\,\citep{visser2007biomixing, wagner2014mixing}.  
At the other extreme, that is, for large $Pe$, quantifying the extent of biogenic mixing requires a detailed study of the drift volume mentioned above. This in turn requires an examination of the disturbance flow field induced by a particle translating vertically in a viscous stably stratified medium. The dominant contributions to the drift volume arise from fluid motion at distances from the particle/swimmer that are large compared to its size\,\citep{varanasi2022motion}, and thus, the interest is in the far-field characteristics of the disturbance velocity in the parameter regime relevant to the oceanic scenario\,(see figure \ref{fig:5_1}d). 

\begin{center}
\footnotesize
\begin{tabular}{p{2.7cm}<{\raggedleft} p{2.25cm}<{\centering} p{2.25cm}<{\centering} p{2.25cm}<{\centering} p{2cm}<{\centering} p{2cm}<{\raggedright}}
        \toprule
        Literature & $Re$ & $Ri$ & $Pe$ & Regime & Focus\\ 
        \midrule 
        \citet{list1971laminar}$^{A}$ & $\ll 1$  & $\ll 1$  & $\ll 1$&  $l_c\ll l_o$ & Flow-field\\
        \citet{zvirin1975settling}$^{A}$ & $\ll 1$  & $\ll 1$ & $\gg 1$ & $l_c\ll l_o$ & Drag\\
        \citet{hanazaki2009jets}$^{E}$ & $\mathcal{O}$($10$) - $\mathcal{O}$($10^3$)  & $\mathcal{O}$($10^{-2}$) - $\mathcal{O}$($10^5$) & $\mathcal{O}$($10^4$) - $\mathcal{O}$($10^6$) &$l_c \gg l_o$& Flow-field \& drag\\
        \citet{ardekani2010stratlets}$^{A}$ & $\ll 1$  & $\ll 1$ & $\ll 1$ & $l_c\ll l_o$ & Flow-field\\
        \citet{candelier2014history}$^{A}$ & $\ll 1$  & $\ll 1$ & $\ll 1$ & $l_c \ll l_o$ & Drag\\
        \citet{wagner2014mixing}$^{A}$ & $\ll 1$  & $\ll 1$ & $\ll 1$ & $l_c \ll l_o$ & Mixing eff.\\
        \citet{fouxon2014convective}$^{A}$ & $\ll 1$  & $\ll 1$ & $\ll 1$ & $l_c \ll l_o$ & Stability of disturbance flow-field\\
        \citet{mehaddi2018inertial}$^{A}$ & $\ll 1$  & $\ll 1$ & - & - & Drag\\
        \citet{zhang2019core}$^{N}$ & $\mathcal{O}$($10^{-2}$) - $\mathcal{O}$($10^2$) & $\mathcal{O}$($10^{-3}$) - $\mathcal{O}$($10^3$) & $\mathcal{O}$($10^{-2}$) - $\mathcal{O}$($10^5$) & - & Flow-field \& drag\\
        \citet{lee2019sedimentation}$^{A, N}$ & $\mathcal{O}$($10^{-3}$) - $\mathcal{O}$($10$) & $\mathcal{O}$($10^{-7}$) - $\mathcal{O}$($10^2$) & $\mathcal{O}$($10^{-3}$) - $\mathcal{O}$($10^4$) & - & Flow-field and drag\\
        \citet{shaik2020drag}$^{A}$ & $\ll 1$  & $\ll 1$ & $\ll 1$ & - & Drag, Flow-field \& drift\\
        \citet{shaik2020far}$^{A}$ & $\ll 1$  & $\ll 1$ & $\ll 1$, $\gg 1$ & $l_c \ll l_o$ & Flow-field \& drift\\
        \citet{varanasi2022rotation} & $\ll 1$ & $\ll 1$ & $\ll 1, \gg 1$ & $l_c \ll l_o $ & Orientation dynamics\\
        \midrule
        \citet{varanasi2022motion}$^{A}$ & $\ll 1$  & $\ll 1$ & $\ll 1$, $\gg 1$ & $l_c \ll l_o$ & Flow-field \\
        Present work$^{A}$ & $\ll 1$  & $\ll 1$ & $\gg 1$ & $l_c \ll l_o$ & Flow-field\\
        \bottomrule
\end{tabular}
\label{tab5p1}
\captionof{table}{The parameter regimes considered in the previous literature. Here, $Re$, $Ri$, and $Pe$ are the Reynolds number, viscous Richardson number, and Peclet number, respectively, and are defined in the third paragraph of \S~1. The flow regime considered by each study is given in the fifth column: The Stokes-stratification regime corresponds to $l_c \ll l_o$ and the inertia-stratification regime to $l_c \gg l_o$. Here, $l_c$ is the stratification screening length and $l_o$ is the Oseen length\,(see para 2, \S~1). The focus of the different studies can be classified broadly into the flow-field description, drag calculation, drift volume calculation, and mixing efficiency estimation. The nature of the work is also indicated via superscripts $A$\,(analytical), $N$\,(numerical) or $E$\,(experimental).}
\normalsize
\end{center}

Particle motion in a stratified fluid ambient is characterized by three non-dimensional numbers: the Reynolds number ($Re = \rho U a/\mu$), the viscous Richardson number ($Ri_v = \gamma a^3 g/\mu U$), and the Peclet number ($Pe = U a/D$). Here, $U$ is the particle translational velocity, $a$ is the particle size\,(radius),  $\rho$ is an appropriate reference fluid density, $\mu$ is the fluid viscosity,  $\gamma$ is the absolute value of the background density gradient, and $D$ may be taken as the salt diffusivity, keeping in mind the oceanic scenario. The viscous Richardson number is related to the Froude number ($Fr = U\sqrt{\rho}/a\sqrt{g\gamma}$) as $Ri_v = Fr^{-2}Re$. In the vicinity of the particle, viscous forces are dominant, yielding the familiar $\mathcal{O}(1/r)$ Stokesian decay of the flow field, as already mentioned in the context of the drift volume discussion above. A length scale ($l_c$, measured from the particle), termed the ``stratification screening length'', can be defined as the distance where the decaying viscous forces become comparable with buoyancy forces \citep{ardekani2010stratlets}, leading to a departure from the $\mathcal{O}(1/r)$ decay above. The stratification screening lengths ($l_c$) for small and large $Pe$ are $\mathcal{O}\left(a(Ri_vPe)^{-1/4}\right)$ and $\mathcal{O}\left(aRi_v^{-1/3}\right)$, respectively. These are analogous to the well-known Oseen length\,($l_o = \mathcal{O}(aRe^{-1})$), where the departure from the Stokesian rate of decay occurs due to viscous and inertia forces becoming comparable in magnitude. The existence of both inertia and buoyancy forces, and their competition with viscous forces, leads to two flow regimes: (a) the ``Stokes-stratification regime'' where $l_c \ll l_o$, and buoyancy effects dictate the flow behavior at distances of order and larger than $l_c$, and (b) the ``Inertia-stratification regime" where $l_c \gg l_o$, and inertial effects influence the flow much before buoyancy does.

For a migrating marine bacterium (for instance, $Pseudoalteromonas~haloplanktis$), the aforementioned non-dimensional numbers are $Re \sim \mathcal{O}(10^{-4})$, $Ri_v \sim \mathcal{O}(10^{-12})$ and $Pe \sim \mathcal{O}(10^{-2})$ \citep{wagner2014mixing}, and the bacterium therefore corresponds to the small-$Pe$ limit of the Stokes-stratification regime. In contrast, typical zooplankton (for instance, copepods), with $Re \sim \mathcal{O}(10^{-1})$, $Ri_v \sim \mathcal{O}(10^{-8})$ and $Pe \sim \mathcal{O}(10^2)$ \citep[henceforth \citetalias{varanasi2022motion}]{kunze2006observations, varanasi2022motion}, correspond to the large-$Pe$ limit of the inertia-stratification regime. Although both scenarios are relevant for oceanic biomass, the latter is more important due to the abundance of zooplankton\,\citep{wang2015biogenic}. In figure \ref{fig:5_2}a, we present typical sizes and speeds of different classes of marine particulate matter and swimming organisms considered in the previous literature, along with estimates of the Reynolds and viscous Richardson numbers shown using blue and red contours. Figure \ref{fig:5_2}b displays the same information on a plot whose coordinates are the Stokes-stratification screening length\,(ordinate) and the Oseen length\,(abscissa). In addition, in table 1, we have organized the earlier efforts in the literature by identifying the parameter space examined in each of these, along with the particular focus of the work. From this table, it is evident that while there have been numerous efforts studying particle/drop motion in stratified fluids, the focus, especially for large-$P\!e$, had until recently been on the drag correction arising from the ambient stratification \citep{zvirin1975settling, mehaddi2018inertial, lee2019sedimentation}; fewer efforts had examined the nature of fluid motion, and the associated distortion of the iso-pycnals. \citep{varanasi2022rotation} studied the orientation dynamics of a settling spheroid in a linearly stratified ambient for both small and large $Pe$ limits, providing predictions for broadside-on and edgewise settling. Recent analytical efforts \citep{shaik2020far, varanasi2022motion} have studied the flow field due to a translating particle in the above limit, in an attempt to understand the resulting fluid drift. The current study has an analogous motivation, and again concerns the disturbance velocity and density fields induced by a vertically translating passive particle in the large-$Pe$ limit of the Stokes-stratification regime (see figure \ref{fig:5_2}b).

\begin{figure}
      \centering
      \includegraphics[width=\textwidth]{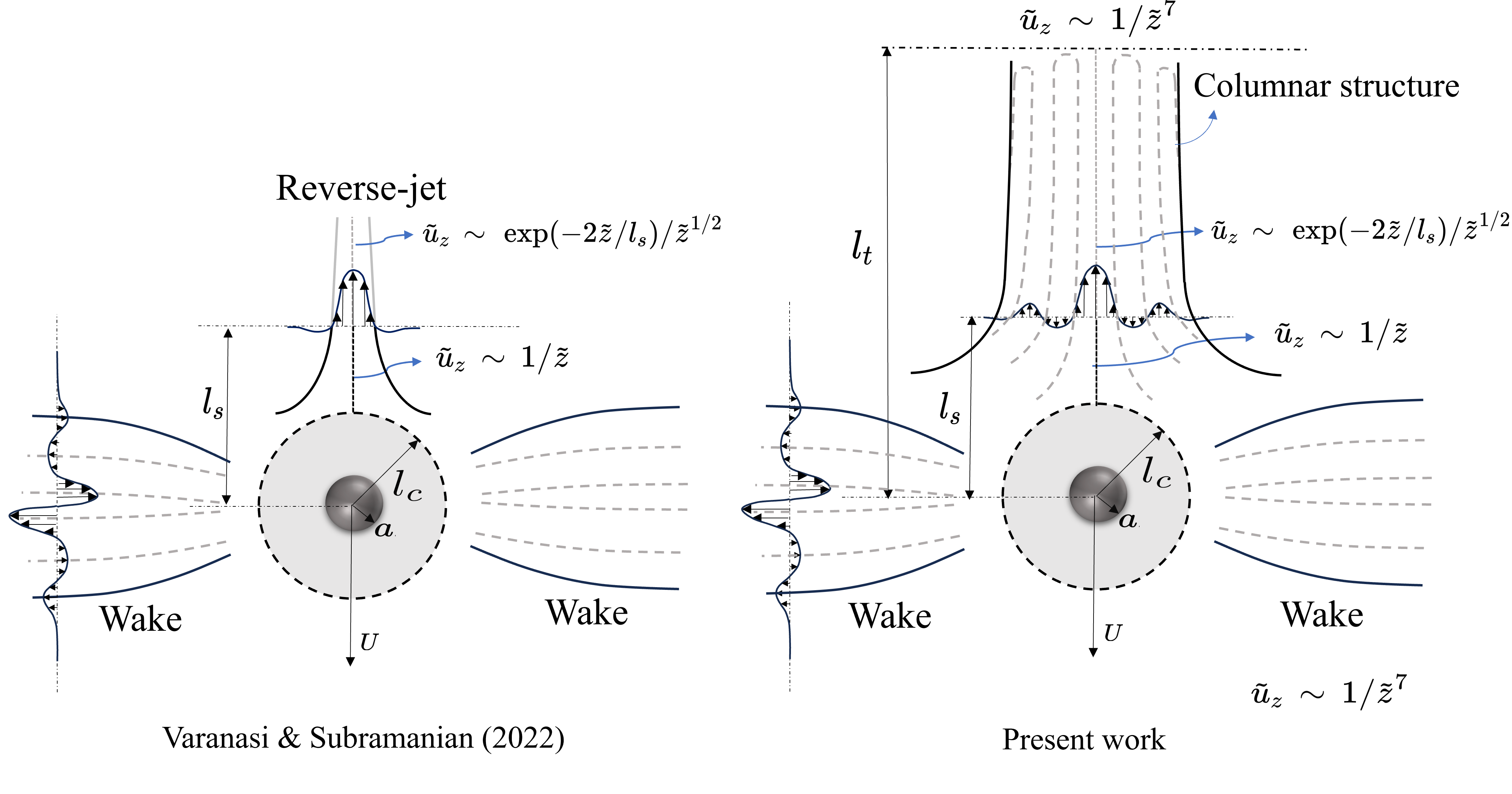}
  \caption{Schematics of the flow field induced by a sphere translating vertically in a viscous density-stratified ambient. The one on the left is based on \citetalias{varanasi2022motion}, and the one on the right is based on the present study. In both figures, the primary screening length appears as a dashed circle with radius $l_c \sim \mathcal{O}(aRi_v^{-1/3})$, and diffusion smears out transverse density variations, downstream of the sphere, at distances beyond the secondary screening length, $l_s \sim \mathcal{O}(aRi_v^{-1/2}Pe^{1/2})$. The tertiary screening length, $l_t \sim \mathcal{O}\left(aRi_v^{-1/2}Pe^{1/2}\left[\zeta + \dfrac{13}{4}\ln{\zeta} + \dfrac{13^2}{4^2}\dfrac{\ln\zeta}{\zeta}\right]\right)$ with $\zeta = \dfrac{1}{2}\ln\left(\dfrac{\sqrt{\pi}Ri_v^{-1}Pe^3}{2160}\right)$, marks the length of the columnar structure on the right.}
  \label{fig:1}
\end{figure}

The disturbance flow field due to a point-force in a stratified medium was first studied by \citet{list1971laminar} and later, for a point-force and a point-dipole, by \citet{ardekani2010stratlets}, with both studies pertaining to the small-$Pe$ limit of the Stokes-stratification regime. The point force in \cite{list1971laminar} represented a momentum jet, while it was taken to be a vertically translating particle in the latter study. In making the point-force approximation, the studies focused on the far-field and followed the Fourier transform approach pioneered by \citet{childress1964slow} and \citet{saffman1965lift}. \citet{list1971laminar} used contour integration to simplify the inverse transform integral, whereas \citet{ardekani2010stratlets} used a fast Fourier transform technique for numerical integration. The resulting flow field was found to be fore-aft symmetric with a series of horizontal recirculating cells induced by buoyancy forces, and that acted to inhibit vertical motion of fluid parcels. In studying the stability of the disturbance flow field of a sphere translating in a viscous stratified ambient to turbulent fluctuations, \citet{fouxon2014convective} also derived expressions for the far-field velocity disturbance at small $Pe$. They too used contour integration to obtain the final integral expression for the streamfunction. Further, by solving the simplified integral along the stagnation streamline, the authors obtained a far-field algebraic decay of $\mathcal{O}(1/|z|^9)$. \citet{zhang2019core} studied the drag enhancement due to stratification, on a vertically translating sphere, using numerical simulations, and also presented a brief description of the flow field for different $Re$, and for small and large $Pe$. Their results showed a fore-aft asymmetry of the large-$Pe$ flow field, caused by the convection of density perturbations. \citet{lee2019sedimentation} considered the flow field generated by both stationary and vertically translating spheres in a stratified ambient. For the latter case, expressions for the disturbance fields, at small $Pe$, were derived based on a Lorentz-type reciprocal identity, and with the sphere replaced by an equivalent singular forcing. 
The authors also numerically computed streamline and isopycnal patterns, albeit over domains of a limited spatial extent. While these patterns pertained to a range of $Re$, $Pe$, and $Ri_v$\,(see table \ref{table2}), and did show the emergence of a fore-aft asymmetry with an increase in $Pe$, the latter increase was accompanied by an analogous increase in $Re$, owing to the Prandtl number being fixed. As a result, one cannot isolate the asymmetry induced due to convection of density perturbations alone\,(as would be the case in the Stokes-stratification regime). \citet{shaik2020far} examined the disturbance flow field and the partial drift volume, induced by a vertically translating sphere and a spherical swimmer at large $Pe$, using the fast Fourier transform technique originally employed in \citet{ardekani2010stratlets}. The disturbance flow field again exhibited a fore-aft asymmetry, but rather surprisingly, appeared to decay at a slower rate in the upstream direction. The partial drift volume for the passive sphere, although finite, was found to be several times higher than the sphere volume.

Recently, \citetalias{varanasi2022motion} studied features of the disturbance velocity and density fields due to a sphere\,(modeled as a point force) moving vertically through a stably stratified medium, in the absence of inertia, over a range of $Pe$. The authors numerically evaluated the inverse transform integrals to obtain the streamlines and isopycnals as a function of  $\beta_\infty = Ri_v^{1/3}Pe^{-1}$, a dimensionless parameter that is the ratio of the large-$Pe$ stratification screening length ($a Ri_v^{-1/3}$) to the convective screening length\,($aPe^{-1}$). The authors also analytically examined the structure of the disturbance fields at distances much larger than the stratification screening length, identifying both a horizontal wake (as suggested by the recirculating cells in the earlier efforts of \citet{zhang2019core} and \citet{shaik2020far}) and a vertical jet behind the sphere, directed opposite to the sphere motion, for $Pe = \infty$. 
While buoyancy forces caused the velocity field to decay rapidly in almost all directions, a much slower $\mathcal{O}(1/r)$ decay was observed within the jet region close to the rear stagnation streamline. Identification of the jet led to the emergence of a new screening length for finite $Pe$\,(termed the ``secondary screening length'', $l_s \sim \mathcal{O}(aRi_v^{-1/2}Pe^{1/2})$), beyond which density diffusion started to smear out the jet and the `reverse-Stokeslet' decay above gave way to an exponential decay of the disturbance velocity and density fields. It is worth noting that the slower decay of the disturbance velocity along the rear stagnation streamline, until distances of $O(l_s)$, is in contrast to \citet{shaik2020far} above, who found the decay downstream to be faster. The latter is likely due to the limited spatial extent of the computational domain considered - Fig.1b in \citet{shaik2020far} shows the downstream axial velocity only until the first zero-crossing (i.e. till $z/a \approx 0.9l_c$). 
\citetalias{varanasi2022motion} also provided estimates for the total drift volume. A discussion of the literature considering drops, active/passive particles, and the drag induced by their motion has been provided 
in the recent review by \citet{more2023motion}.

While \citetalias{varanasi2022motion} was one of the first efforts to characterize the asymptotic structure of the far-field velocity disturbance at large $Pe$, herein we show that this characterization is nevertheless incomplete, and that the disturbance fields have a richer asymptotic structure. The reverse jet identified in \citetalias{varanasi2022motion} is not an isolated feature, but merely the central part of a configuration of concentric near-vertical recirculating cells that decrease in thickness, and increase in number, with increasing downstream distance. Identification of this columnar structure leads to the emergence of a tertiary screening length, corresponding to the length of the said cells. In \citetalias{varanasi2022motion}, the velocity along the rear stagnation streamline transitioned from a Stokeslet to a reverse-Stokeslet, and then to an exponential decay, for any non-zero $\beta_\infty$. Herein, we show that the tertiary screening length leads to a further transition from the exponential to an eventual algebraic\,($1/z^7$) decay; this algebraic decay being the same as that at sufficiently long distances along the front stagnation streamline, albeit with a different numerical prefactor. Finally, although \citetalias{varanasi2022motion} presented streamline and isopycnal patterns for a range of $\beta_\infty$, the spatial extents of these plots were limited. The limitation was quite severe for the smallest $\beta_\infty$'s, due to convergence issues in the numerical integration technique. For example, for $\beta_\infty = 10^{-5}$, isopycnals were only provided in a domain with dimensions of 16$l_c$ and 5$l_c$ in the vertical and horizontal directions, respectively; the flow field for $\beta_\infty = 0$ was not provided for the same reason. As already mentioned above, the nature of the $\beta_\infty = 0$ flow field at large distances is particularly important for the drift volume calculation. As part of the current effort, we simplify the inverse transform integrals using contour integration, enabling us to obtain the flow field and iso-pycnals over much larger distances, and over a wide range of $Pe$, including $Pe = \infty\,(\beta_\infty = 0)$. The salient features of the flow field based on the analysis of VS22, and those emerging from the analysis in the following sections, are summarized in the pair of schematics in figure \ref{fig:1}.\\

In \S~\ref{sec:Theory}, we first provide the governing equations, and then, the integral expressions for the velocity and density fields resulting from the linearized Fourier-transformed governing equations. Next, we discuss the reduction of these expressions to double integrals using contour integration, and the numerical method used to evaluate the double integrals. In \S~\ref{sec:3.1}, we briefly present the main results of \citetalias{varanasi2022motion} which provide the context for the investigation here. In \S~\ref{sec:3.2}, using asymptotic evaluations of the integrals, valid at large distances behind the translating sphere, we move beyond the findings of \citetalias{varanasi2022motion} and characterize the columnar structure downstream of the translating sphere, providing expressions for its radial extent, the number of cells within, and the tertiary screening length\,($l_t$). The columnar structure ends at a distance of $O(l_t)$, leading to the disturbance fields transitioning from exponential to algebraically decaying functions of the downstream distance. The significance of the various screening lengths are illustrated via plots of the streamlines and isco-pycnals, in the downstream region, over a range of small $\beta_\infty$ values, including $\beta_\infty = 0$. In \S~\ref{sec:3.3}, similar to \citetalias{varanasi2022motion}, we present the streamline and isopycnal patterns for a wide range of $\beta_\infty$, but over a much larger domain - the domain extends to 25$l_c$ in the radial direction, and to $\max(l_t,100l_c)$ in the vertical\,(axial) direction. We present a summary of our main findings in \S~\ref{sec:conclusions}, and end with a pair of figures, each with two halves, that help contrast the streamline and iso-pycnal patterns for large and small $\beta_\infty$.

\section{Governing equations and the disturbance fields}\label{sec:Theory}

We consider a sphere of radius $a$ settling with velocity $U$ across a uniformly (stably)\,stratified fluid, that is, with $d\rho/dz = -\gamma$ ($\gamma$ being a positive constant in the absence of the sphere). The non-dimensional governing equations, with the Boussinesq approximation, are given by \citep{zvirin1975settling, ardekani2010stratlets, varanasi2022motion}:
\begin{equation}
\label{eq:2.1}
    \nabla\cdot\textbf{u} = 0
\end{equation}
\begin{equation}
\label{eq:2.2}
    Re[\textbf{u}\cdot\nabla\textbf{u}] = -\nabla p + \nabla^2\textbf{u} - Ri_v\rho_f\textbf{1}_3,
\end{equation}
\begin{equation}
\label{eq:2.3}
    1-w+\textbf{u}\cdot\nabla\rho_f = \dfrac{1}{Pe}\nabla^2\rho_f,
\end{equation}
in a reference frame translating with the sphere. Here, $\rho_f$ is the density disturbance field, and $\textbf{u}$ is the velocity field, $w$ being its vertical component. The length, velocity and density scales used to non-dimensionalize the relevant variables in \eqref{eq:2.1}-\eqref{eq:2.3} are $a$, $U$ and $\gamma a$, respectively. The no-slip and no-flux boundary conditions are
\begin{equation}
\label{eq:2.4}
    \textbf{u} = 0\hspace{0.5cm}\textrm{and}\hspace{0.5cm}\textbf{n}\cdot\nabla\rho_f = -\textbf{n}\cdot\textbf{1}_3\hspace{0.5cm}\textrm{at}\hspace{0.5cm} r = 1,
\end{equation}
and the far-field boundary conditions are
\begin{equation}
\label{eq:2.5}
    \textbf{u} \to \textbf{1}_3\,(w \to 1), \hspace{0.5cm} p \to p_\infty, \hspace{0.5cm}\textrm{and}\hspace{0.5cm}\rho_f \rightarrow 0\hspace{0.5cm}\textrm{for}\hspace{0.5cm} r  \to \infty,
\end{equation}
where $r = |\boldsymbol{x}|$, $\boldsymbol{x}$ being the position relative to the sphere center. The non-dimensional parameters in \eqref{eq:2.1}-\eqref{eq:2.3} have already been defined in the introduction section. 
In writing \eqref{eq:2.1}-\eqref{eq:2.5}, we assume a quasi-steady state to have developed in the region of interest surrounding the translating sphere. 

Our primary focus is on large $Pe$ in the inertialess limit\,($Re = 0$). For $Pe\to\infty$, the diffusive term in \eqref{eq:2.3} is asymptotically small compared to the convective one. Hence, using $\textbf{u}\cdot\nabla\rho_f\sim (1-w)$, along with the Stokes equations, the large-$Pe$ stratification screening length can be derived as $l_c \sim \mathcal{O}(aRi_v^{-1/3})$. Using this as the characteristic length scale, and further, restricting our attention to the far-field\,(the outer region) where the sphere appears as a point force, \eqref{eq:2.1}-\eqref{eq:2.3} can be written in the form
\begin{equation}
\label{eq:2.6}
    \tilde{\nabla}\cdot\textbf{\~{u}} = 0,
\end{equation}
\begin{equation}
\label{eq:2.7}
    -\alpha_\infty\dfrac{\partial{\textbf{\~u}}}{\partial\tilde{z}} = -\tilde{\nabla} \tilde{p} + \tilde{\nabla}^2\textbf{\~{u}} - [\tilde{\rho}_f+6\pi\delta(\textbf{\~{r}})]\textbf{1}_3,
\end{equation}
\begin{equation}
\label{eq:2.8}
    -\textbf{1}_3\cdot\textbf{\~{u}}+\dfrac{\partial \tilde{\rho}_f}{\partial\tilde{z}} = \beta_\infty\tilde{\nabla}^2\tilde{\rho}_f.
\end{equation}
Here, $\textbf{\~{r}} = Ri_v^{1/3}\textbf{r}$ is the outer-region coordinate, $\textbf{\~{u}} = Ri_v^{-1/3}(\textbf{u}-\textbf{1}_3)$, $\tilde{p} = Ri_v^{-2/3}(p-p_\infty)$ and $\tilde{\rho}_f = \rho_f$. The delta-function forcing on the RHS of (\ref{eq:2.7}) replaces the no-slip condition above, and the system (\ref{eq:2.6}-\ref{eq:2.8}) is only required to satisfy far-field decay conditions. There are three screening lengths associated with the dimensionless parameters in (\ref{eq:2.1}-\ref{eq:2.3}) - the inertial screening length $aRe^{-1}$, the stratification screening length $aRi_v^{-1/3}$ and the screening length $aPe^{-1}$ associated with the convection of density perturbations (for small $Pe$). The ratios of these taken pairwise,  viz. $\alpha_\infty = \frac{Re}{Ri_v^{1/3}}$ and $\beta_\infty = \frac{Ri_v^{1/3}}{Pe}$, are the parameters in (\ref{eq:2.6}-\ref{eq:2.8}). We study the nature of the disturbance flow field and iso-pycnals, as a function of $\beta_\infty$, assuming $\alpha_\infty = 0$.

Fourier transforming (\ref{eq:2.6}-\ref{eq:2.8}), one obtains the disturbance fields in terms of the following Fourier integrals: 
\begin{equation}
\label{eq:2.9}
    \textbf{\~{u}}(\textbf{\~{r}}) = \dfrac{-3}{4\pi^2}\displaystyle\int\dfrac{(ik_3+\beta_\infty k^2)k^2(\textbf{1}_3-\frac{k_3\textbf{k}}{k^2})}{(ik_3+\beta_\infty k^2)k^4+k_t^2}e^{i\textbf{k}\cdot\textbf{\~{r}}}d\textbf{k}, \vspace{-0.5cm}
\end{equation}
\begin{equation}
\label{eq:2.10}
    \tilde{\rho}_f(\textbf{\~{r}}) = \dfrac{-3}{4\pi^2}\displaystyle\int\dfrac{k_t^2}{(ik_3+\beta_\infty k^2)k^4+k_t^2}e^{i\textbf{k}\cdot\textbf{\~{r}}}d\textbf{k}.
\end{equation}
Here, $k^2 = k_1^2+k_2^2+k_3^2$ and $k_t^2 = k_1^2+k_2^2$. Exploiting axisymmetry with respect to the vertical direction (the direction of particle translation) in physical space and the absence of a swirl, one may define a Stokes streamfunction $\tilde{\psi}_s$, with the radial\,($\tilde{u}_r$) and axial\,($\tilde{u}_z$) velocity components related to $\tilde{\psi}_s$ as $\tilde{u}_r = -\dfrac{1}{\tilde{r}_t}\dfrac{\partial \tilde{\psi}_s}{\partial \tilde{z}}$ and $\tilde{u}_z = \dfrac{1}{\tilde{r}_t}\dfrac{\partial \tilde{\psi}_s}{\partial \tilde{r}_t}$ in cylindrical coordinates ($\tilde{r}_t$,$\tilde{z}$). The expression for the Stokes streamfunction can then be written as
\begin{equation}
\label{eq:psis}
    \tilde{\psi}_s(\textbf{\~{r}}) = \dfrac{3\tilde{r}_t i}{4\pi^2}\displaystyle\int\dfrac{(ik_3+\beta_\infty k^2)k_2}{(ik_3+\beta_\infty k^2)k^4+k_t^2}e^{i\textbf{k}\cdot\textbf{\~{r}}}d\textbf{k}.
\end{equation}

The triple integrals in $\textbf{k}$ in (\ref{eq:2.9}-\ref{eq:psis}) are first reduced to double integrals, which can then be evaluated numerically to obtain the velocity\,(or Stokes streamfunction) and density fields. This is done using spherical polar coordinates $(k,\theta, \phi)$ in Fourier space with the polar axis along $\textbf{1}_3$. \citetalias{varanasi2022motion} first evaluated the integral over $\phi$, but this led to an integrand with an oscillatory dependence on $k$, in turn leading to numerical difficulties. To circumvent this issue, we first evaluate the $k$-integral using contour integration. The resulting double integral over $\theta$ and $\phi$ has finite integration limits, and the associated integrand is non-oscillatory. The expressions for $\tilde{\psi}_s$ and density $\tilde{\rho}_f$ obtained in this manner are given by
\begin{equation}
    \label{eq:2.11}
    \tilde{\psi}_s (\tilde{r}_t,\tilde{z}) = \dfrac{3\tilde{r}_t}{2\pi}(T_{\psi 1} + T_{\psi 2}),\hspace{1cm} \tilde{\rho}_f(\tilde{r}_t,\tilde{z}) = \dfrac{-3}{2\pi}(T_{\rho 1} + T_{\rho 2}).
\end{equation}
where, for $\beta_\infty \neq 0$, $T_{\psi 1}, T_{\psi 2}, T_{\rho 1}$ and $T_{\rho 2}$ are defined as follows: 
\begin{equation}
\label{eq:2.12}
    \begin{rcases}
        & T_{\psi 1} = \pi\displaystyle\int_0^1dy(1-y^2)^{\frac{3}{2}}\sum_{m=p}^{p+1}\dfrac{\left(i s J_1(\tilde{r}_t \kappa_m\sqrt{1-y^2})+H_{-1}(\tilde{r}_t \kappa_m\sqrt{1-y^2})\right)e^{i \kappa_m\tilde{z}y}}{\kappa_m^3(3iy+4\beta_\infty \kappa_m)},\\ \vspace{0.3cm}
        & T_{\psi 2} = 2\displaystyle\int_0^1dx\bigg[\int_0^{y_c}dy\sum_{m=q}^{q+1}-\int_{y_c}^1dy\sum_{m=p}^{p+1}\bigg]\dfrac{(1-y^2)^{\frac{3}{2}}e^{-i \kappa_m\delta}}{\kappa_m^3(3iy+4\beta_\infty \kappa_m)},\\ \vspace{0.3cm}
        & T_{\rho 1} = \pi i \displaystyle\int_0^1dy(1-y^2)\sum_{m=p}^{p+1}\bigg[\dfrac{\big[sJ_0(\kappa_m\tilde{r}_t\sqrt{1-y^2})+ i H_0(\kappa_m\tilde{r}_t\sqrt{1-y^2})\big]e^{i\kappa_m\tilde{z}y}}{3iy+ 4\beta \kappa_m}\bigg],\\ \vspace{0.3cm}
        & T_{\rho 2} = 2 i \displaystyle\int_0^1\dfrac{dx}{\sqrt{1-x^2}}\bigg[-\int_0^{y_c}dy\sum_{m=q}^{q+1}+\int_{y_c}^1dy\sum_{m=p}^{p+1}\bigg]\dfrac{s(1-y^2)e^{-i\kappa_m\delta}}{3iy+ 4\beta \kappa_m},
    \end{rcases}
\end{equation}
with $\delta = s \tilde{r}_t \sqrt{1-x^2} \sqrt{1-y^2} - \tilde{z} y$ and $y_c = \tilde{r}_t\sqrt{1-x^2}\Big/\sqrt{\tilde{z}^2+\tilde{r}_t^2(1-x^2)}$; note that the $\phi$-integral for one of the terms in the integrand can be evaluated analytically, and leads to the single-integral contributions\,($T_{\psi 1}$ and $T_{\rho 1}$) above. 
$J_n$ and $H_n$ in (\ref{eq:2.12}) are, respectively, the Bessel and Struve functions of the first kind of order $n$ \citep{abramowitz1988handbook}. The variables $x$ and $y$, used in \eqref{eq:2.12} for purposes of notational simplicity, are given by $x = \cos\phi$ and $y = \cos\theta$.  The $\kappa_m$'s\,($m=1\!-\!4$) in \eqref{eq:2.12} denote the roots of the quartic polynomial, $ik^3 y + \beta_\infty k^4 + (1-y^2)$ that appears in the denominator of the Fourier integrals, and arise as part of the contour integration. 
The indices, that appear in the summations in (\ref{eq:2.12}), are defined as $(p,q,s) \equiv (1,3,1)$ for $z > 0$ and $(p,q,s) \equiv (3,1,-1)$ for $z<0$; the indices number the roots in terms of their position in the complex$-k$ plane, taken in an anticlockwise manner starting from the first quadrant. A detailed explanation of the contour integration is given in Appendix \ref{sec:AppA}.

The limit $\beta_\infty \to 0$ is a singular one when the quartic polynomial in $k$ degenerates to a cubic one, with one of the aforementioned four roots receding to infinity. This prevents an explicit evaluation of the associated residue, and we therefore directly substitute $\beta_\infty = 0$ in (\ref{eq:2.9}-\ref{eq:2.10}), and then implement the contour integration. Subsequent simplifications lead to the following alternate expressions for $T_{\psi 1}$, $T_{\psi 2}$, $T_{\rho 1}$ and $T_{\rho2}$ for $\beta_\infty = 0$:
\begin{equation}
\label{eq:2.13}
    \begin{rcases}
        & T_{\psi 1} = \dfrac{1}{3}\displaystyle \int_0^1dx\int_0^1dy\sqrt{1-y^2}\left(3-\Psi_1(\bar{\delta})\Theta(\tilde{z})-\Psi_2(\delta)\Theta(-\tilde{z})\right),\\ \vspace{0.2cm}
        & T_{\psi 2} = \dfrac{1}{3}\displaystyle \int_0^1dx \bigg[\int_0^{y_c}dy\sqrt{1-y^2}\left(3-\Psi_1(\bar{\delta})\Theta(-\tilde{z})-\Psi_2(\delta)\Theta(\tilde{z})\right)\\ &\hspace{3cm}-\displaystyle\int_{y_c}^1dy\sqrt{1-y^2}\left(3-\Psi_1(\delta)\Theta(\tilde{z})-\Psi_2(|\bar{\delta}|)\Theta(-\tilde{z})\right)\bigg],\\ \vspace{0.2cm}
        & T_{\rho 1} = \dfrac{1}{3}\displaystyle \int_0^1dx\int_0^1dy\dfrac{(1-y^2)}{y\sqrt{1-x^2}}\left(\Psi_1(\bar{\delta})\Theta(\tilde{z})-\Psi_2(\delta)\Theta(-\tilde{z})\right),\\ \vspace{0.2cm}
        & T_{\rho 2} = \dfrac{1}{3}\displaystyle \int_0^1dx \bigg[\int_0^{y_c}dy\dfrac{(1-y^2)}{y\sqrt{1-x^2}}\left(\Psi_1(\bar{\delta})\Theta(-\tilde{z})-\Psi_2(\delta)\Theta(\tilde{z})\right)\\ &\hspace{3cm}\displaystyle+\int_{y_c}^1dy\dfrac{(1-y^2)}{y\sqrt{1-x^2}}\left(\Psi_1(\delta)\Theta(\tilde{z})-\Psi_2(|\bar{\delta}|)\Theta(-\tilde{z})\right)\bigg],
    \end{rcases}
\end{equation}
where, 
\begin{equation*}
    \Psi_1(\Delta) = 4\exp\left(-\frac{\Delta}{2}\sqrt[3]{\frac{1-y^2}{y^2}}\right)\cos\left(\frac{\Delta \sqrt{3}}{2}\sqrt[3]{\frac{1-y^2}{y^2}}\right), \hspace{1cm} \Psi_2(\Delta) = 2\exp\left(-\Delta\sqrt[3]{\frac{1-y^2}{y^2}}\right)
\end{equation*}
\begin{equation*}
    \bar{\delta} = s \tilde{r}_t \sqrt{1-x^2} \sqrt{1-y^2} + \tilde{z} y, \hspace{1cm} {\delta} = s \tilde{r}_t \sqrt{1-x^2} \sqrt{1-y^2} - \tilde{z} y,
\end{equation*}
and $\Theta(\tilde{z})$ is the Heaviside function. Other variables, including $s$ in particular, have the same definitions as in \eqref{eq:2.12}. In obtaining (\ref{eq:2.13}), the $k-$integrals were evaluated using relations 3.738.1 and 3.738.2 of  \cite{gradstein2007table}. 

We perform the numerical integrations in \eqref{eq:2.12} and \eqref{eq:2.13}, using the in-built integration routine ``Clenshaw-Curtis rule'' in {\it Mathematica} and vary the `Working Precision' option to obtain the desired precision of the result. Because of the smallness of the result at large $\tilde{z}$, the variable precision arithmetic in {\it Mathematica} is well-suited to this problem.

\section{Towards a complete flow picture}\label{sec:CompletePicture}

In this section, we first present the important results of \citetalias{varanasi2022motion} (in \S~3.1), which helps set the stage for reporting our findings starting from \S~3.2. In \S~3.2, we derive expressions for the boundary of, and the number of recirculating cells within, the vertical columnar structure that arises behind the translating sphere in the non-diffusive ($\beta_\infty = 0$) limit. Next, for non-zero $\beta_\infty$, we identify an additional length scale, the tertiary screening length, beyond which the columnar structure ceases to exist. In \S~3.3, we put together the complete picture using streamline and isopycnal patterns as a function of $\beta_\infty$.

\subsection{Varanasi and Subramanian (2022)} \label{sec:3.1}
For $Re \ll Ri_v^{1/3}$ and $Pe \gg 1$, corresponding to the large-$Pe$ limit of the Stokes-stratification regime, \citetalias{varanasi2022motion} evaluated the $\phi-$integral in (\ref{eq:2.10}) and (\ref{eq:psis}) to obtain the following double integrals for the Stokes streamfunction and density fields\,(see \citetalias{varanasi2022motion}-3.4 and \citetalias{varanasi2022motion}-3.2): 
\begin{equation}
    \tilde{\psi}_s = \displaystyle \dfrac{-3\tilde{r}_t^2}{4\sqrt{\tilde{r}_t^2+\tilde{z}^2}} + \dfrac{3\tilde{r}_t}{2\pi}\int_0^\infty dk \int_0^\pi d\theta \dfrac{\sin^4\theta J_1\left(k\tilde{r}_t\sin\theta\right)e^{ik\tilde{z}\cos\theta}}{k\left(ik^3\cos\theta + \beta_\infty k^4 +\sin^2\theta\right)}, \label{eq:3.1}
\end{equation}
\begin{equation}
    \tilde{\rho}_f = \displaystyle \dfrac{-3}{2\pi}\int_0^\infty dk \int_0^\pi d\theta \dfrac{k^2\sin^3\theta J_0\left(k\tilde{r}_t\sin\theta\right)e^{ik\tilde{z}\cos\theta}}{\left(ik^3\cos\theta + \beta_\infty k^4 +\sin^2\theta\right)}. \label{eq:3.2}
\end{equation}
The first term in \eqref{eq:3.1} is the Stokeslet contribution written out separately to facilitate numerical integration of the second term. The integrands in \eqref{eq:3.1}\,(second term) and \eqref{eq:3.2} decay as $1/k^{9/2}$ and $1/k^{5/2}$ for $k \to \infty$, for any non-zero $\beta_\infty$; the slower decay of the density integrand increases the difficulty of evaluation at larger spatial distances. The presence of $\tilde{r}_t$ and $\tilde{z}$ in the arguments of the Bessel and complex exponential functions implies an increasingly oscillatory integrand at large spatial distances, again contributing to an increased difficulty of calculation.

The need to retain the full complex exponential in the integrands of  \eqref{eq:3.1} and \eqref{eq:3.2} is consistent with a fore-aft asymmetry of the disturbance fields. This was evident in the different numbers of recirculating cells in front of, and behind the translating sphere, in the streamline and isopycnal patterns obtained from numerically evaluating \eqref{eq:3.1} and \eqref{eq:3.2} - refer to figure 2 of \citetalias{varanasi2022motion}, and to figures \ref{fig:4} and \ref{fig:6} below. Since the method used made numerical integration difficult at large distances, the authors also conducted an asymptotic analysis of the disturbance fields on scales larger than $\mathcal{O}(a Ri_v^{-\frac{1}{3}})$. In the wake region where the flow is predominantly horizontal, the approximation $k_t \ll k_3$ was used\,(see \eqref{eq:2.10} and \eqref{eq:2.11}), rendering one of the integrals readily evaluable via contour integration. The remaining 1D integral revealed a self-similar wake structure with a width that grows as $\tilde{z} \sim \tilde{r}_t^{2/5}$. Evaluating this integral led to a quantitative characterization of the fore-aft asymmetry, and in particular, the different rates of decay of the disturbance fields in the upstream and downstream regions outside the wake; these algebraic asymptotes have been given in table \ref{table2}.

\begin{center}
\captionof{table}{the large-$Pe$ wake region asymptotes; the wake region has a self-similar structure with the similarity variable defined by $\tilde{\eta} = \tilde{z}/\tilde{r}_t^{2/5}$.}
    \begin{tabular}{cccc}
    \hline
    $\tilde{z}$ & $\tilde{u}_r$ & $\tilde{u}_z$ & $\tilde{\rho}_f$\\
    \hline 
    +ve  & $11340/\tilde{r}_t^{11/5} \tilde{\eta}^8$ & $3240/\tilde{r}_t^{14/5}\tilde{\eta}^7$ & $-540/\tilde{r}_t^{12/5} \tilde{\eta}^6$ \\
    -ve  & $-7560/\tilde{r}_t^{11/5} \tilde{\eta}^8$ & $-2160/\tilde{r}_t^{14/5} \tilde{\eta}^7$ & $360/\tilde{r}_t^{12/5} \tilde{\eta}^6$\\
    \hline
    \label{table2}
\end{tabular}
\end{center}
The authors used the complementary approximation of the flow being predominantly vertical, corresponding to $k_t\gg k_3$, to examine what they termed the `reverse-jet' behind the sphere. The resulting 1D integral permitted analytical evaluation only along the rear stagnation streamline $(\tilde{r}_t = 0, \tilde{z} > 0)$, giving the axial velocity and density disturbance fields in terms of modified Bessel functions of the second kind, of orders 1\,($\tilde{u}_z \approx 3\beta_\infty^{1/2}K_1[2\beta_\infty^{1/2}\tilde{z}]$) and 0\,($\tilde{\rho}_f \approx -3K_0[2\beta_\infty^{1/2}\tilde{z}]$), respectively. The small-argument limiting form of $K_1$ gave rise to the reverse-Stokeslet decay $(\tilde{u}_z \sim 3/2\tilde{z}$), with its large-argument limiting form leading to an exponentially decaying velocity field\,($\tilde{u}_z \sim \exp(- 2\beta_\infty^{1/2} \tilde{z})/\tilde{z}^{1/2})$. The transition between these decay regimes occurred at $\tilde{z} \sim \mathcal{O}(\beta_\infty^{-1/2})$ which, in dimensional terms, corresponds to the secondary screening length $l_s \sim \mathcal{O}(aRi_v^{-1/2}Pe^{1/2})$. While the large-argument form of $K_0$ again led to an exponential decay of the density field\,($\tilde{\rho}_f \sim \exp(- 2\beta_\infty^{1/2} \tilde{z})/\tilde{z}^{1/2}$), interestingly, the small-argument limit revealed a logarithmic behavior; the density disturbance is logarithmically singular along the rear stagnation streamline for $\beta_\infty = 0$. The exponential decay for both $\tilde{u}_z$ and $\tilde{\rho}_f$ arises from diffusive effects that start to smear out the jet.

\subsection{The vertical columnar structure behind the translating sphere}\label{sec:3.2}
\begin{figure}
    \hspace{-0.5cm}
    \begin{minipage}[t]{0.56\textwidth}
      \centering
      \includegraphics[width=\textwidth]{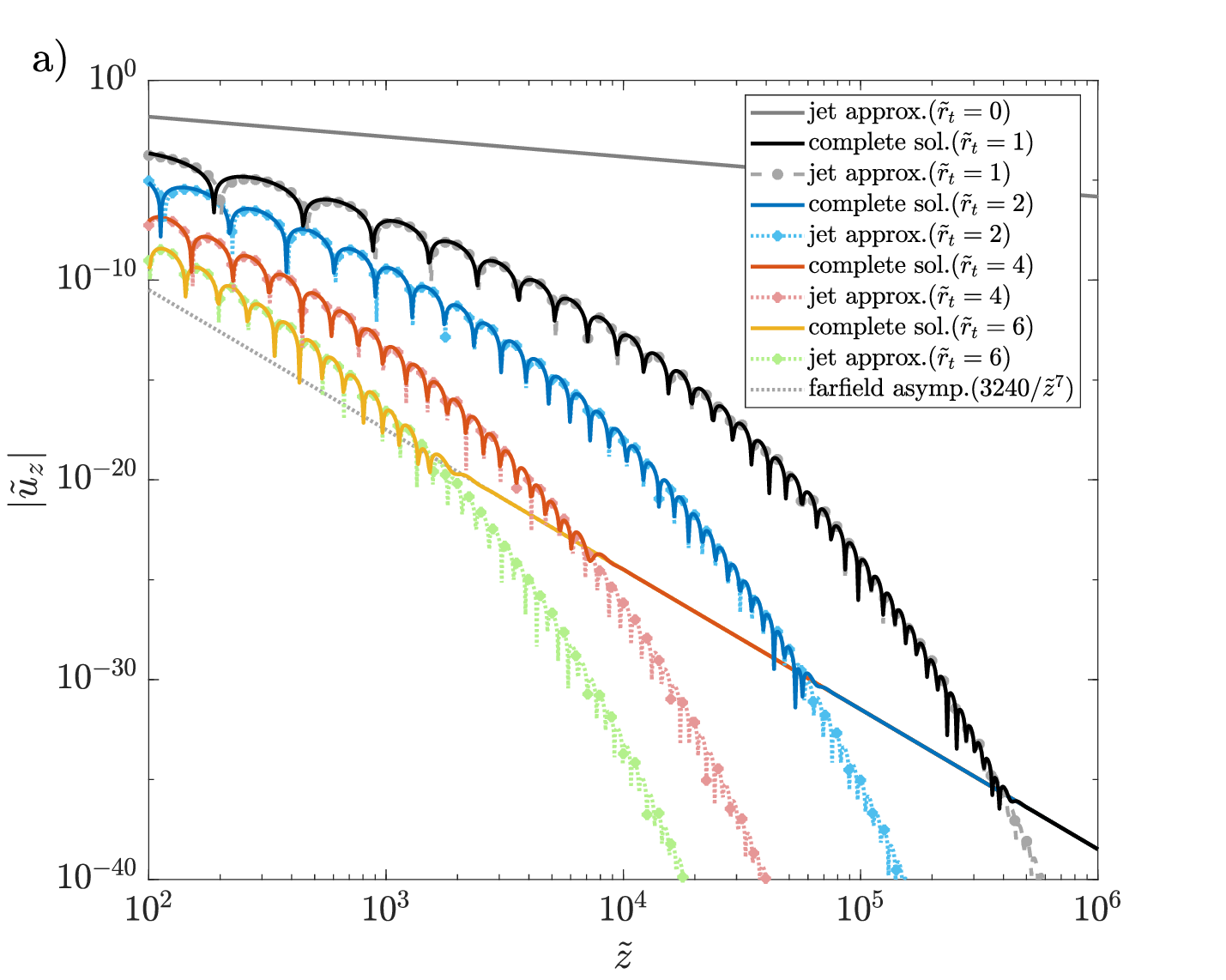}
    \end{minipage}
     \hspace{-0.4cm}
    \begin{minipage}[t]{0.54\textwidth}
      \centering
      \includegraphics[width=\textwidth]{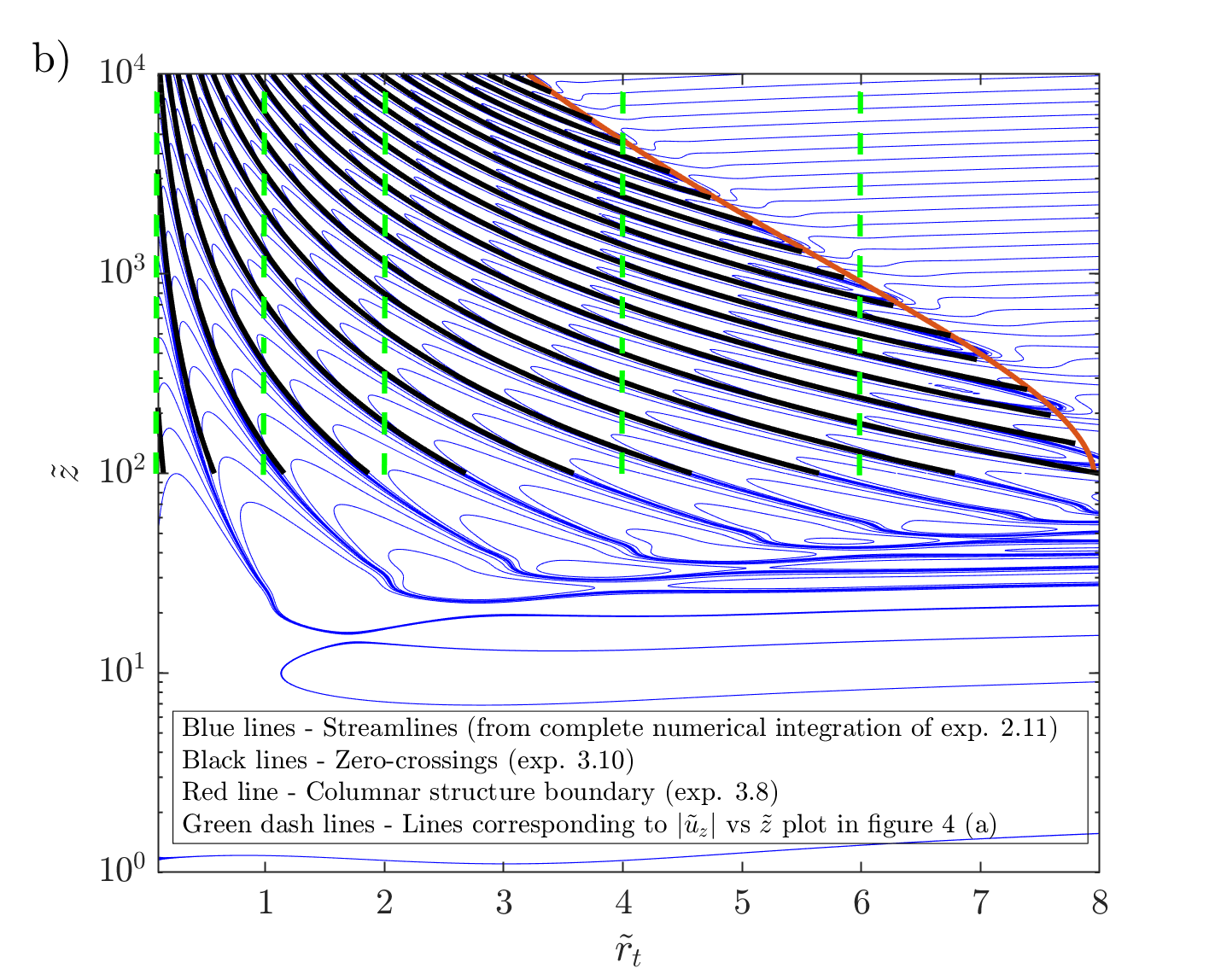}
    \end{minipage}
  \caption{The downstream axial velocity disturbance \,($\tilde{u}_z$) and streamline pattern for $\beta_\infty = 0$. (a) $|\tilde{u}_z|$ plotted as a function of $\tilde{z}\,(> 0)$. The numerical results\,(bright-colored solid curves), obtained from evaluating the inverse transform integral in (\ref{eq:2.11}) are compared to the asymptotic approximation derived from \eqref{eq:3.6}\,(pale-colored dashed curves with dots added for better differentiation) for different non-zero $\tilde{r}_t$. The reverse-Stokeslet decay for $\tilde{r}_t = 0$ is shown as a solid gray line, while the (common)\,algebraic asymptote, $3240/\tilde{z}^7$, apppears as a dashed gray line.; (b) The streamline pattern for $\tilde{z} > 1$, depicting the columnar structure behind the translating sphere. The boundary of the columnar structure, which is the red curve connecting the ends of the cell boundaries\,(black curves) above the wake region\,($\tilde{z} \gtrsim 100$), is given by by \eqref{eq:3.7}.}
  \label{fig:2}
\end{figure}
As mentioned in the previous subsection, \citetalias{varanasi2022motion} obtained analytical forms for the disturbance fields along the rear stagnation streamline, allowing them to infer contrasting behavior on either side of the secondary screening length. For non-zero $\tilde{r}_t$ and $\tilde{z} > 0$, the 1D integrals for the Stokes streamfunction and the density field, with the aforementioned jet approximation ($k_t \gg k_3$), are given by:

\begin{equation}
    \tilde{\psi}_s= \displaystyle 3\tilde{r}_t\int_0^\infty dk \dfrac{J_1(k \tilde{r}_t)e^{-\tilde{z}(\beta_\infty k_t^2 + 1/k_t^2)}}{k^4}, \label{eq:3.3}
\end{equation}
\begin{equation}
    \tilde{\rho}_f= -\displaystyle 3\int_0^\infty dk \dfrac{J_0(k \tilde{r}_t)e^{-\tilde{z}(\beta_\infty k_t^2 + 1/k_t^2)}}{k}. \label{eq:3.4}
\end{equation}

While the structure of the flow field away from the rear stagnation streamline can be understood from numerically evaluating (\ref{eq:3.3}), as was done in \citetalias{varanasi2022motion}\,(see Figure 7 therein), important features are better understood using an analytical expression. The integral in (\ref{eq:3.3}) can be evaluated in closed form 
for $\beta_\infty = 0$, and for $\tilde{r}_t\tilde{z}^{1/2} \gg 1$, using a steepest descent method\,\citep{bender2013advanced}; details of this calculation are given in Appendix \ref{sec:AppB1}. One obtains the following expression for the Stokes streamfunction:
\begin{align}
\label{eq:3.5}
    \tilde{\psi}_s|_{\beta_\infty = 0} &= 3\tilde{r}_t\displaystyle\int_0^\infty dk \dfrac{J_1(k \tilde{r}_t)e^{-\tilde{z}/k^2}}{k^4},  \\ &\approx \dfrac{\sqrt{3}}{\sqrt[3]{2}}\left(\tilde{r}_t^{4/3}\tilde{z}^{-4/3}\right)\exp\left(-\dfrac{3}{2\sqrt[3]{4}} \left(\tilde{r}_t\tilde{z}^{1/2}\right)^{2/3}\right)\sin\left(\dfrac{3\sqrt{3}}{2\sqrt[3]{4}}\left(\tilde{r}_t\tilde{z}^{1/2}\right)^{2/3}+\dfrac{\pi}{3}\right).\label{eq:3.6}
\end{align}
The sinusoidal term in (\ref{eq:3.6}) is indicative of a cellular structure with the flow direction alternating between adjacent cells, and with the exponential pre-factor determining the manner of decay of these oscillations from the central to the peripheral cells. 
The cell widths 
may be correlated to the distance between successive zero-crossings of the sine, and therefore scale as $\tilde{z}^{-1/2}$. The argument, $(\tilde{r}_t\tilde{z}^{1/2})^{2/3}$, of the sine implies that, at a fixed $\tilde{z}$, the cell widths, when projected onto the horizontal plane, increase with increasing $\tilde{r}_t$; on the other hand, a given cell becomes increasingly columnar\,(that is, approaches a vertical orientation) with increasing $\tilde{z}$. Only the central portion of the above columnar structure, corresponding to radial distances less than that of the first zero-crossing, was identified in \citetalias{varanasi2022motion} as the reverse jet. Although formally valid only for $\tilde{r}_t \tilde{z}^{1/2} \gg 1$, the location of the first zero crossing obtained from (\ref{eq:3.6}), given by $\tilde{r}_t\tilde{z}^{1/2}  = (10\pi\sqrt[3]{4}/9\sqrt{3})^{3/2}$, nevertheless gives a good approximation for the reverse-jet boundary.

In figure \ref{fig:2}a, the axial velocity field obtained from the steepest descent approximation, (\ref{eq:3.6}), is compared to a numerical evaluation of the original inverse transform integral in (\ref{eq:2.11}) using (\ref{eq:2.13}) for $\beta_\infty = 0$. The comparison is as a function of $\tilde{z}$ for different $\tilde{r}_t$, starting from $\tilde{z} = 100$. The latter choice ensures that $\tilde{r}_t\tilde{z}^{1/2}$ remains large for the chosen non-zero $\tilde{r}_t$'s, and (\ref{eq:3.6}) therefore remains valid; for $\tilde{r}_t$ = 0, (\ref{eq:3.6}) remains inapplicable regardless of $\tilde{z}$, but the decay is known to have a reverse-Stokeslet character in this case. The numerical results, for non-zero $\tilde{r}_t$, are seen to agree with the analytical prediction up to a critical $\tilde{z}$\,(that decreases with increasing $\tilde{r}_t$), there being a subsequent departure
due to the numerical results transitioning to an algebraic decay for larger $\tilde{z}$\,(implying the absence of recirculating cells at these distances). This algebraic decay is the same as the far-field asymptote obtained using the wake approximation, and independent of $\tilde{r}_t$ - see table \ref{table2}. In other words, for a given $\tilde{r}_t \neq 0$, the disturbance velocity field shifts from the oscillatory behavior given by the jet approximation in (\ref{eq:3.6}), to an algebraic decay given by the far-field asymptote\,($\tilde{u}_z = 3240/\tilde{z}^7$) in table \ref{table2}. Interestingly, 
the sum of the aforementioned approximations turns out to match very well with the numerical results over the entire range of $\tilde{z}$, and we exploit this fact to find the boundary of the columnar structure comprising the recirculating cells. This boundary is where the flow changes from being predominantly vertical\,(for smaller $\tilde{z}$) to predominantly horizontal\,(for larger $\tilde{z}$), and is identified by equating the amplitude of the oscillations in \eqref{eq:3.6} to the algebraic asymptote. That is,
\begin{equation}
    \dfrac{1620 \tilde{r}_t^2}{\tilde{z}^7} = \dfrac{\sqrt{3}}{\sqrt[3]{2}}\dfrac{(\tilde{r}_t\tilde{z}^{1/2})^{4/3}}{\tilde{z}^2}\exp\left(-\dfrac{3}{2\sqrt[3]{4}} (\tilde{r}_t\tilde{z}^{1/2})^{2/3}\right). \label{eq:3.66}
\end{equation}
The critical $\tilde{r}_t$ corresponding to the boundary 
can be obtained by solving \eqref{eq:3.66} perturbatively, in the limit 
$\tilde{r}_t\tilde{z}^{1/2} \gg 1$, after taking a logarithm on both sides. One obtains 
(see Appendix \ref{sec:AppB2} for details):
\begin{equation}
\label{eq:3.7}
    \tilde{r}_J = \dfrac{1}{\sqrt{\tilde{z}A^3}}\left[\ln(\tilde{z}^6/B) - \ln\left(\dfrac{1}{A}\left(\ln(\tilde{z}^6/B)\right)\right)\right]^{3/2},
\end{equation}
to second order in the perturbation expansion, with $A = 3/2\sqrt[3]{4}$ and $B = 1620\sqrt[3]{2}/\sqrt{3}$. The radial distance to the boundary\,($\propto \tilde{z}^{-1/2}(\ln \tilde{z})^{3/2}$ at leading logarithmic order), at a fixed downstream distance, is seen to be logarithmically larger than that of an individual cell\,($\propto \tilde{z}^{-1/2}$), which points to the number of recirculating cells $N(\tilde{z})$ being a logarithmically increasing function of $\tilde{z}$. $N$ can be obtained by equating the argument of 
the sinusoidal term in (\ref{eq:3.6}) to $(N+1)\pi$, recognizing that the number of zero crossings is one more than the cell number. Substituting $\tilde{r}_J$ from (\ref{eq:3.7}) leads to:
\begin{equation}
\label{eq:3.77}
    N(\tilde{z}) = \dfrac{\sqrt{3}}{\pi\sqrt{A}}\left[\ln(\tilde{z}^6/B) - \ln\left(\dfrac{1}{A}\ln(\tilde{z}^6/B) \right)\right]^{3/2}-\dfrac{2}{3},
\end{equation}
again to second order in the logarithmic expansion. As expected, (\ref{eq:3.77}) increases logarithmically with downstream distance. 

Expressions \eqref{eq:3.7} and \eqref{eq:3.77} serve to characterize the columnar structure downstream that, despite narrowing down with increasing $\tilde{z}$, nevertheless encloses an increasing number of recirculating cells. This is depicted via the downstream portion of the streamline pattern\,(for $\tilde{z} \geq 1$) for $\beta_\infty = 0$ in figure \ref{fig:2}b, again obtained from a numerical integration of \eqref{eq:2.11} using \eqref{eq:2.13}. The red curve in this figure corresponds to \eqref{eq:3.7}, and evidently demarcates the columnar structure from the predominantly horizontal streamline pattern outside. The black curves in the figure correspond to the zero-crossings of the sine function in \eqref{eq:3.6}, obtained by equating its argument to $(m+1)\pi\,(m=1,2,3,..,)$; this leads to $\tilde{r}_t\tilde{z}^{1/2}  = \left[\left(m\pi+\dfrac{2\pi}{3}\right)\dfrac{2\sqrt[3]{4}}{3\sqrt{3}}\right]^{3/2}$.

\begin{figure}
\centering
    \hspace{-0.5cm}
    \begin{minipage}[t]{0.8\textwidth}
      \centering
      \includegraphics[width=\textwidth]{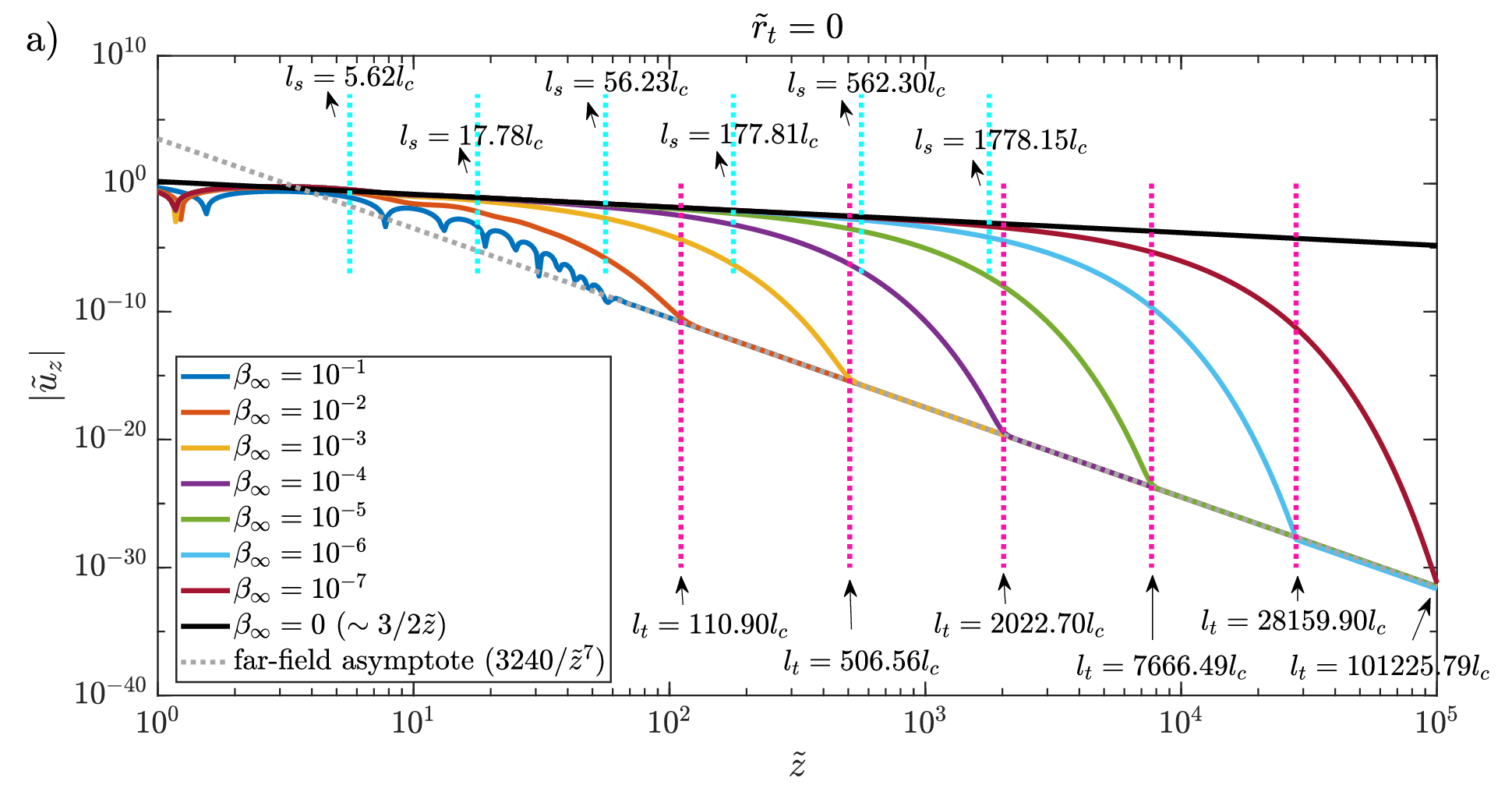}
    \end{minipage}
    
     \hspace{-0.4cm}
    \begin{minipage}[t]{0.8\textwidth}
      \centering
      \includegraphics[width=\textwidth]{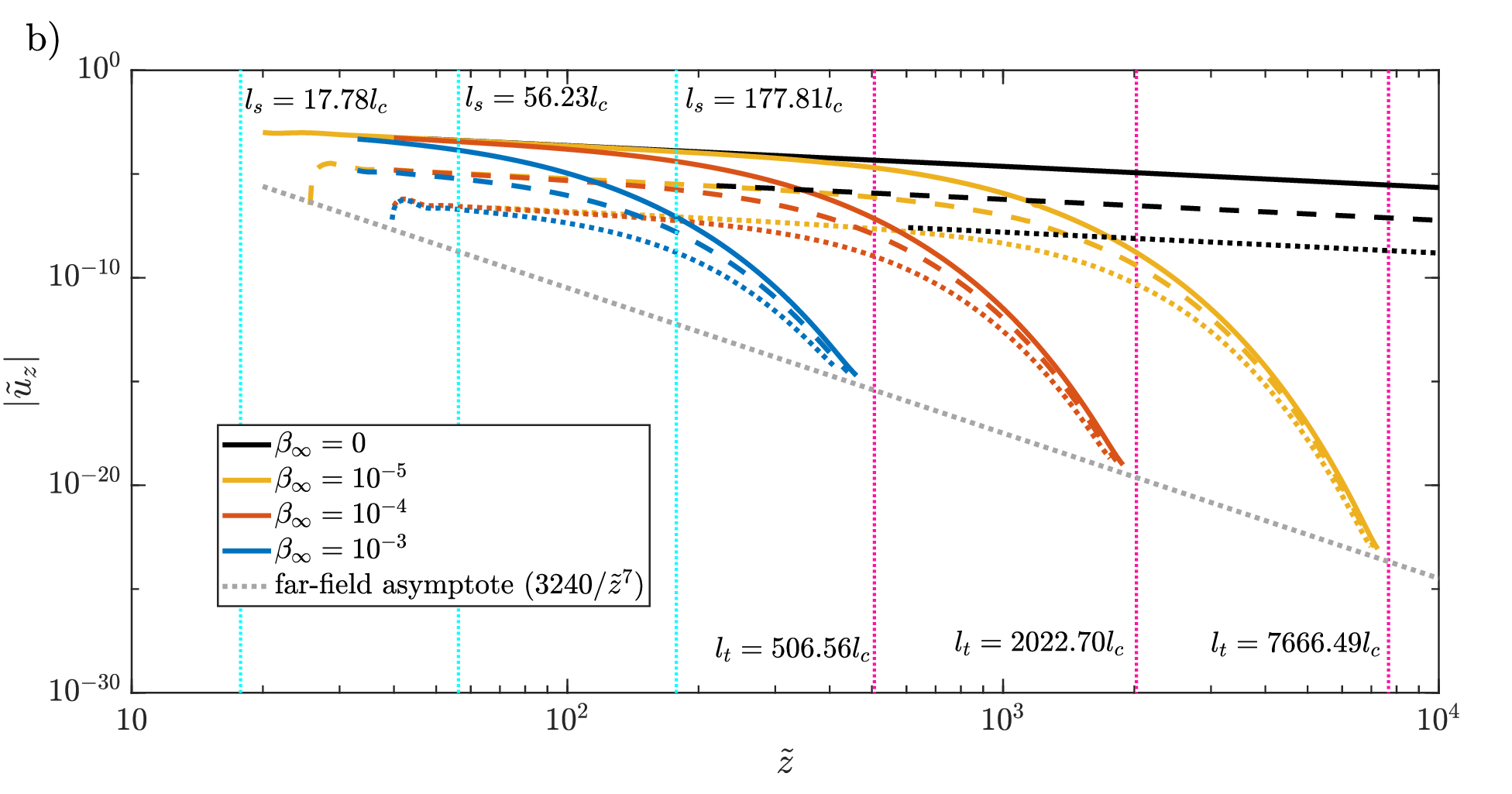}
    \end{minipage}
  \caption{The absolute value of the axial velocity ($|\tilde{u}_z|$) (a) along the rear stagnation streamline; and (b) along the third (continuous), fifth (dashed), and seventh (dotted) zero-crossings (of the Stokes streamfunction), for different $\beta_\infty$, including $\beta_\infty = 0$. For all non-zero $\beta_\infty$, the plots depict the transition from a reverse-Stokeslet decay ($3/2\tilde{z}$) to the far-field asymptote ($3240/\tilde{z}^7$) at large $\tilde{z}$. The secondary and tertiary screening lengths, marking the beginning and end of the exponential decay phase, are shown for each $\beta_\infty$.}
  \label{fig:3}
\end{figure}

For $\beta_\infty = 0$, the axial velocity along the rear stagnation streamline exhibits the reverse-Stokeslet behavior for all distances (much)\,greater than the primary screening length\,($\tilde{z} \gg 1$), and this is indicated by the solid gray line in figure \ref{fig:2}a. For any non-zero $\tilde{r}_t$, however, there is an eventual transition to the $O(1/\tilde{z}^7)$ wake asymptote, corresponding to the dashed gray line in the said figure. In contrast, for $\beta_\infty \neq 0$, the axial velocity along the rear stagnation streamline changes from the reverse-Stokeslet decay to an exponential decay beyond the secondary screening length\,($\tilde{z} \sim \beta_\infty^{-1/2}$), and then at still larger distances, reverts to the aforementioned wake asymptote. The plot of the axial velocity along the rear stagnation streamline in figure \ref{fig:3}a shows the aforementioned pair of transitions for different small $\beta_\infty$. If we take the second transition to define a tertiary screening length\,($l_t$), then the exponential decay is essentially an intermediate asymptotic regime acting to connect the $O(1/\tilde{z})$-asymptote at distances smaller than $l_s$, to the $O(1/\tilde{z}^7)$-asymptote at distances larger than $l_t$. 
An estimate of $l_t$ can be obtained by equating the large-argument form of $K_1(2\beta_\infty^{1/2}\tilde{z})$, that characterizes the exponential decay phase, to the algebraic wake asymptote. That is, $3\beta_\infty^{1/2}K_1(2\beta_\infty^{1/2}\tilde{z}_t) = \dfrac{3240}{\tilde{z}_t^7}$ with $\tilde{z}_t \beta_\infty^{1/2} \gg 1$, $\beta_\infty \ll 1$; here, $\tilde{z}_t = l_t/l_c$ is the axial location corresponding to the tertiary screening length in units of $l_c$. Using $\lim_{\tilde{z}_t\beta_\infty^{1/2} \gg 1} K_1(2\beta_\infty^{1/2}\tilde{z}_t) = \sqrt{\pi}\exp(- 2\beta_\infty^{1/2} \tilde{z})/(4\beta_\infty^{1/2}\tilde{z})^{1/2}$, one may solve the resulting transcendental equation in a manner similar to $\tilde{r}_J$ above, by taking logarithms on both sides. This gives:
\begin{equation}
\label{eq:3.9}
    l_t = l_c\beta_\infty^{-1/2}\left(\zeta + \dfrac{13}{4}\ln\zeta + \dfrac{13^2}{4^2}\dfrac{\ln\zeta}{\zeta}\right),
\end{equation}
to third order in the small parameter $\zeta^{-1}$, where $\zeta = \dfrac{1}{2}\ln\left(\dfrac{\sqrt{\pi}}{2160\beta_\infty^3}\right) \gg 1$; see Appendix \ref{sec:AppC}. Using the original dimensionless parameters with $l_c = aRi_v^{-1/3}$, one has $l_t \sim \mathcal{O}\left(aRi_v^{-1/2}Pe^{1/2}\ln\left[\sqrt{\pi}Ri_v^{-1}Pe^3/2160\right] \right)$ to leading logarithmic order. 
We have verified that the three-term expansion in (\ref{eq:3.9}) is necessary for a precise estimate of the second exponential-to-algebraic transition mentioned above; the first term alone leads to a significant underestimate. 
In figure \ref{fig:3}a, the tertiary screening lengths, obtained from (\ref{eq:3.9}), are indicated by (vertical)\, dashed gray lines, and correlate very well to the transition from an intermediate exponential to the eventual algebraic decay. The secondary screening lengths, given by $l_s = 0.5623\beta_\infty^{-1/2}l_c$ and marking the beginning of the exponential decay phase, are also shown. The numerical prefactor in the said expression is obtained by finding the difference between the axial velocities for non-zero $\beta_\infty$ and $\beta_\infty = 0$, along the rear-stagnation streamline. This difference increases with $\tilde{z}$ to begin with, owing to the onset of diffusion effects, attains a maximum, and then decreases. The latter decrease is because the behavior for large $\tilde{z}$ is dictated by the reverse-Stokeslet decay for $\beta_\infty = 0$, the contribution due to the non-zero $\beta_\infty$ becoming exponentially small. The location of the aforementioned maximum gives the numerical prefactor.

Figure \ref{fig:3}b shows the behavior of the axial velocity field, as a function of distance, along curves corresponding to the third, fifth, and seventh zero crossings of $\psi_s$; this is done for different small $\beta_\infty$, including $\beta_\infty = 0$. In the latter case, the loci of the zero crossings are the black curves shown earlier in figure \ref{fig:2}b. The first zero crossing corresponds to the rear stagnation streamline which serves as the axis of the downstream columnar structure. The flow within this structure reverses direction from one cell to the next, and the zero crossing numbers above correspond therefore to the `centerlines' of every alternate cell, starting from the central reverse jet, with flow directed away from the sphere. Interestingly, for a given $\beta_\infty$, the decay has a similar character for the different cells, although its nature is essentially different for zero and non-zero $\beta_\infty$. For $\beta_\infty = 0$, $\tilde{u}_z$ decays as the inverse of the distance along the centerline for all three cells shown, although with smaller numerical pre-factors for the peripheral ones. For $\beta_\infty \neq 0$, the inverse-centerline-distance decay only occurs until the secondary screening length, and an exponential decay ensues thereafter. Note that all curves in figure \ref{fig:3}b start from a finite downstream distance, with this distance being larger for the peripheral cells - this is consistent with the nature of the columnar structure for $\beta_\infty = 0$ as evident from the streamline pattern in figure \ref{fig:2}b above; and for non-zero $\beta_\infty$ as shown in \S~\ref{sec:3.3} below\,(see figures \ref{fig:3} and \ref{fig:5} below). Note also that all the curves for non-zero $\beta_\infty$ terminate at the tertiary screening length, as must be the case. Since the latter length scale diverges for $\beta_\infty \rightarrow 0$, all three black lines in Fig.\ref{fig:3}b continue to $\tilde{z} = \infty$. 

\subsection{The complete streamline and isopycnal patterns}\label{sec:3.3}
\begin{figure}
    \hspace{-0.5cm}
    \begin{minipage}[t]{0.55\textwidth}
      \centering
      \includegraphics[width=\textwidth]{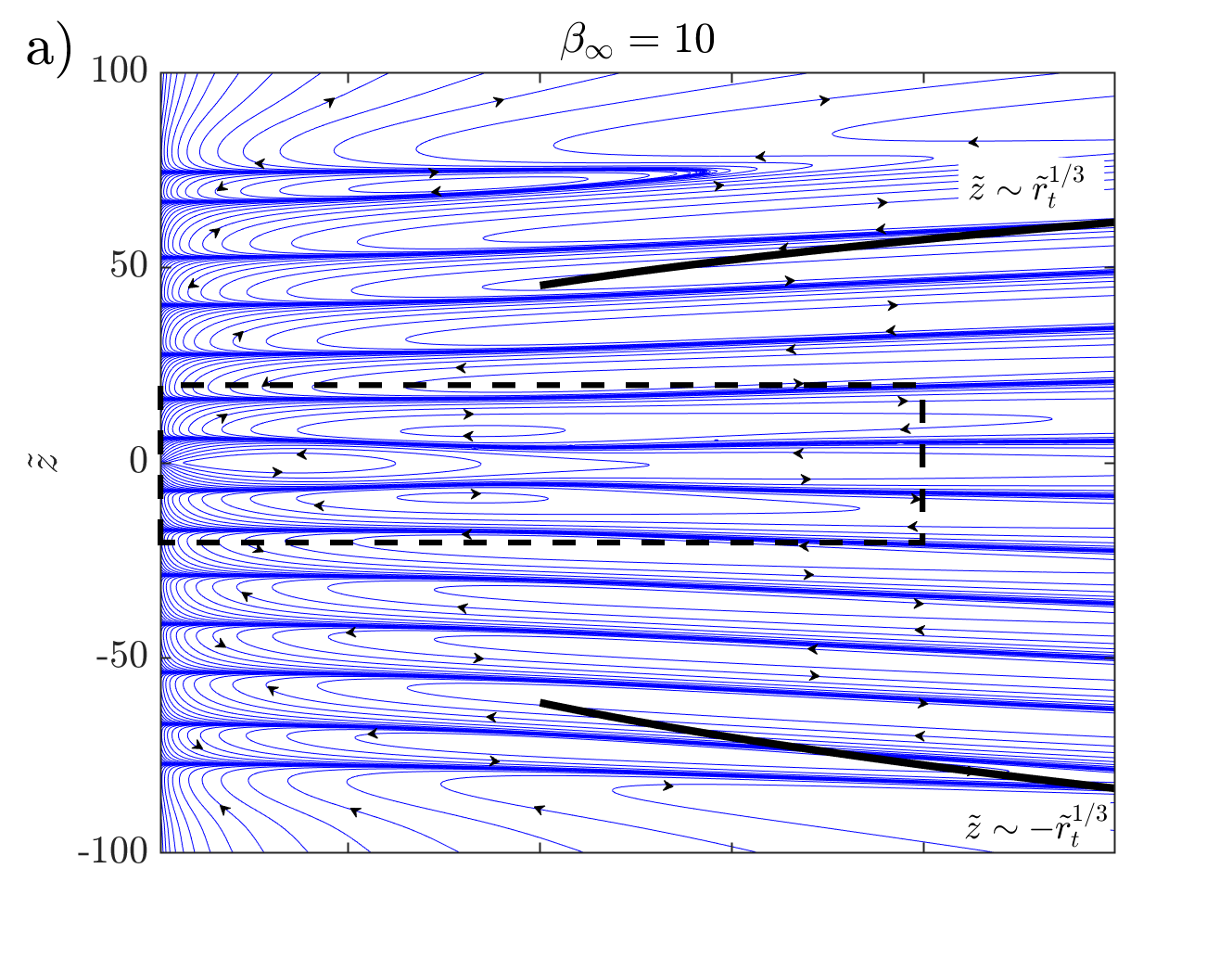}
    \end{minipage}
     \hspace{-0.4cm}
    \begin{minipage}[t]{0.55\textwidth}
      \centering
      \includegraphics[width=\textwidth]{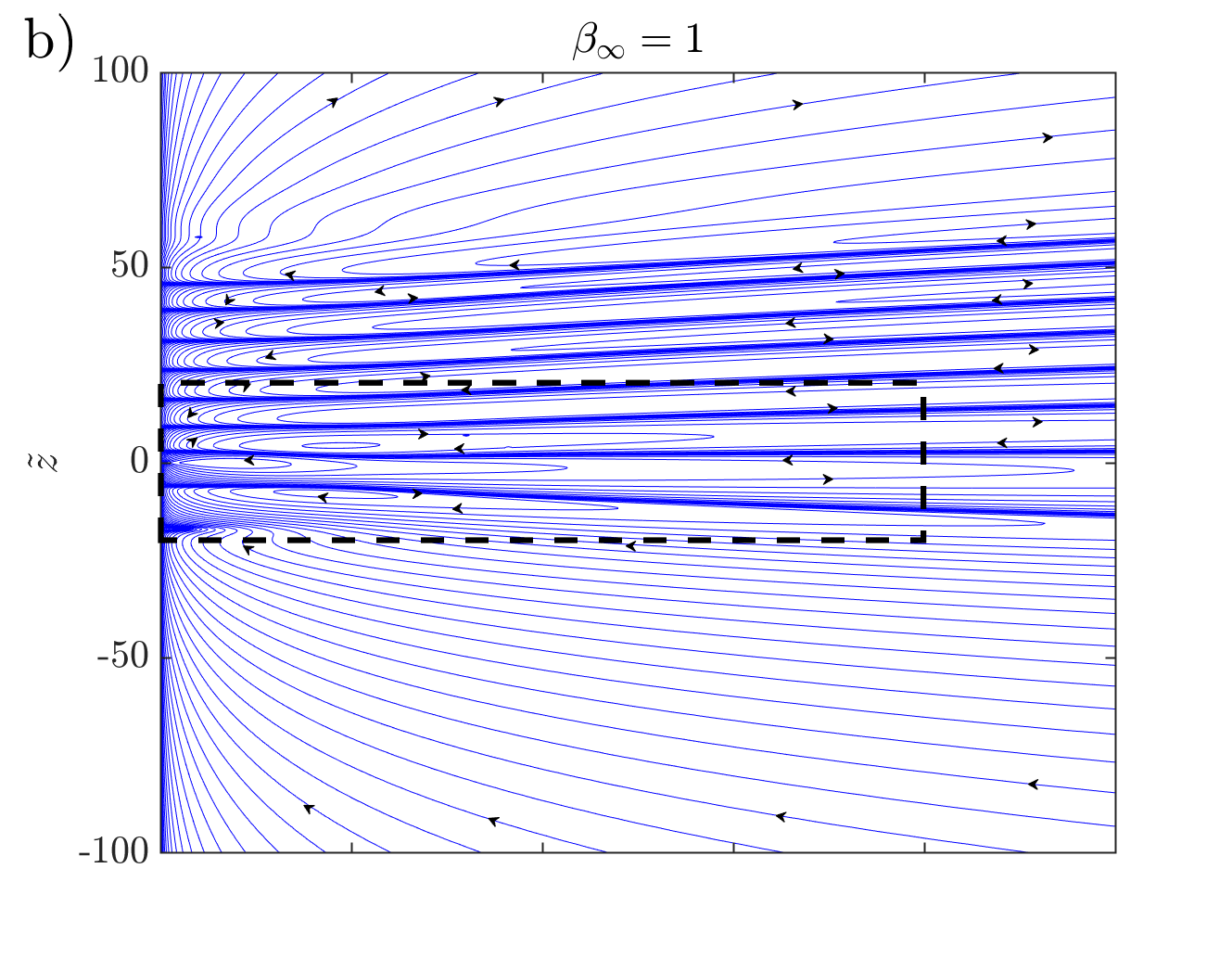}
    \end{minipage}
    
    \vspace{-0.5cm}
    \hspace{-0.5cm}
    \begin{minipage}[t]{0.55\textwidth}
      \centering
      \includegraphics[width=\textwidth]{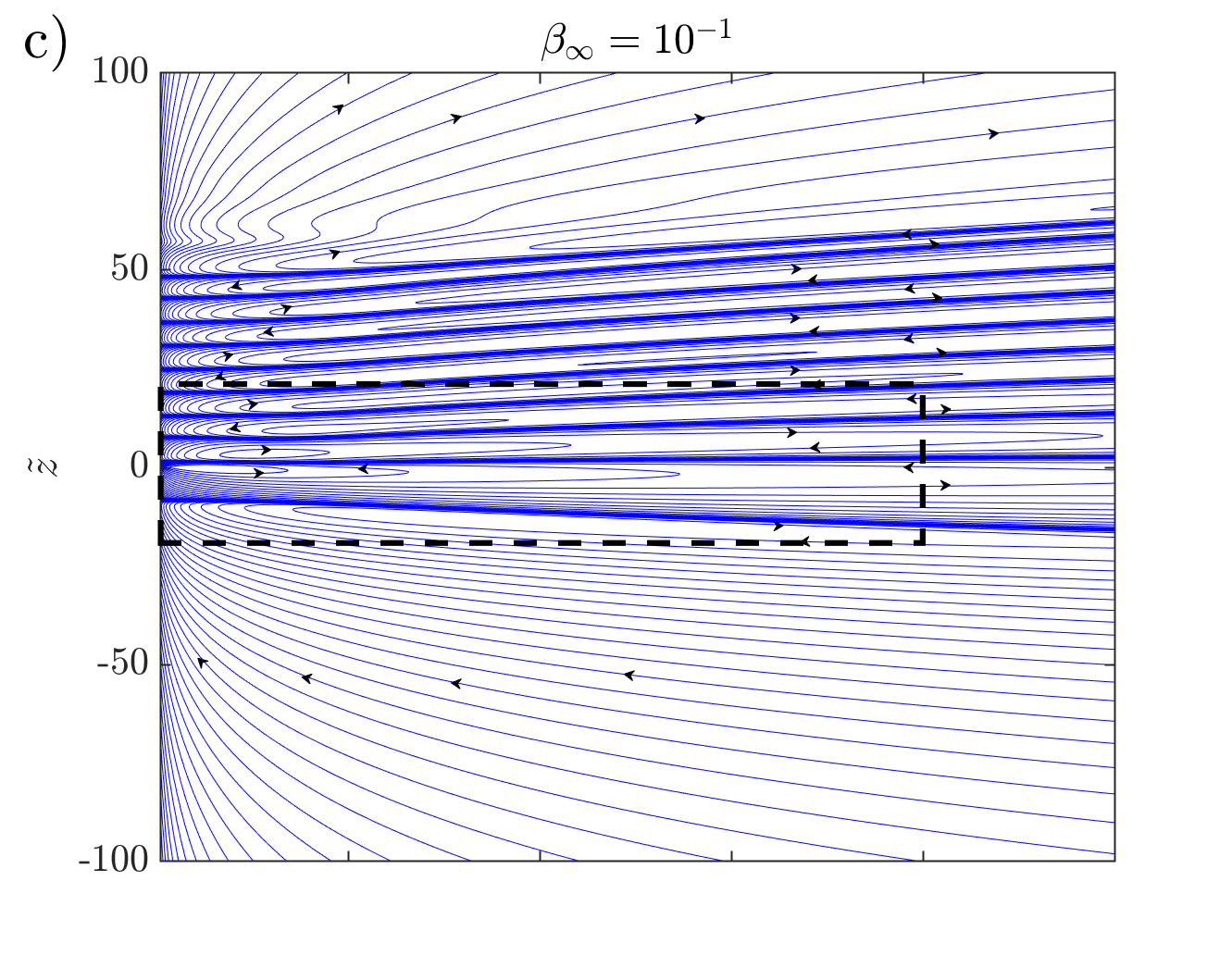}
    \end{minipage}
    \hspace{-0.4cm}
    \begin{minipage}[t]{0.55\textwidth}
      \centering
      \includegraphics[width=\textwidth]{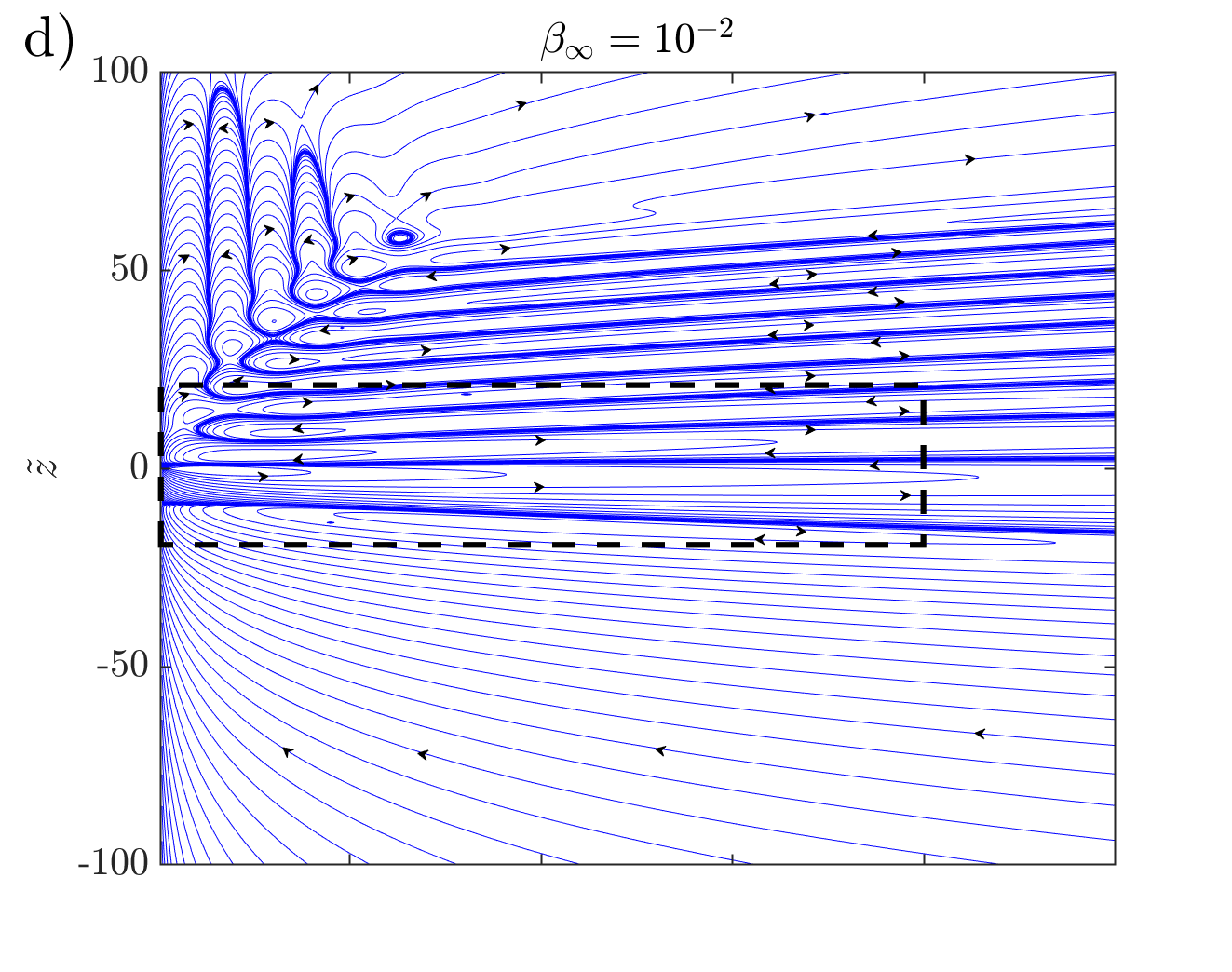}
    \end{minipage}

    \vspace{-0.5cm}
    \hspace{-0.5cm}
    \begin{minipage}[t]{0.55\textwidth}
      \centering
      \includegraphics[width=\textwidth]{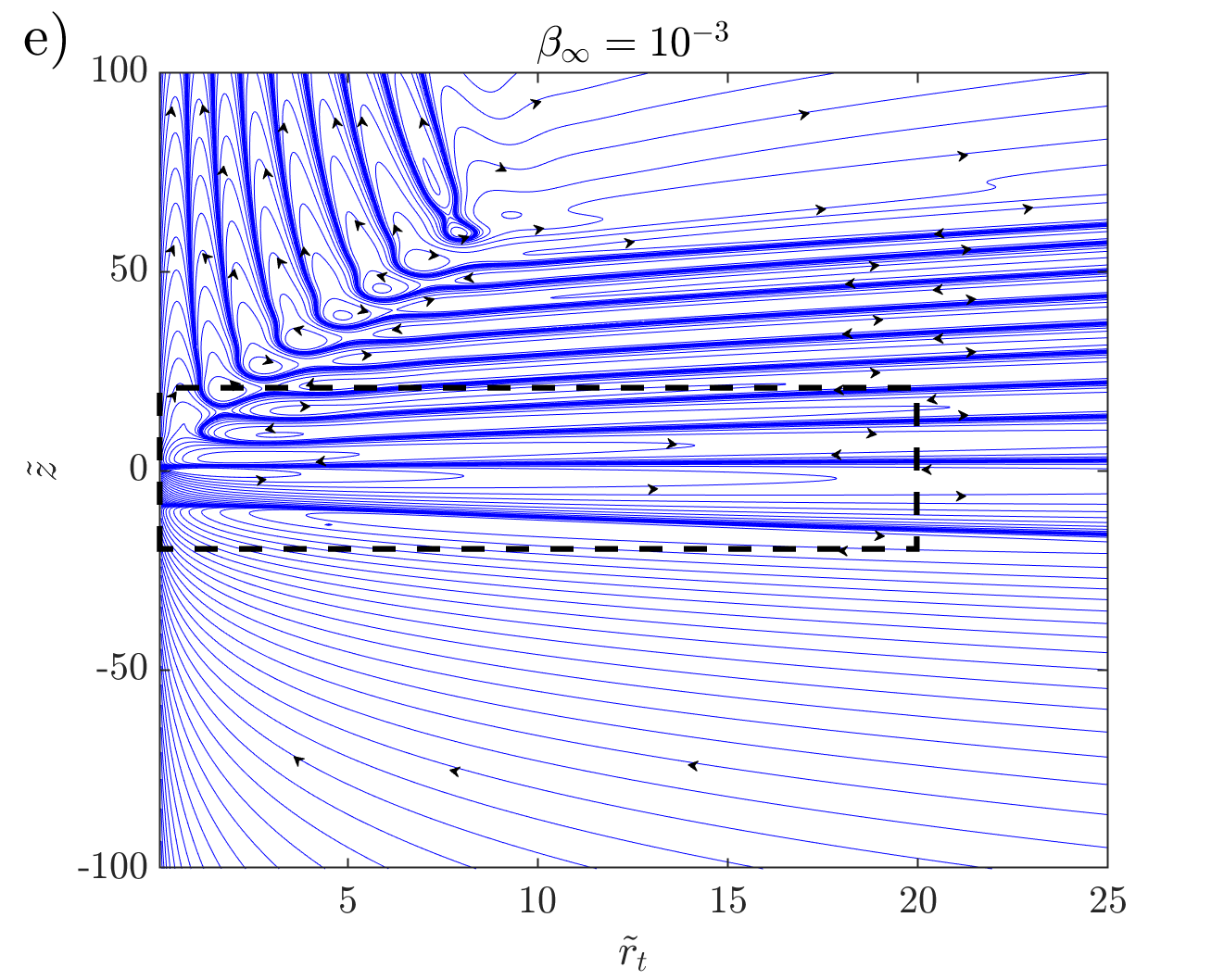}
    \end{minipage}
    \hspace{-0.4cm}
    \begin{minipage}[t]{0.55\textwidth}
      \centering
      \includegraphics[width=\textwidth]{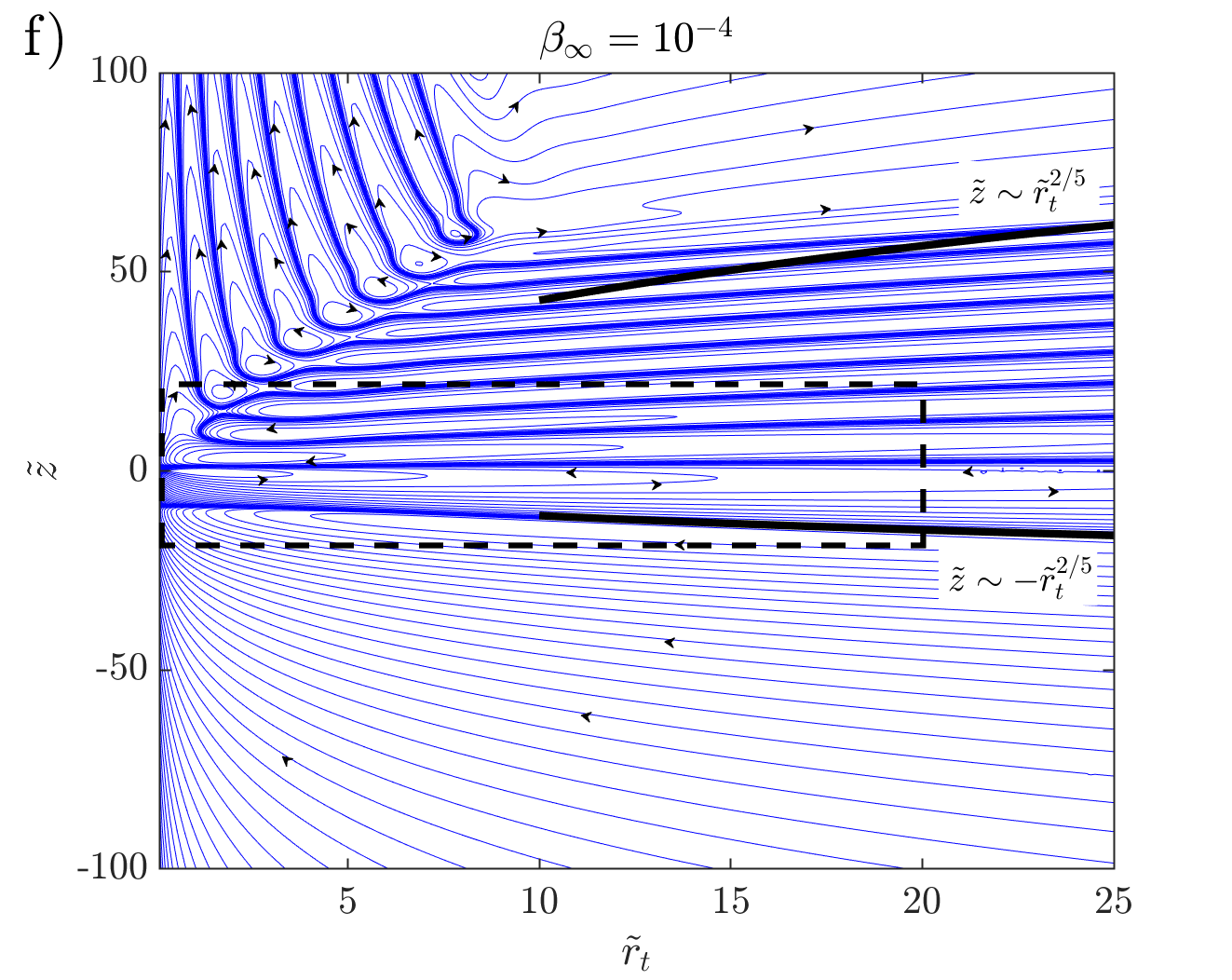}
    \end{minipage}
  \caption{Streamline patterns in an axial (half)\,plane for $\beta_\infty = $ (a) $10$, (b) $1$, (c) $10^{-1}$, (d) $10^{-2}$, (e) $10^{-3}$, and (f) $10^{-4}$. The sphere is located at $(\tilde{r}_t,\tilde{z}) \equiv (0,0)$, with $\tilde{r}_t$ and $\tilde{z}$ being scaled by the primary screening length $l_c = \mathcal{O}(aRi_v^{-1/3})$. Here, the dense blue bands mark the zero-crossings of the Stokes streamfunction. The streamline patterns within the central regions, enclosed by the black dashed rectangles, are those originally obtained by \citetalias{varanasi2022motion}. The black continuous curves in (a) and (f) are wake boundaries calculated from the similarity solution corresponding to small$-Pe$ and large$-Pe$ limits, respectively.}
  \label{fig:4}
\end{figure}
\begin{figure}
    \hspace{-0.5cm}
    \begin{minipage}[t]{0.55\textwidth}
      \centering
      \includegraphics[width=\textwidth]{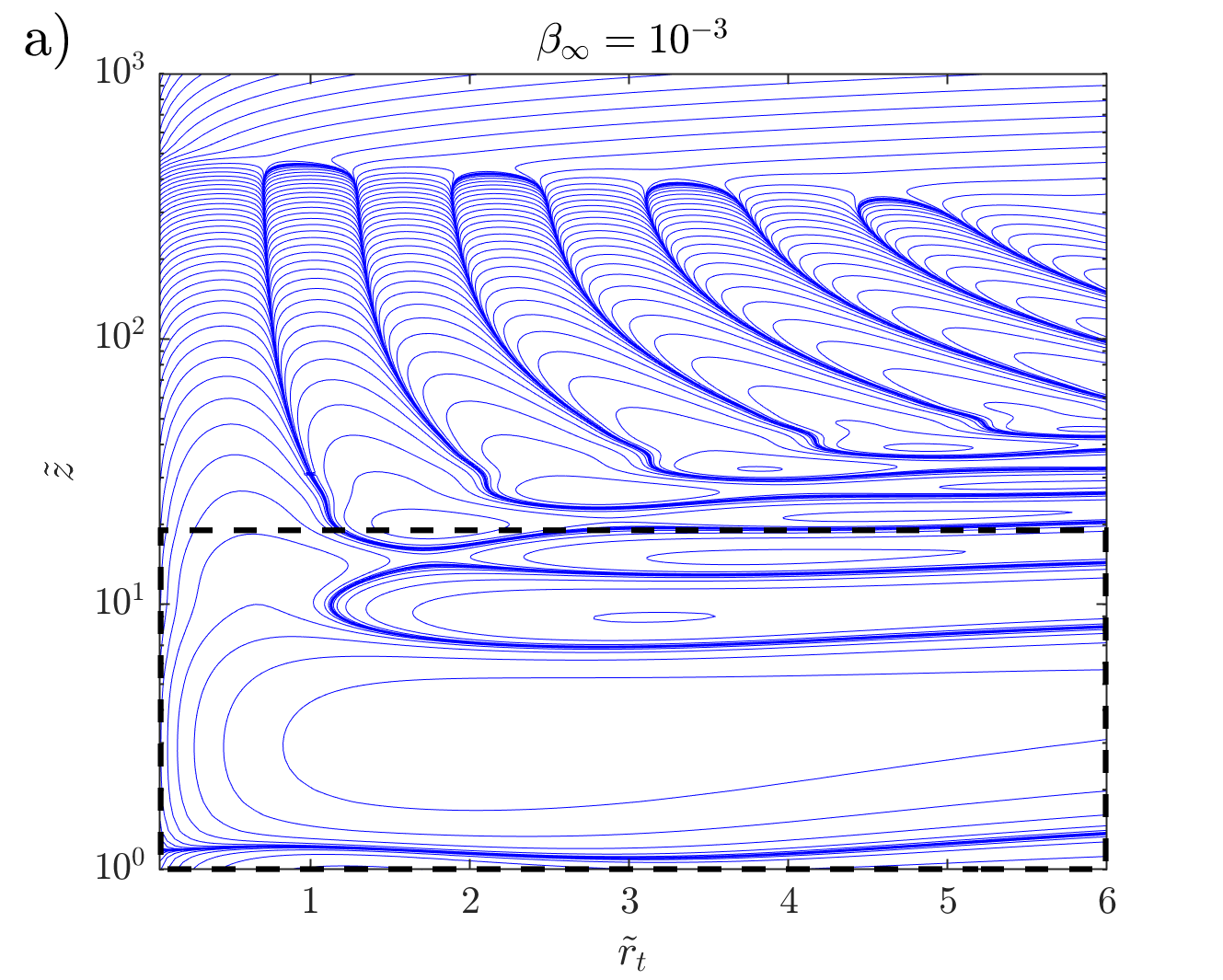}
    \end{minipage}
     \hspace{-0.4cm}
    \begin{minipage}[t]{0.55\textwidth}
      \centering
      \includegraphics[width=\textwidth]{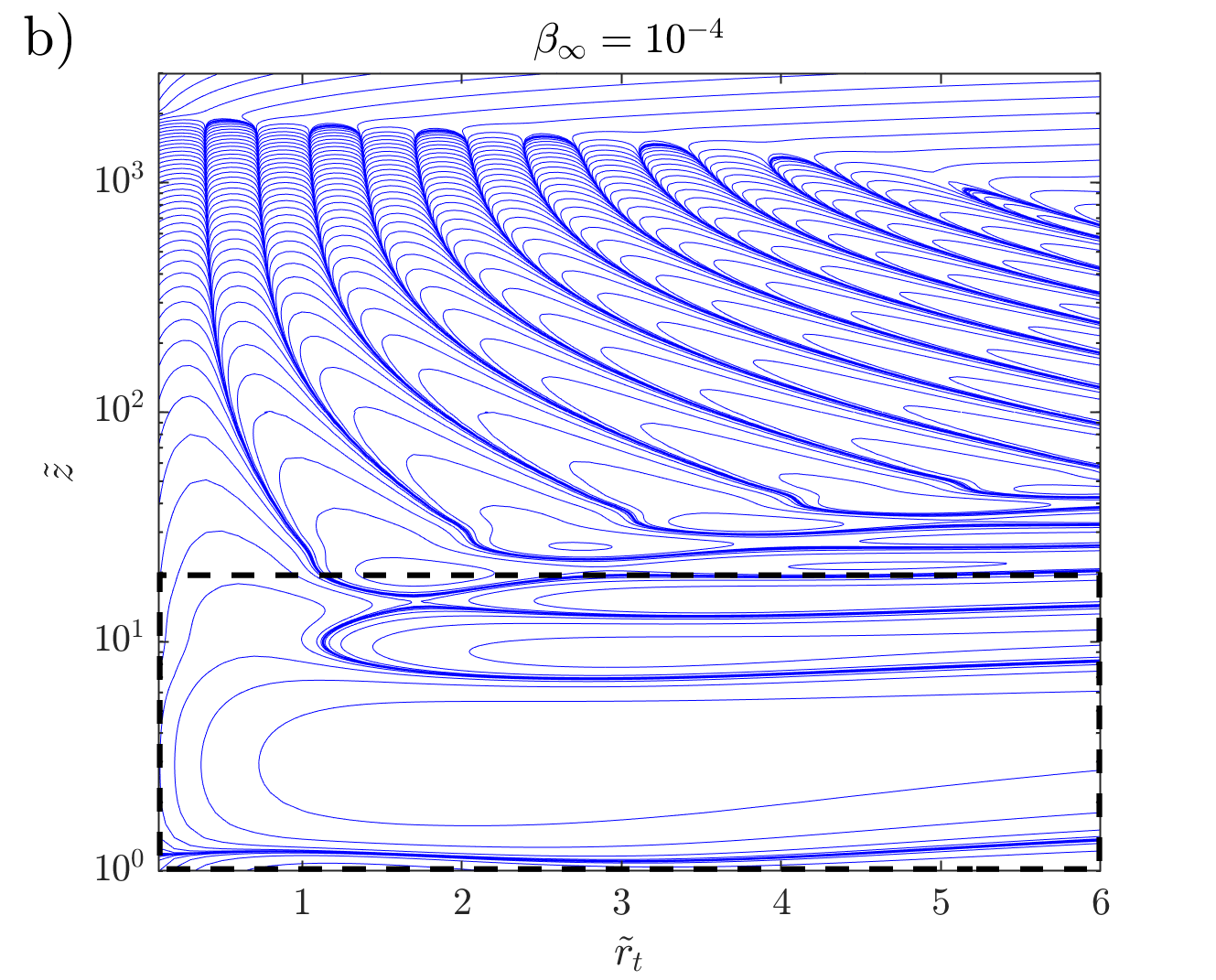}
    \end{minipage}
    
    \vspace{-0.31cm}
    \hspace{-0.5cm}
    \begin{minipage}[t]{0.55\textwidth}
      \centering
      \includegraphics[width=\textwidth]{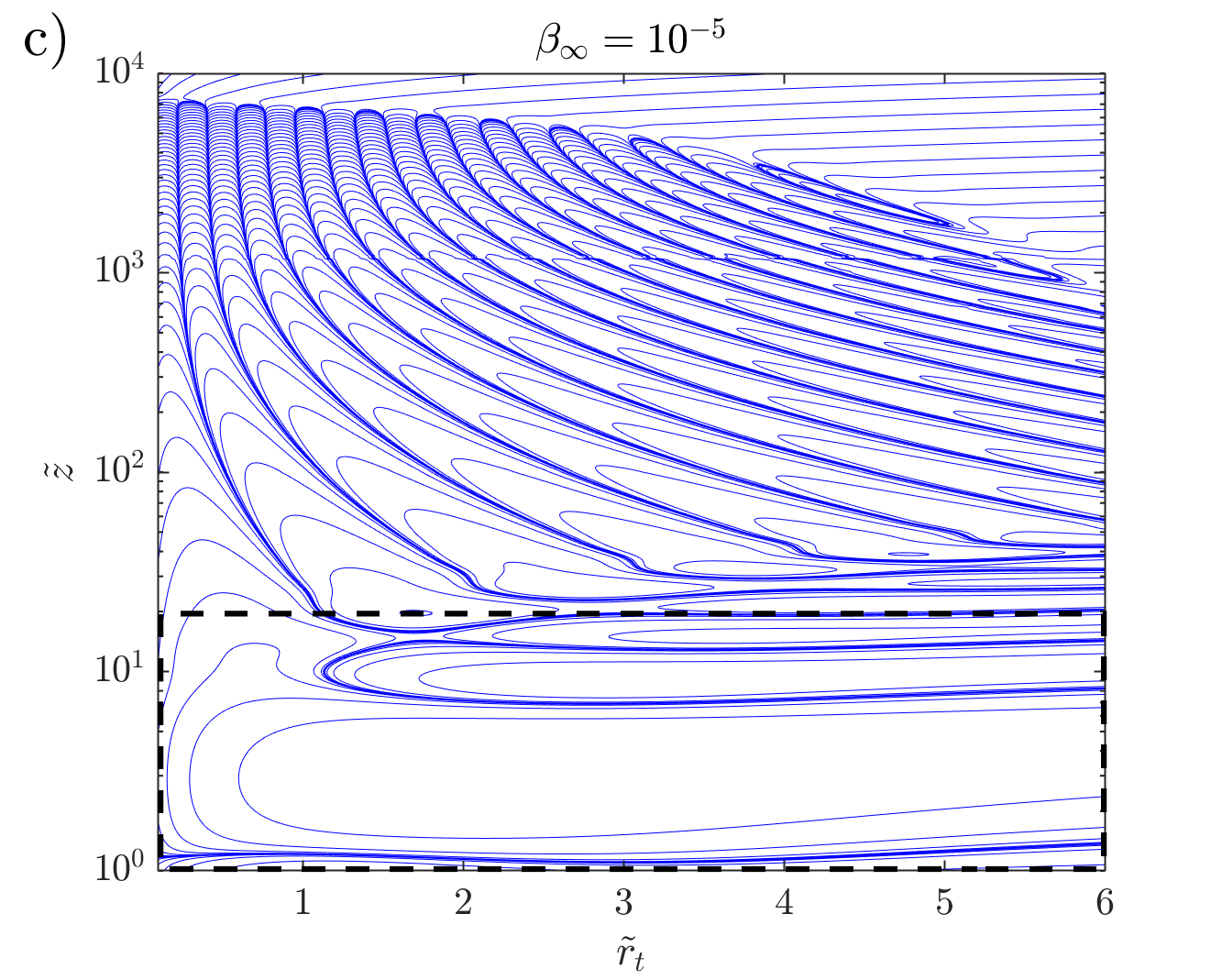}
    \end{minipage}
    \hspace{-0.4cm}
    \begin{minipage}[t]{0.55\textwidth}
      \centering
      \includegraphics[width=\textwidth]{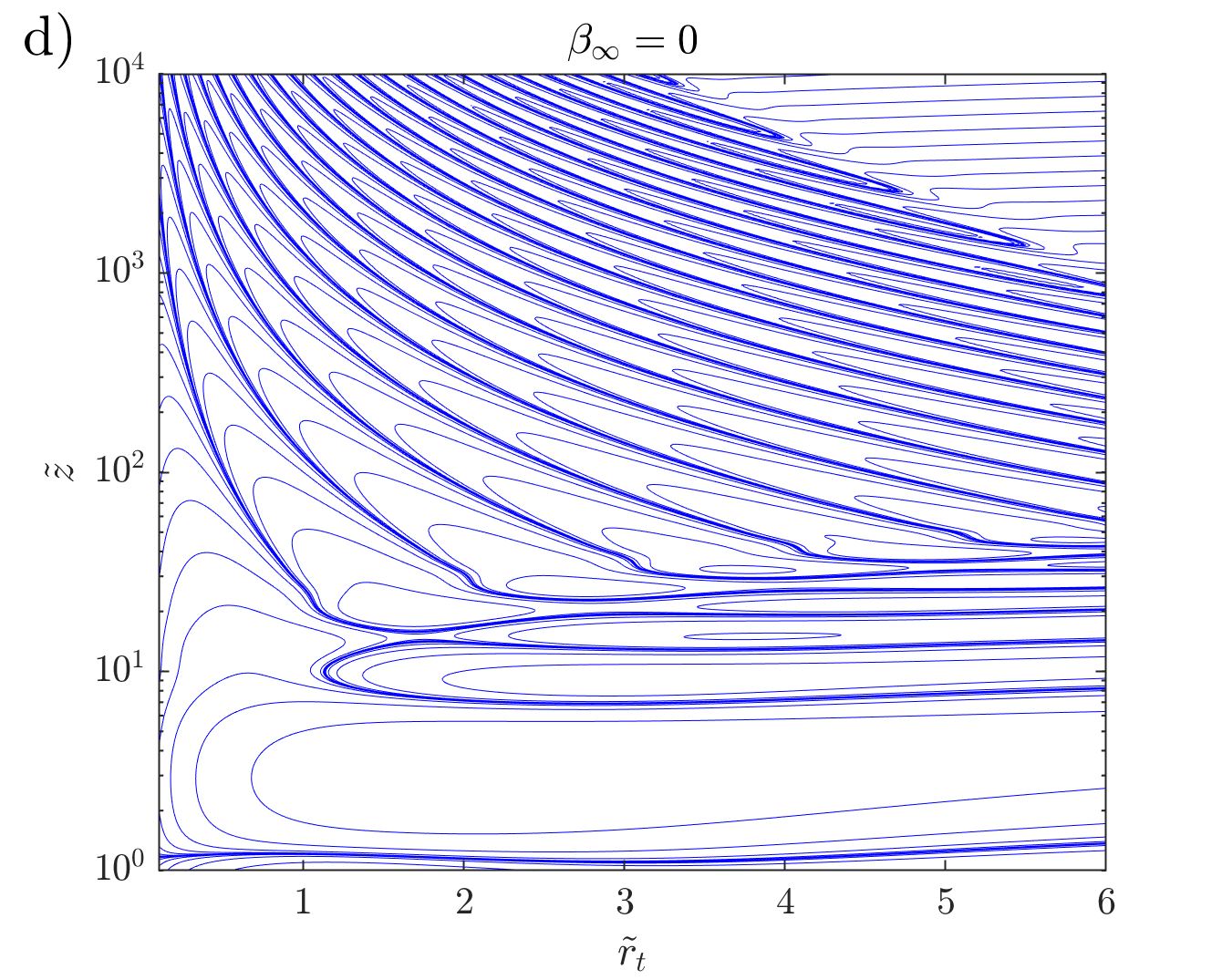}
    \end{minipage}

    \vspace{0.31cm}
    \hspace{-0.5cm}
    \centering
    \begin{minipage}[t]{0.6\textwidth}
      \centering
      \includegraphics[width=\textwidth]{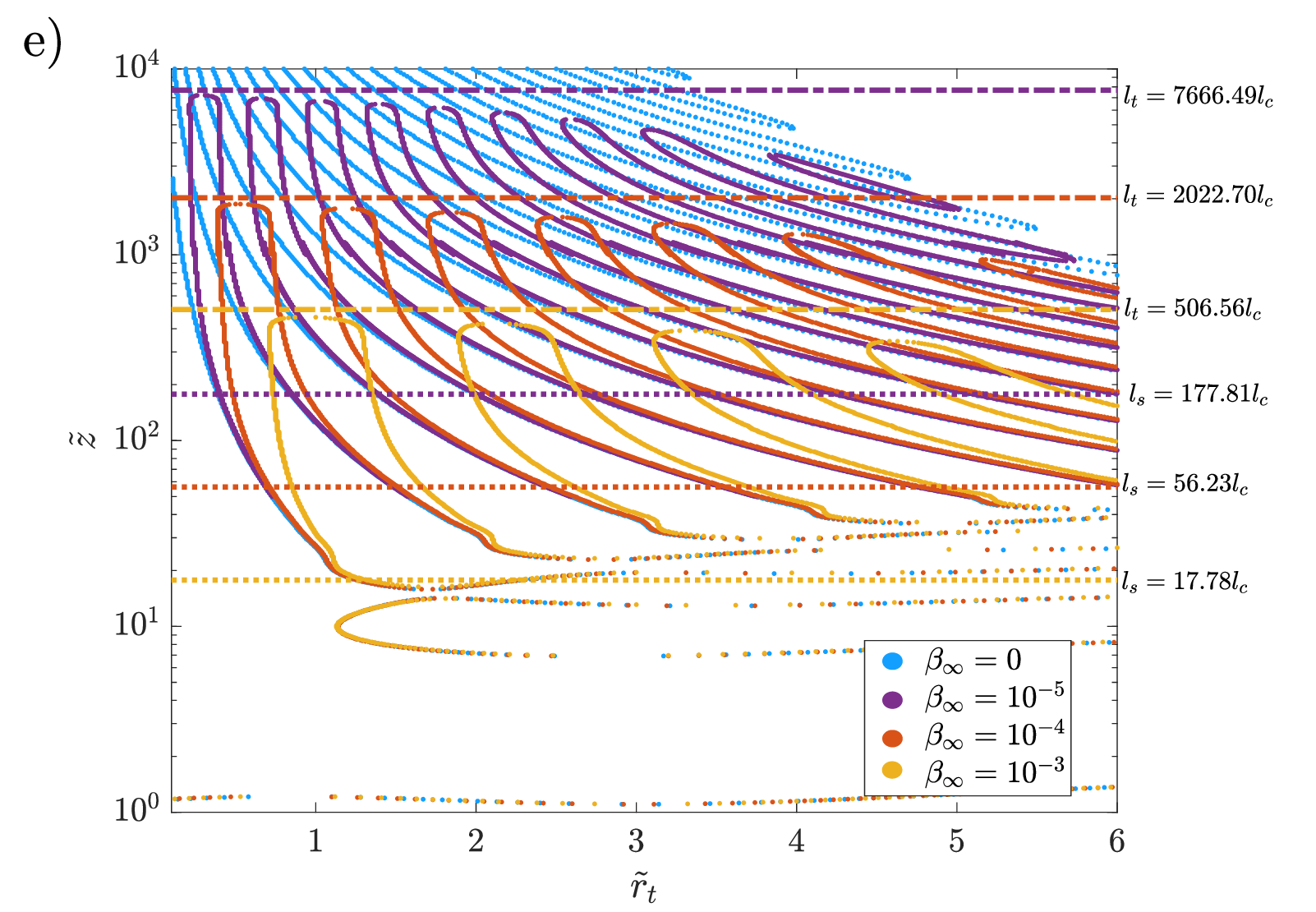}
    \end{minipage}
  \caption{Streamline patterns in an axial (half)\,plane, and over an extended downstream region\,($1 < \tilde{z} < 10^4$), showing the columnar structure for $\beta_\infty = $ (a) $10^{-3}$, (b) $10^{-4}$, (c) $10^{-5}$, and (d) $0$. The sphere is at $(\tilde{r}_t,\tilde{z}) \equiv (0,0)$, with $\tilde{r}_t$ and $\tilde{z}$ being scaled using the primary screening length $l_c = \mathcal{O}(aRi_v^{-1/3})$. The dense blue bands mark the zero-crossings of the Stokes streamfunction; the region enclosed by black dashed rectangles correspond to the downstream patterns accessed in \citetalias{varanasi2022motion}. (e) Features of the columnar structures, for different non-zero $\beta_\infty$, compared to the limiting case of $\beta_\infty = 0$; horizontal dashed and dot-dashed lines indicate secondary($l_s$) and tertiary ($l_t$) screening lengths, respectively. }
  \label{fig:5}
\end{figure}

Although different flow features at large distances from the sphere have been predicted using asymptotic methods, as illustrated in \S~\ref{sec:3.2}, the extent to which these approximations are applicable can only be ascertained by a full numerical integration. Towards this end, we now present the complete streamline and isopycnal patterns from a numerical evaluation of the original inverse transform integrals in \eqref{eq:2.9} and \eqref{eq:2.10} using \eqref{eq:2.11}-\eqref{eq:2.13}.

Figure \ref{fig:4} shows the streamline patterns around the translating sphere\,(a point force at the origin) over a range of $\beta_\infty$, from the diffusion dominant limit\,($\beta_\infty = 10$; figure \ref{fig:4}a) up until the convection dominant one\,($\beta_\infty = 10^{-4}$; figure \ref{fig:4}f); black arrows attached to select streamlines in each pattern indicate the flow direction. The portions of the patterns in the rather limited central rectangular region, demarcated by black dashed lines in each of figures \ref{fig:4}a-f, were the ones obtained earlier by \citetalias{varanasi2022motion} (see figures 8 and 9 therein). The increasing importance of convective effects, accompanying a decrease in $\beta_\infty$, leads to a progressive departure from the fore-aft symmetry that characterizes the streamline pattern in the limit $\beta_\infty \rightarrow \infty$. One manifestation of this asymmetry is a significant reduction in the number of horizontal recirculating cells upstream of the translating sphere\,(corresponding to negative $\tilde{z}$) - from $7$ for $\beta_\infty = 10$, to 3 for $\beta_\infty = 1$, to $1$ for $\beta_\infty < O(10^{-2})$. The radial extent of all streamline patterns shown is $25 aRi_v^{-1/3}$ which is sufficient for $\beta_\infty \lesssim 10^{-1}$ in the sense that the number of (horizontal)\,recirculating cells should remain unchanged for $\tilde{r}_t \gtrsim 25$. This constancy of cell number is consistent with the self-similar structure of the large-$Pe$ horizontal wake at these radial distances; the wake grows as $\tilde{z} \propto \tilde{r}_t^{2/5}$ for large $Pe$, as shown in \citetalias{varanasi2022motion}, with its structure being a function of the similarity variable $\tilde{z}/\tilde{r}_t^{2/5}$. For $\beta_\infty \lesssim 10^{-1}$, the number of recirculating cells at larger distances therefore corresponds to the (fixed)\,number of zero crossings of the similarity solution, when plotted as a function of the similarity variable above.  Further, the far-field asymptotes of the large-$Pe$ similarity solution, given in table \ref{table2}, give the decay rates outside of the wake recirculating cells, in both the upstream and downstream directions; 
for the Stokes streamfunction, the upstream and downstream asymptotes are given by $1620\tilde{r}_t^2/\tilde{z}^7$ and $-1080\tilde{r}_t^2/\tilde{z}^7$, respectively. 

The wake grows differently for small $Pe$, as $\tilde{z} \propto \tilde{r}_t^{1/3}$ for distances larger than $a(Ri_vPe)^{-1/4}$. An implication of the different screening lengths in the small and large-$Pe$ limits, for the streamline patterns in Figure \ref{fig:4}, is that, for a fixed $\tilde{r}_t$, the effective size of the domain decreases as $\beta_\infty^{-\frac{1}{4}}$ for $\beta_\infty > 1$\,($\beta_\infty^{\frac{1}{4}}$ being the ratio of the two screening lengths). As a result, for a given choice of the maximum $\tilde{r}_t$ value, the domain will become small enough for sufficiently large $\beta_\infty$, in the sense of the numerically generated streamline pattern not connecting to the pattern consistent with the farfield wake-similarity solution. This insufficiency is starting to be evident in Figure \ref{fig:4}a, for $\beta_\infty = 10$, where the black curves\,($\tilde{z} \propto \tilde{r}_t^{\frac{1}{3}}$) denoting the self-similar growth of the small-$Pe$ wake are seen to depart noticeably from the contours that mark the final zero-crossings in the upstream and downstream directions. The analogous curves for large $Pe$\,($\tilde{z} \propto \tilde{r}_t^{\frac{2}{5}}$) exhibit a better match with the zero crossing contours in Figure \ref{fig:4}f. 
    
Starting from $\beta_\infty = 10^{-2}$, corresponding to figure \ref{fig:4}d, one sees the emergence of an additional set of nearly vertical recirculating cells behind the sphere. Only a limited portion of the central cell in Fig.\ref{fig:4}f, for $\beta_\infty = 10^{-4}$, falls within the domain accessed in \citetalias{varanasi2022motion}\,(the dashed black rectangle, as mentioned above), which gave the impression of an isolated rearward jet-like structure. The larger domains in figure \ref{fig:4}d-f make it evident that this reverse-jet is only the central portion of an ensemble of concentrically arranged nearly vertical cells. With a decrease in $\beta_\infty$ from $10^{-2}$ to $10^{-4}$, the number of vertical recirculating cells in the columnar structure increases from $4$ to $8$ over the axial extent\,($100 l_c$) of the domain examined. 

To better understand the structure and spatial extent of the downstream columnar structure in figures \ref{fig:4}d-f, we focus next on the streamline patterns for the smaller $\beta_\infty$'s \,($=10^{-3}$, $10^{-4}$, $10^{-5}$ and $0$), and on a much longer region downstream of the sphere\,($\tilde{z}$ extending up to $10^4$). The depiction of the patterns in figures \ref{fig:5}a-d is now via a semi-log plot, with the $\tilde{z}$-axis on a logarithmic scale. For $\tilde{z} \lesssim 20$, the streamline patterns in figures \ref{fig:5}a-d show a largely identical pattern of horizontal recirculating cells, implying that the wake structure has converged to its limiting form for $\beta_\infty \rightarrow 0$. In contrast, the emergence of vertical recirculating cells with increasing $\tilde{z}$, and the associated pattern of streamlines, continues to be sensitively dependent on $\beta_\infty$ even in the said limit. Figure \ref{fig:5}d, for $\beta_\infty = 0$, shows a continuous increase in the number of recirculating cells with increasing $\tilde{z}$. The radial extents of the individual cells decrease as $\tilde{z}^{-1/2}$, with the number of cells increasing in a manner consistent with the prediction given by (\ref{eq:3.77}) - from 11 cells at $\tilde{z} = 100$ to 24 cells at $\tilde{z} = 10^4$. For nonzero $\beta_\infty$, as the effects of density diffusion start to become important at the secondary screening length, $l_s = \mathcal{O}(l_c\beta_\infty^{-1/2})$, the recirculating cell boundaries begin to deviate from those for $\beta_\infty = 0$, with the cells, especially the ones closer to the rear stagnation streamline, becoming vertical and having a nearly constant width at larger distances. This change due to density diffusion occurs at a smaller downstream distance for the larger $\beta_\infty$'s, as a result of which the recirculating cells are wider in these cases. All the recirculating cells for non-zero $\beta_\infty$ end at $\tilde{z} \sim O(l_t)$ with $l_t$ given by \eqref{eq:3.9}. It is worth reiterating that, both secondary and tertiary screening lengths diverge in the non-diffusive limit\,($\beta_\infty = 0$). Fig.\ref{fig:5}e superposes the columnar structures for different $\beta_\infty$, validating the aforementioned significance of the secondary and tertiary screening lengths. 
\begin{figure}
    \hspace{-0.5cm}
    \begin{minipage}[t]{0.55\textwidth}
      \centering
      \includegraphics[width=\textwidth]{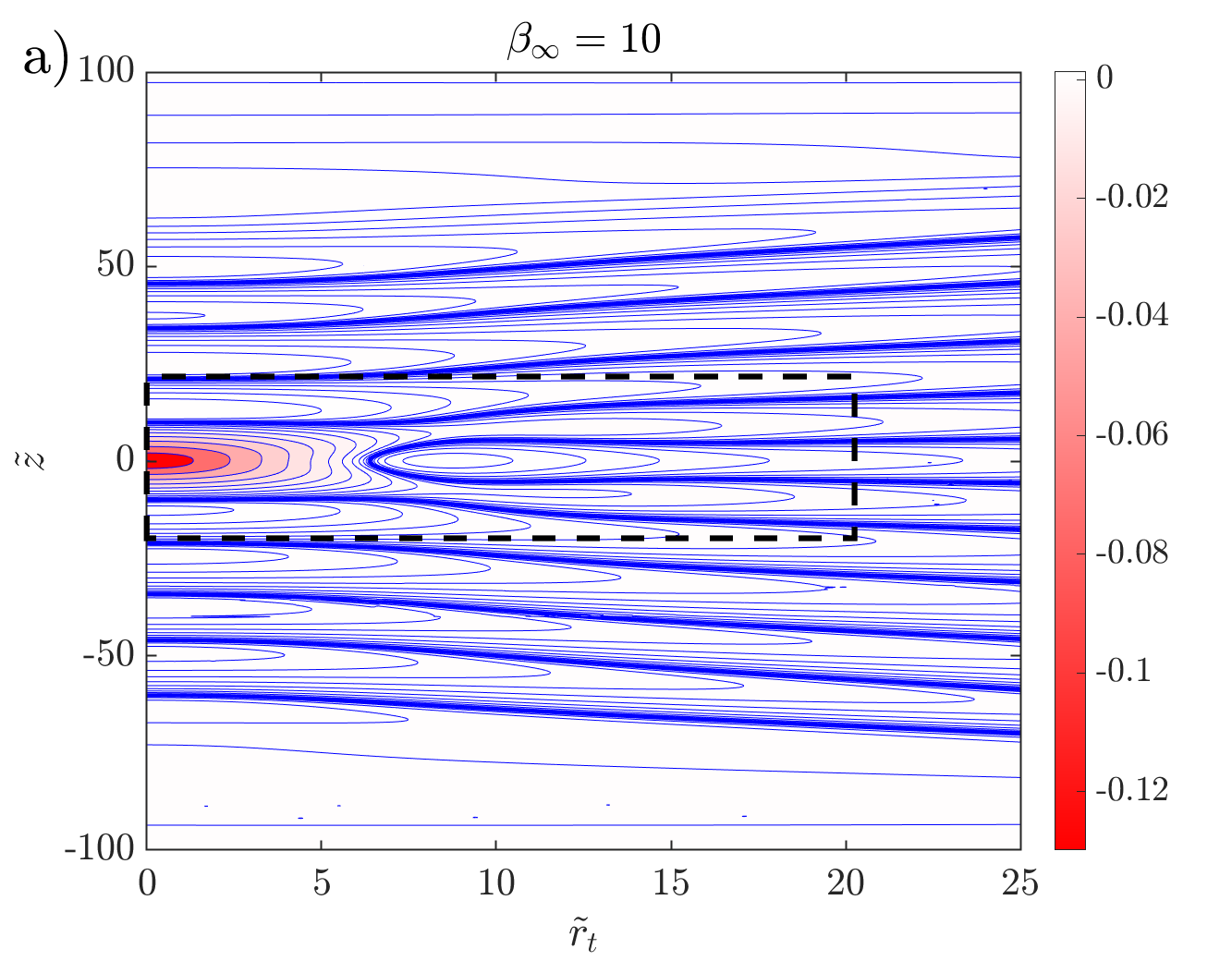}
    \end{minipage}
     \hspace{-0.4cm}
    \begin{minipage}[t]{0.55\textwidth}
      \centering
      \includegraphics[width=\textwidth]{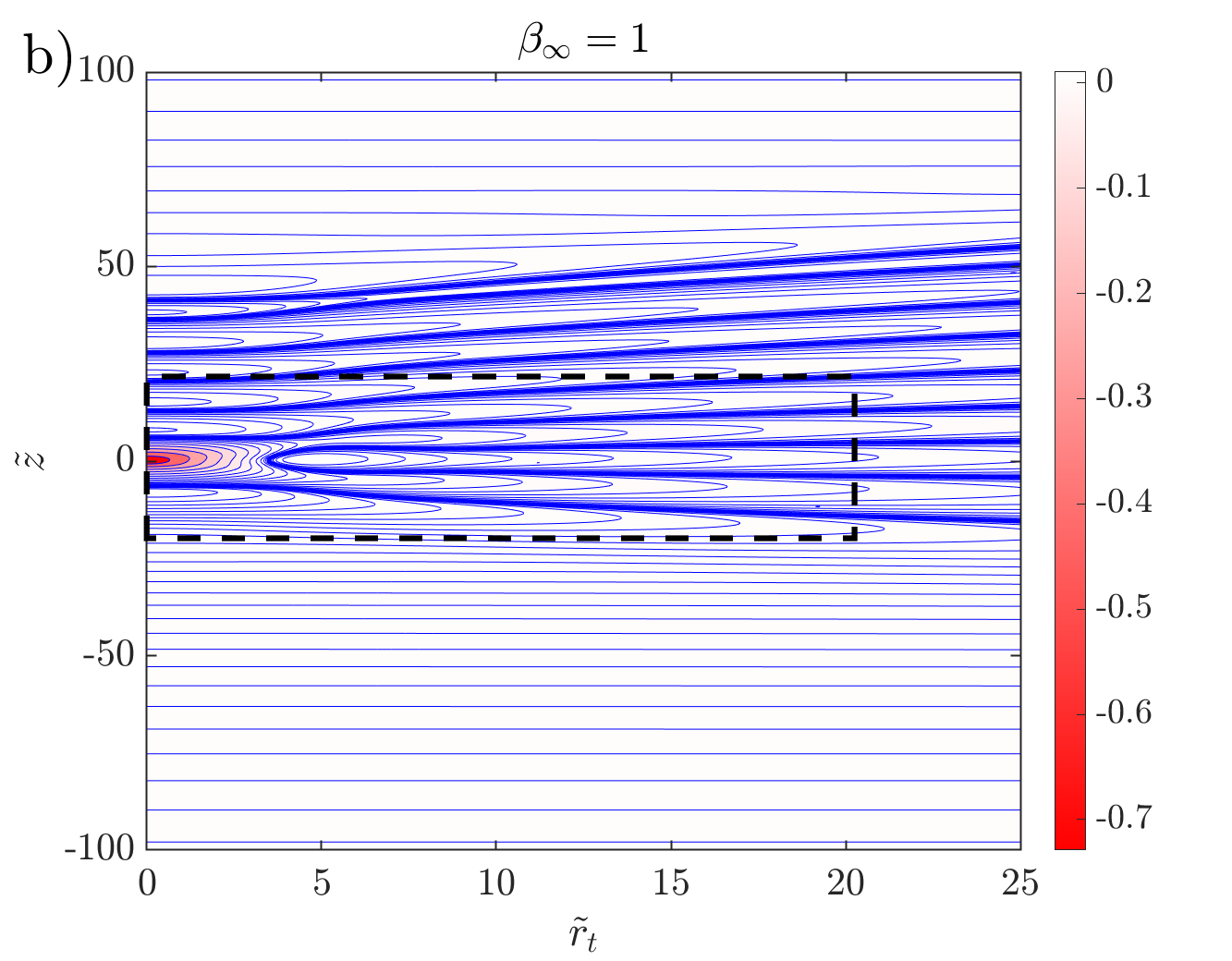}
    \end{minipage}
    
    \vspace{-0.6cm}
    \hspace{-0.5cm}
    \begin{minipage}[t]{0.55\textwidth}
      \centering
      \includegraphics[width=\textwidth]{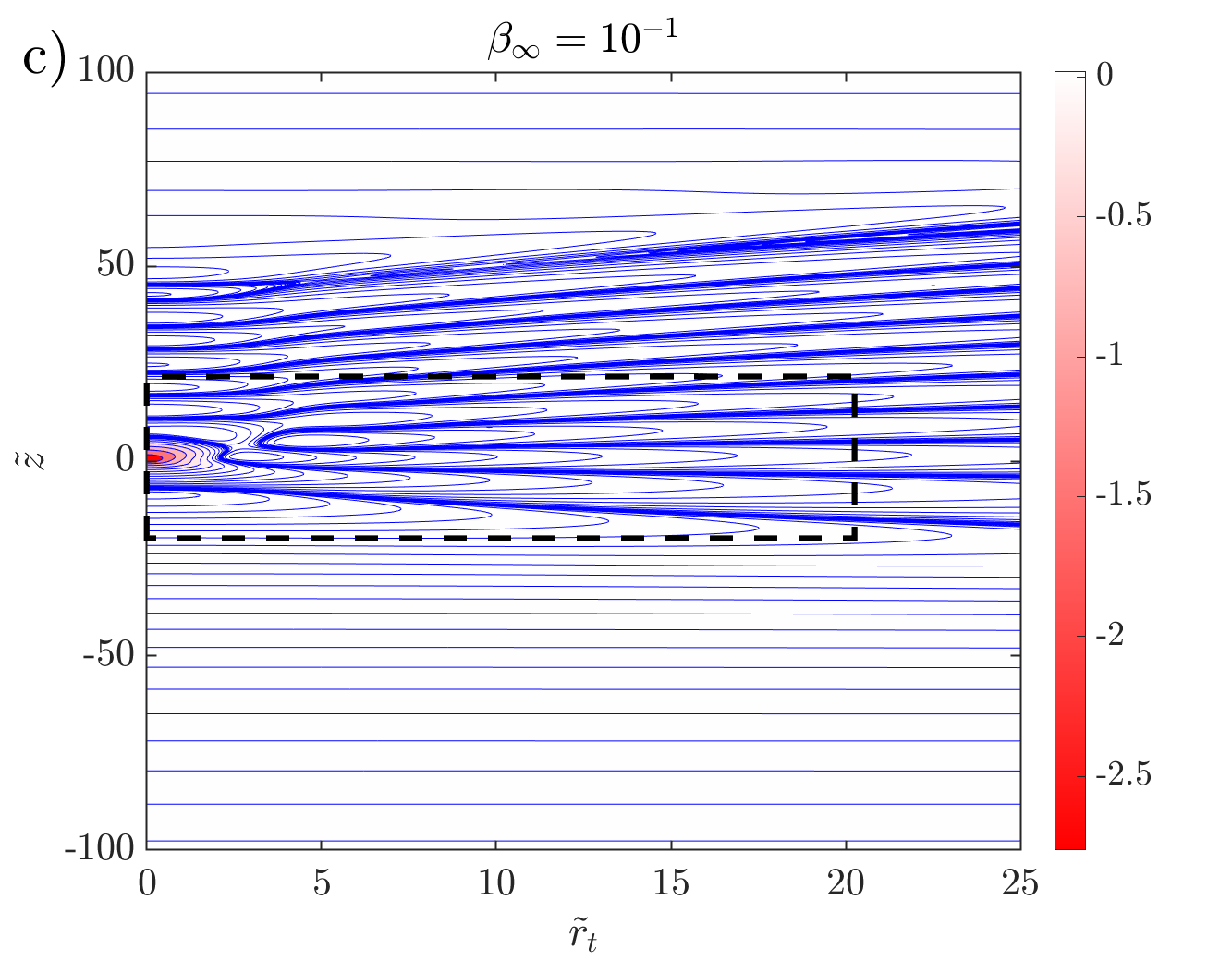}
    \end{minipage}
    \hspace{-0.4cm}
    \begin{minipage}[t]{0.55\textwidth}
      \centering
      \includegraphics[width=\textwidth]{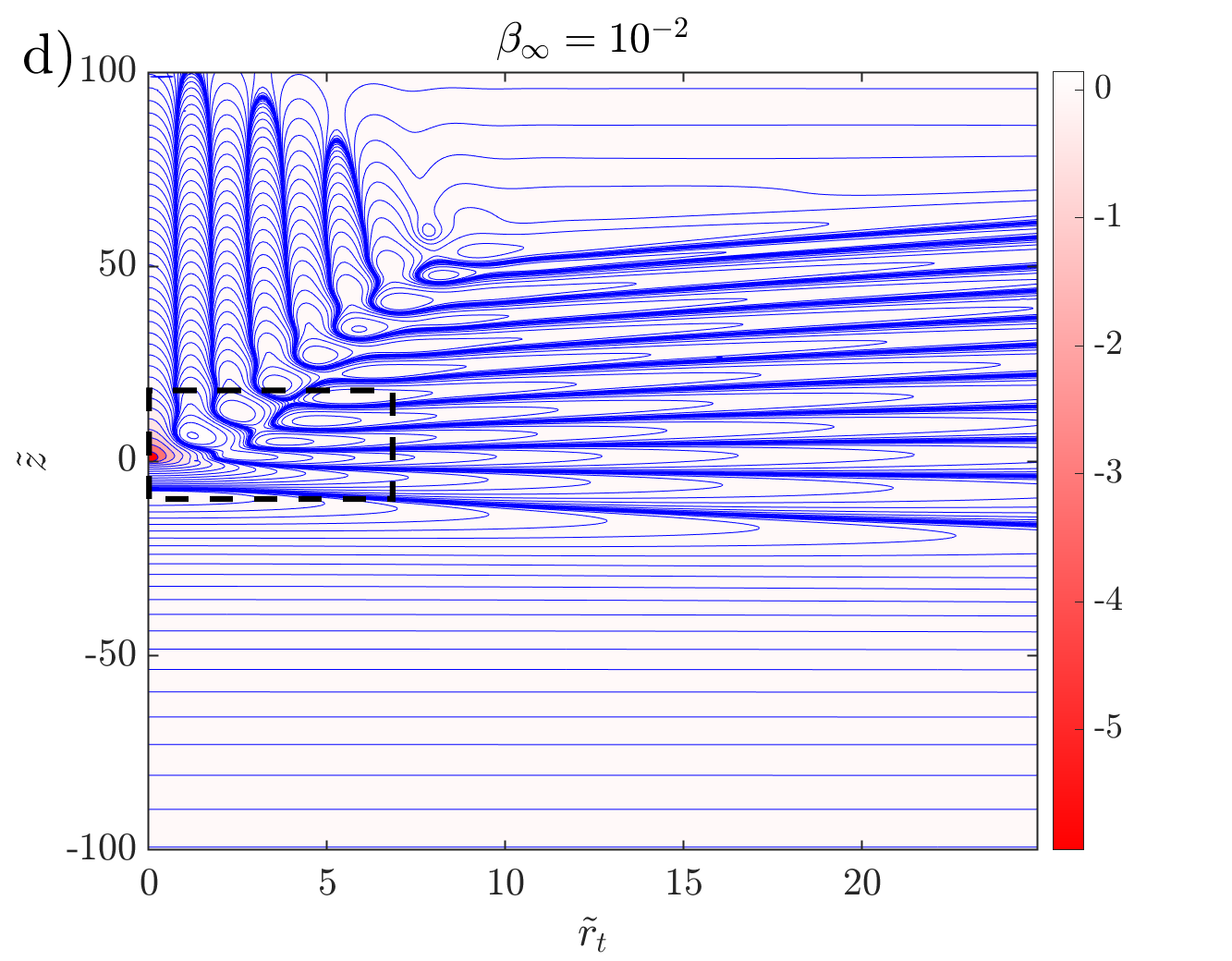}
    \end{minipage}

    \vspace{-0.6cm}
    \hspace{-0.5cm}
    \begin{minipage}[t]{0.55\textwidth}
      \centering
      \includegraphics[width=\textwidth]{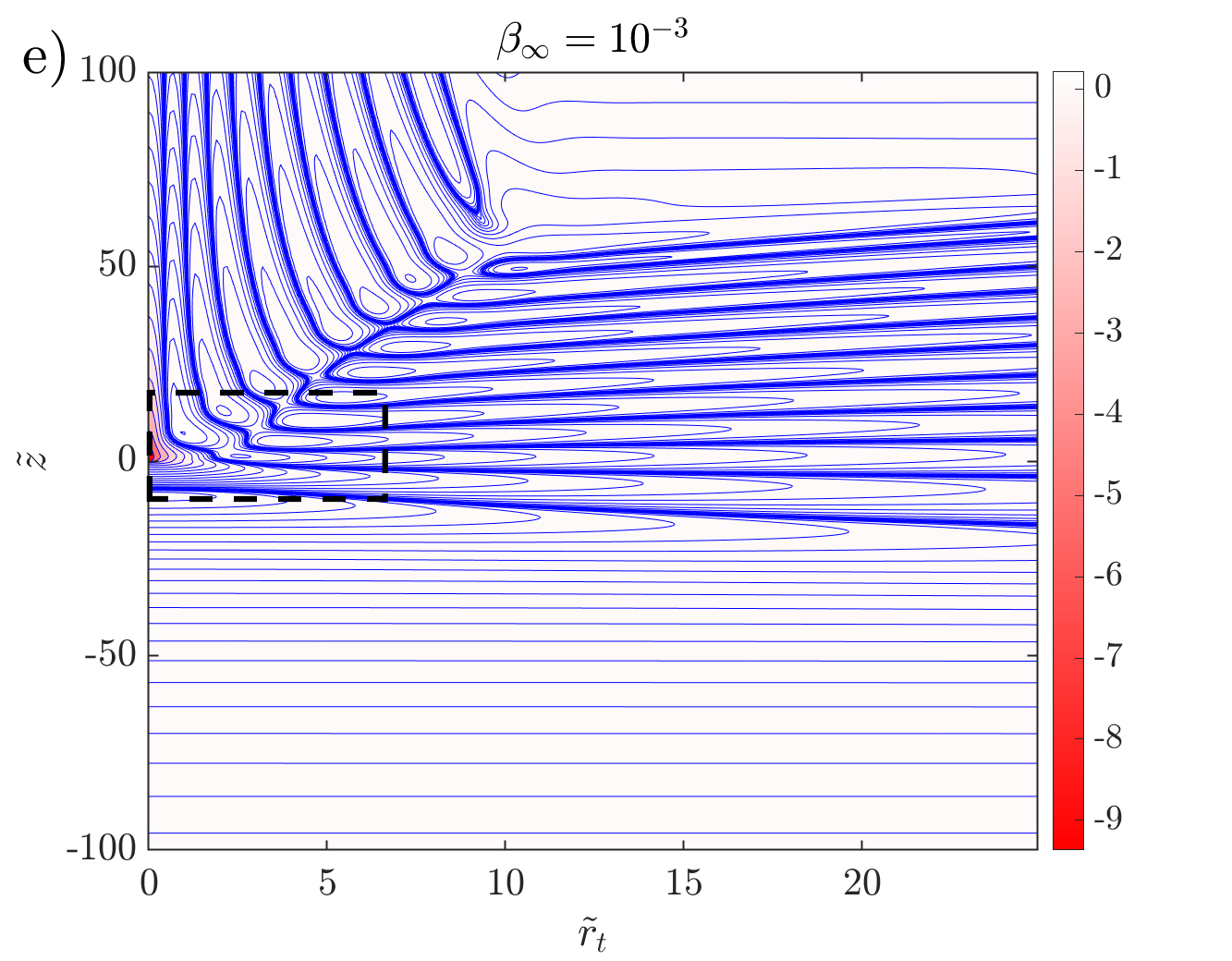}
    \end{minipage}
    \hspace{-0.4cm}
    \begin{minipage}[t]{0.55\textwidth}
      \centering
      \includegraphics[width=\textwidth]{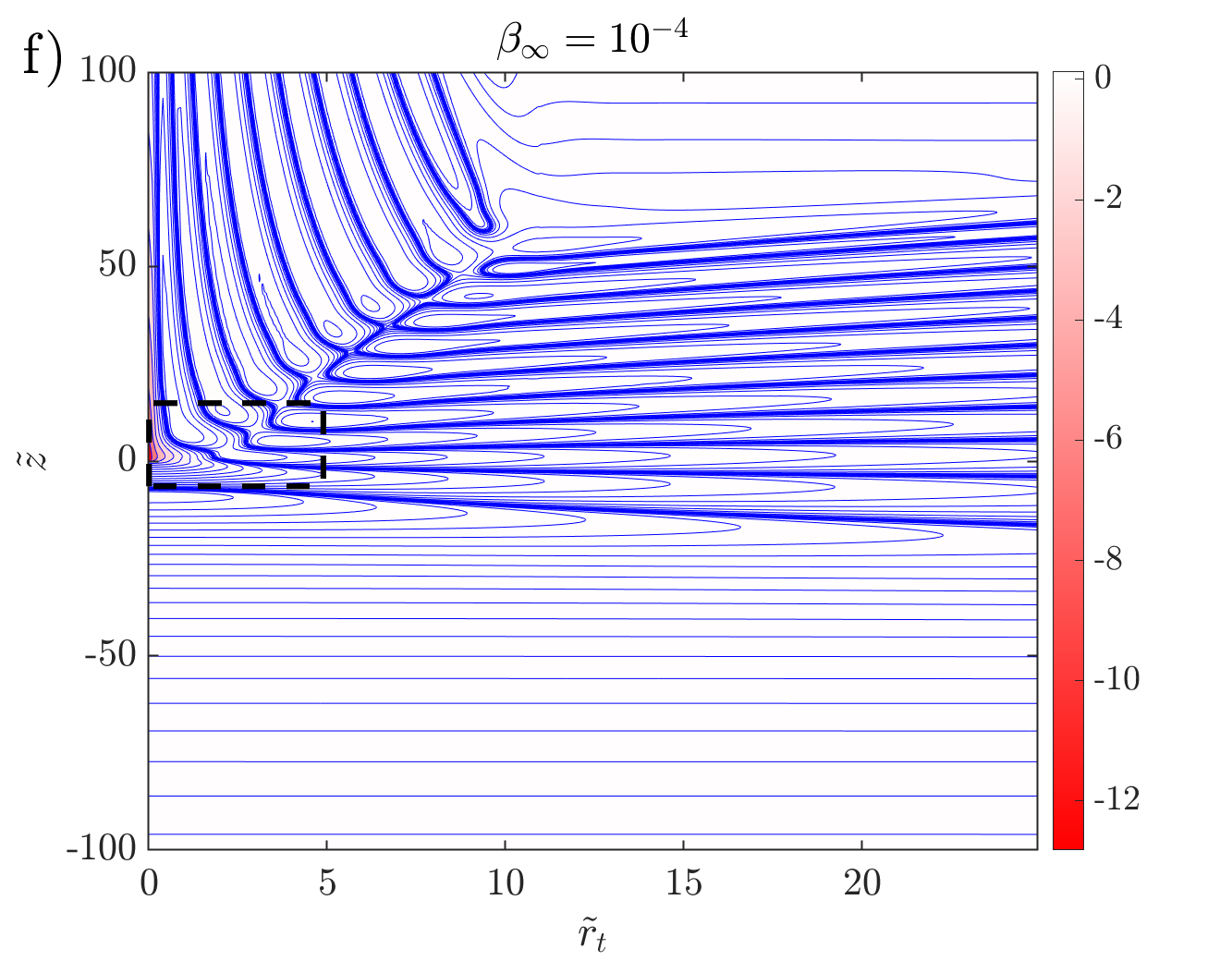}
    \end{minipage}
  \caption{Isopycnal patterns in an axial (half)\,plane at $\beta_\infty = $ (a) $10$, (b) $1$, (c) $10^{-1}$, (d) $10^{-2}$, (e) $10^{-2}$, and (f) $10^{-3}$. The sphere is located at the origin $(\tilde{r}_t,\tilde{z}) = (0,0)$. $\tilde{r}_t$ and $\tilde{z}$ are nondimensionalized using the primary screening length $l_c = \mathcal{O}(aRi_v^{-1/3})$. Here, the dense bands correspond to the zero-crossings (i.e., $\tilde{\rho}_f = 0$) of the density. The portions of the patterns in the region enclosed by the black dashed rectangles correspond to those obtained by \citetalias{varanasi2022motion}.}
  \label{fig:6}
\end{figure}

\begin{figure}
    \hspace{-0.5cm}
    \begin{minipage}[t]{0.55\textwidth}
      \centering
      \includegraphics[width=\textwidth]{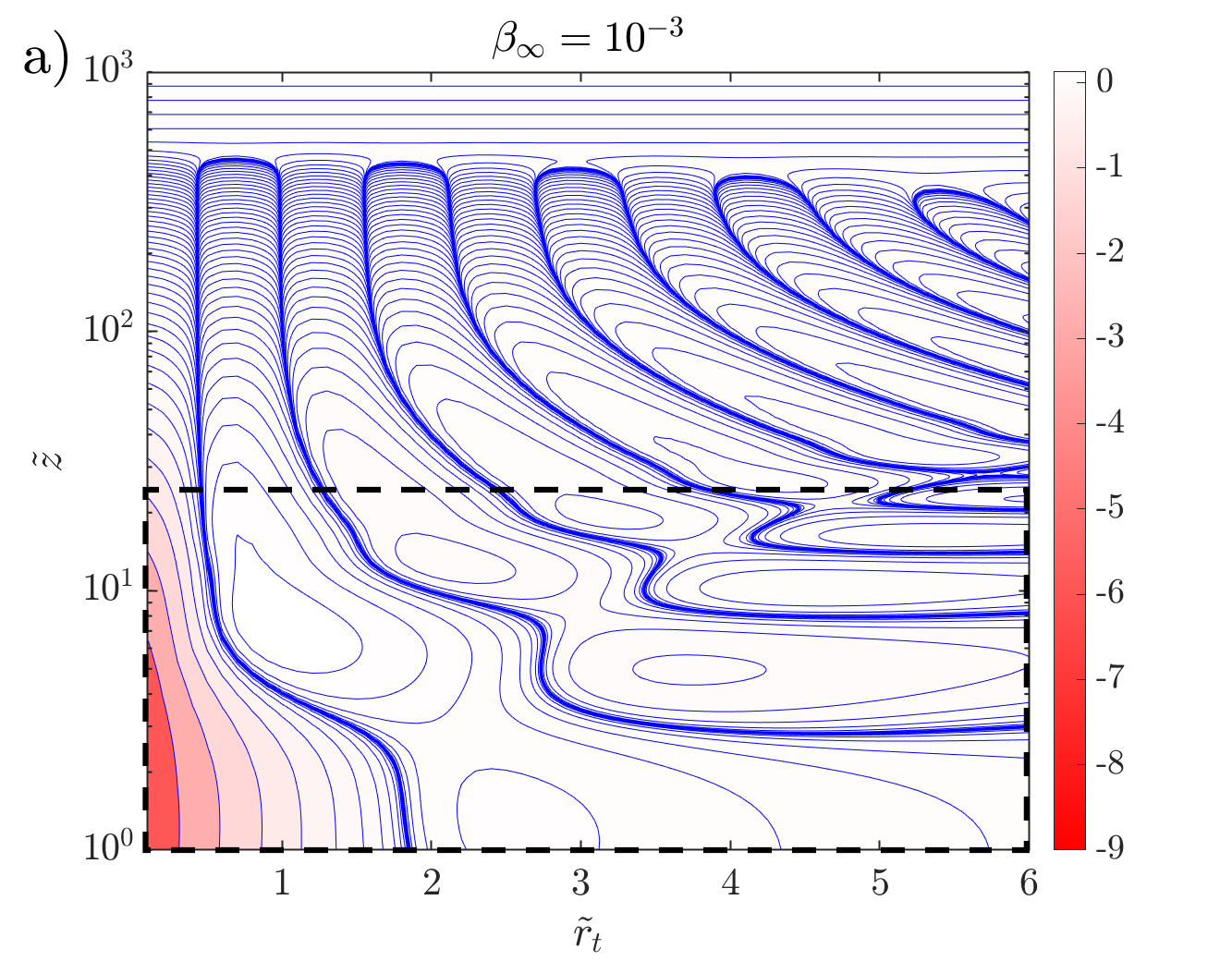}
    \end{minipage}
     \hspace{-0.4cm}
    \begin{minipage}[t]{0.55\textwidth}
      \centering
      \includegraphics[width=\textwidth]{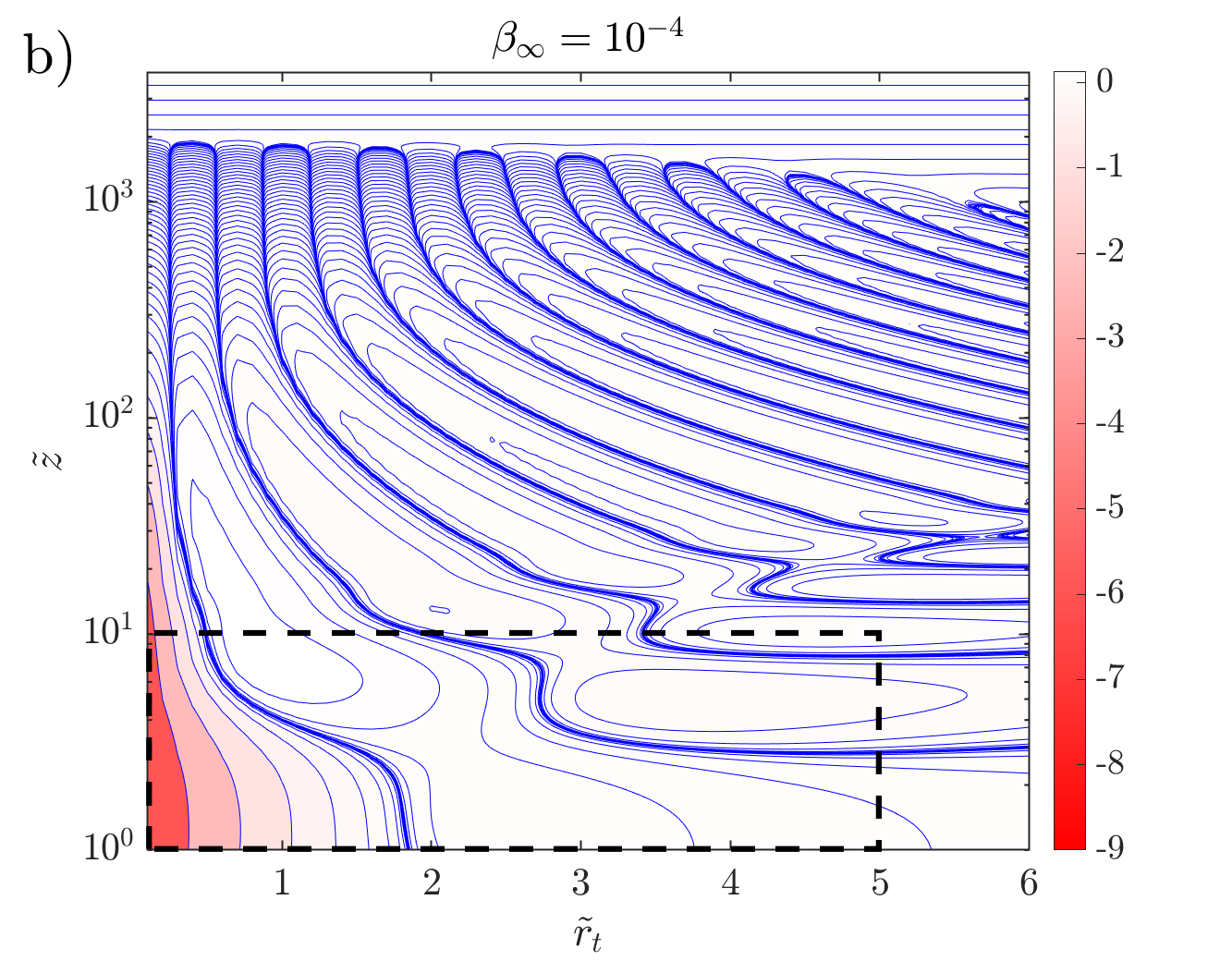}
    \end{minipage}
    
    \vspace{-0.31cm}
    \hspace{-0.5cm}
    \begin{minipage}[t]{0.55\textwidth}
      \centering
      \includegraphics[width=\textwidth]{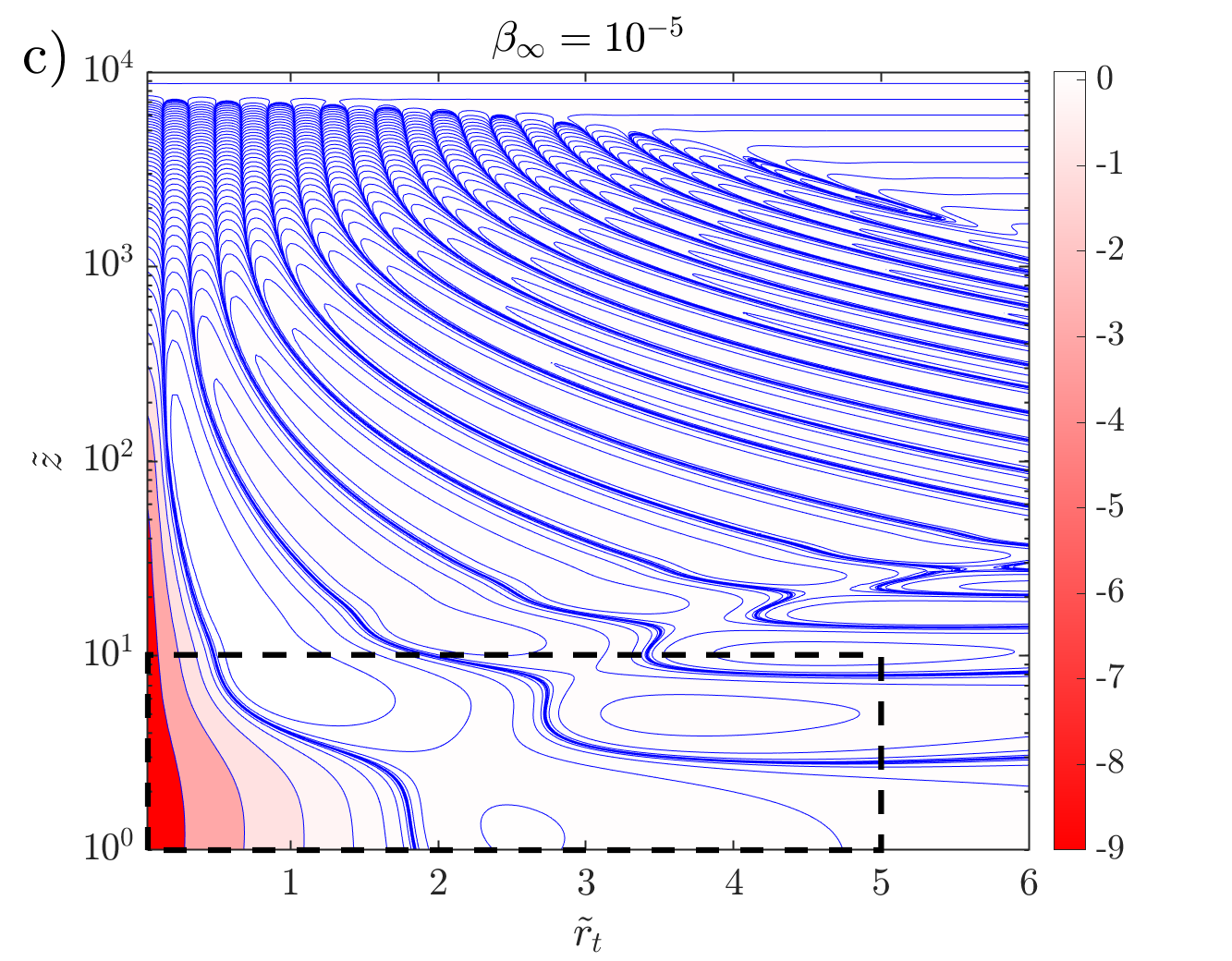}
    \end{minipage}
    \hspace{-0.4cm}
    \begin{minipage}[t]{0.55\textwidth}
      \centering
      \includegraphics[width=\textwidth]{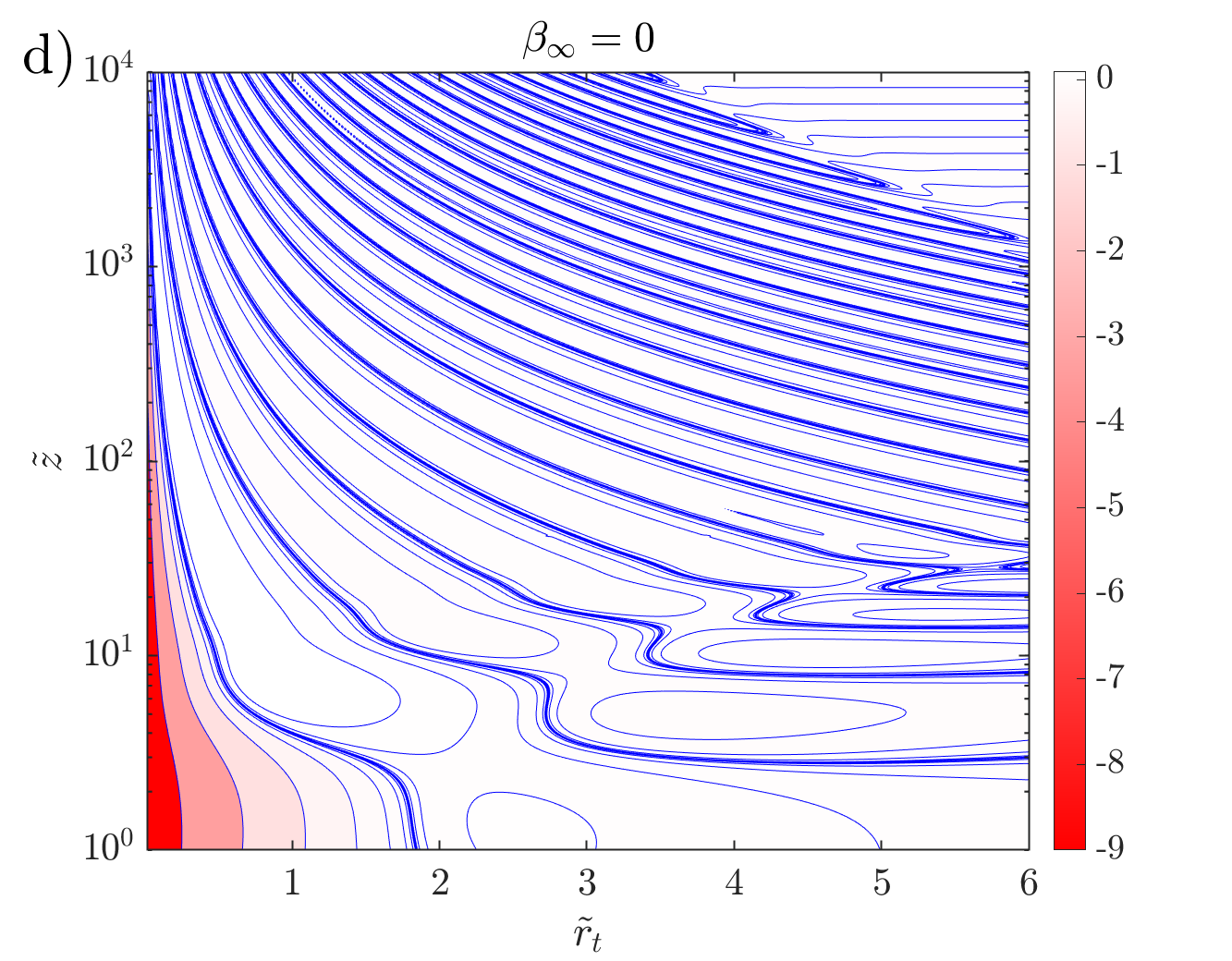}
    \end{minipage}

    \vspace{0.31cm}
    \hspace{-0.5cm}
    \centering
    \begin{minipage}[t]{0.6\textwidth}
      \centering
      \includegraphics[width=\textwidth]{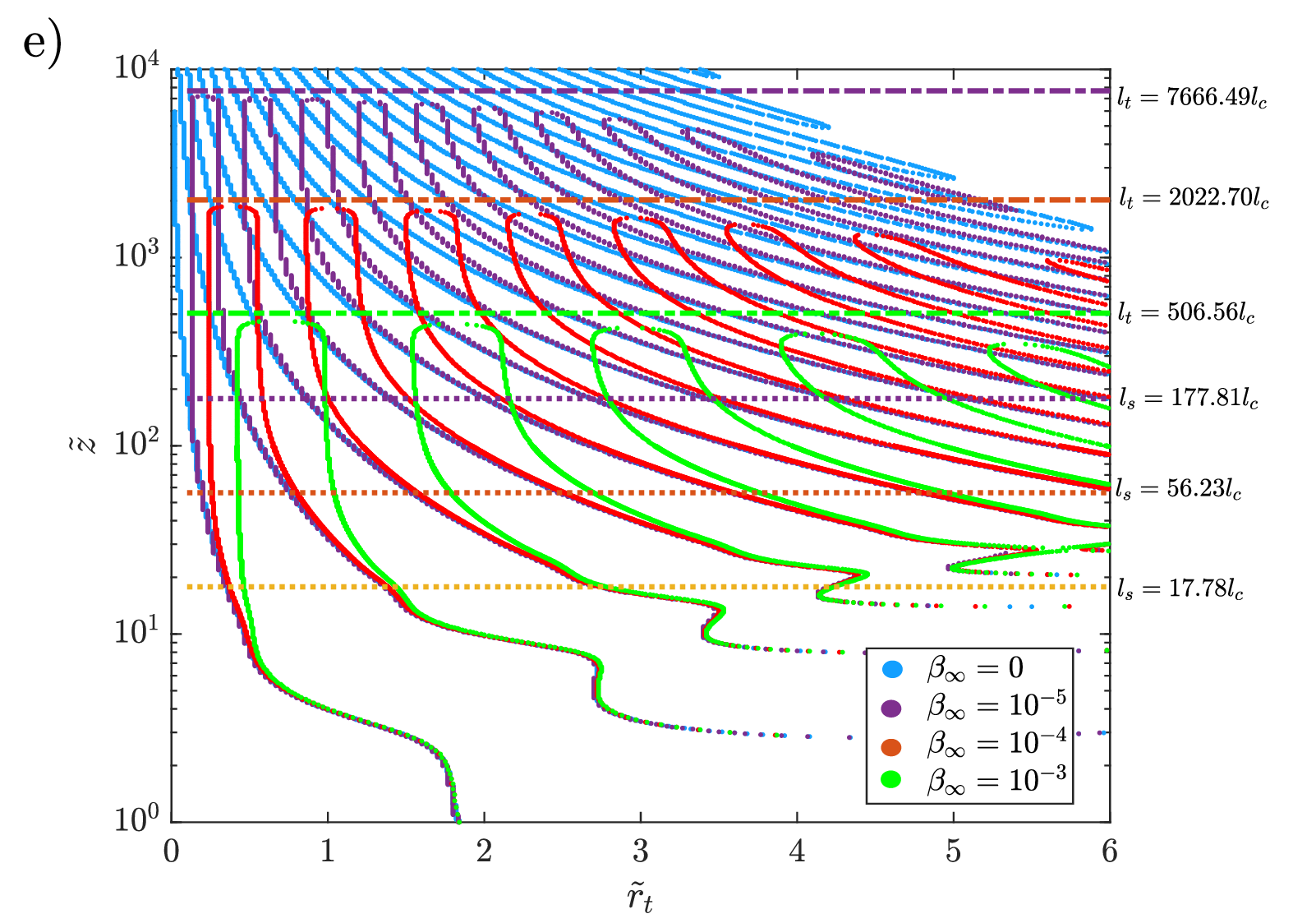}
    \end{minipage}
  \caption{Isopycnal patterns in an axial half\,(plane) (and positive $\tilde{z}$) showing the downstream columnar structure at $\beta_\infty = $ (a) $10^{-3}$, (b) $10^{-4}$, (c) $10^{-5}$, and (d) $0$. The sphere is located at the origin $(\tilde{r}_t,\tilde{z}) = (0,0)$. $\tilde{r}_t$ and $\tilde{z}$ are nondimensionalized using the primary screening length $l_c = \mathcal{O}(aRi_v^{-1/3})$. Here, the dense bands correspond to the zero-crossings of perturbation density (i.e., $\tilde{\rho}_f = 0$). The isopycnal patterns in the regions enclosed by the black dashed rectangles are those obtained by \citetalias{varanasi2022motion}. (e) The columnar structures at various non-zero $\beta_\infty$, compared with the case of $\beta_\infty = 0$. The horizontal dashed and dot-dashed lines indicate secondary ($l_s$) and tertiary ($l_s$) screening lengths for different $\beta_\infty$.}
  \label{fig:7}
\end{figure}
Figure \ref{fig:6} shows color plots (in red) of the perturbation density ($\tilde{\rho}_f$), on which are superimposed blue isopycnal contours. The isopycnal patterns correspond to the same $\beta_\infty$ values as in figure \ref{fig:4} and have the same spatial extents; dashed black rectangles again mark out the domains accessed previously in \citetalias{varanasi2022motion}, with the size of these regions reducing starting from $\beta_\infty =10^{-2}$, owing to numerical convergence issues already mentioned in the introduction. Similar to the streamline patterns, there exist a set of horizontal cells comprising the wake region in each of figures \ref{fig:6}a-f, that is largely insensitive to $\beta_\infty$ in the limit $\beta_\infty \rightarrow 0$ (compare figures \ref{fig:6}e-f); and a set of vertical cells comprising the downstream columnar structure that emerge for $\beta_\infty \lesssim O(10^{-2})$, and that continue to lengthen and increase in number even as $\beta_\infty$ decreases to zero. The decay of the density perturbation outside the aforementioned cellular regions is algebraic, and given in table \ref{table2}. The color plots in figure \ref{fig:6} also help identify the buoyant envelope of fluid\,(bright red) that surrounds the sphere\,(the point force at the origin). While this envelope remains localized for larger $\beta_\infty$\,(see figures \ref{fig:6}a, b and c), it starts to get stretched out in the downstream direction with decreasing $\beta_\infty$, indicative of the sphere dragging along a column of buoyant fluid behind it. 

The structure of the buoyant tail mentioned above is better seen in figures \ref{fig:7}a-d where, on account of the domain extending until the tertiary screening length in the downstream direction, one can also see the columnar structure that surrounds the buoyant tail, in its entirety, down until $\beta_\infty  = 10^{-5}$; features of this structure are analogous to those seen in the streamline patterns in figure \ref{fig:5}. The elongated buoyant tail above, with a length of $O(l_s)$ is one of the most important manifestations of the fore-aft asymmetry that develops in the isopycnal patterns with decreasing $\beta_\infty$. 
Further, as mentioned earlier, the maximum negative value of density in this region
should diverge logarithmically along the entire rear stagnation streamline for $\beta_\infty \rightarrow  0$, as may also be inferred from the expression for the density field obtained within the jet approximation\,($\tilde{\rho}_f = -3K_0[2\beta_\infty^{1/2}\tilde{z}] $; see\citetalias{varanasi2022motion}).

\section{Conclusions}\label{sec:conclusions}

In this paper, we consider the disturbance flow field due to a sphere translating vertically in a viscous stratified fluid ambient, the emphasis being on the convection-dominant limit\,($Pe \gg 1$) within the Stokes-stratification regime; the latter regime is where buoyancy forces are dominant over inertial forces, and therefore first balance viscous forces beyond a (primary)\,screening length\,($a Ri_v^{-1/3}$). \citetalias{varanasi2022motion} showed that the flow field in this limit, at large distances from the sphere, consists of a set of horizontal recirculating cells (the wake) surrounding the sphere in the equatorial plane, and a vertical jet in the rear, directed opposite to the sphere translation. Interestingly, the authors speculated on the existence of additional recirculating cells in the region downstream, and the possibility of the reverse Stokeslet being part of a more elaborate structure behind the translating sphere\,(see bottom of p.18 in \citetalias{varanasi2022motion}). Our results, both analytical and numerical, confirm this speculation, and help quantitatively characterize this more elaborate columnar structure downstream of the sphere. We show that, although this structure extends to downstream infinity in the absence of diffusion\,($\beta_\infty = 0$), for any nonzero $\beta_\infty$, it has a finite length, which we term the tertiary screening length, and that is $\mathcal{O}[Ri_v^{-1/2}Pe^{1/2}\ln(Ri_v^{-1}Pe^3)]$ to leading logarithmic order. The analysis in \citetalias{varanasi2022motion} revealed an exponential decay of the disturbance fields for any non-zero $\beta_\infty$ that, at sufficiently large distances, would have led to the downstream influence of the sphere being smaller than its upstream influence; a feature that defies intuition\,(at least one based on homogeneous fluids at finite Reynolds numbers). However, the aforementioned tertiary screening mechanism intervenes, so to speak, at the right distance, the result being that the aforementioned exponential decay transitions to an algebraic one at larger distances. Further, the rate of algebraic decay is the same as that upstream, albeit with a larger numerical prefactor\,($3240$ as opposed to $2160$). Thus, the resulting downstream influence is still larger than that upstream, although the difference between the two is only a constant of order unity\,(the ratio $3240/2160 = 1.5$). A more profound upstream-downstream asymmetry will have to await consideration of inertial forces.
\begin{figure}
    \hspace{-0.5cm}
    \begin{minipage}[t]{\textwidth}
      \centering
      \includegraphics[width=\textwidth]{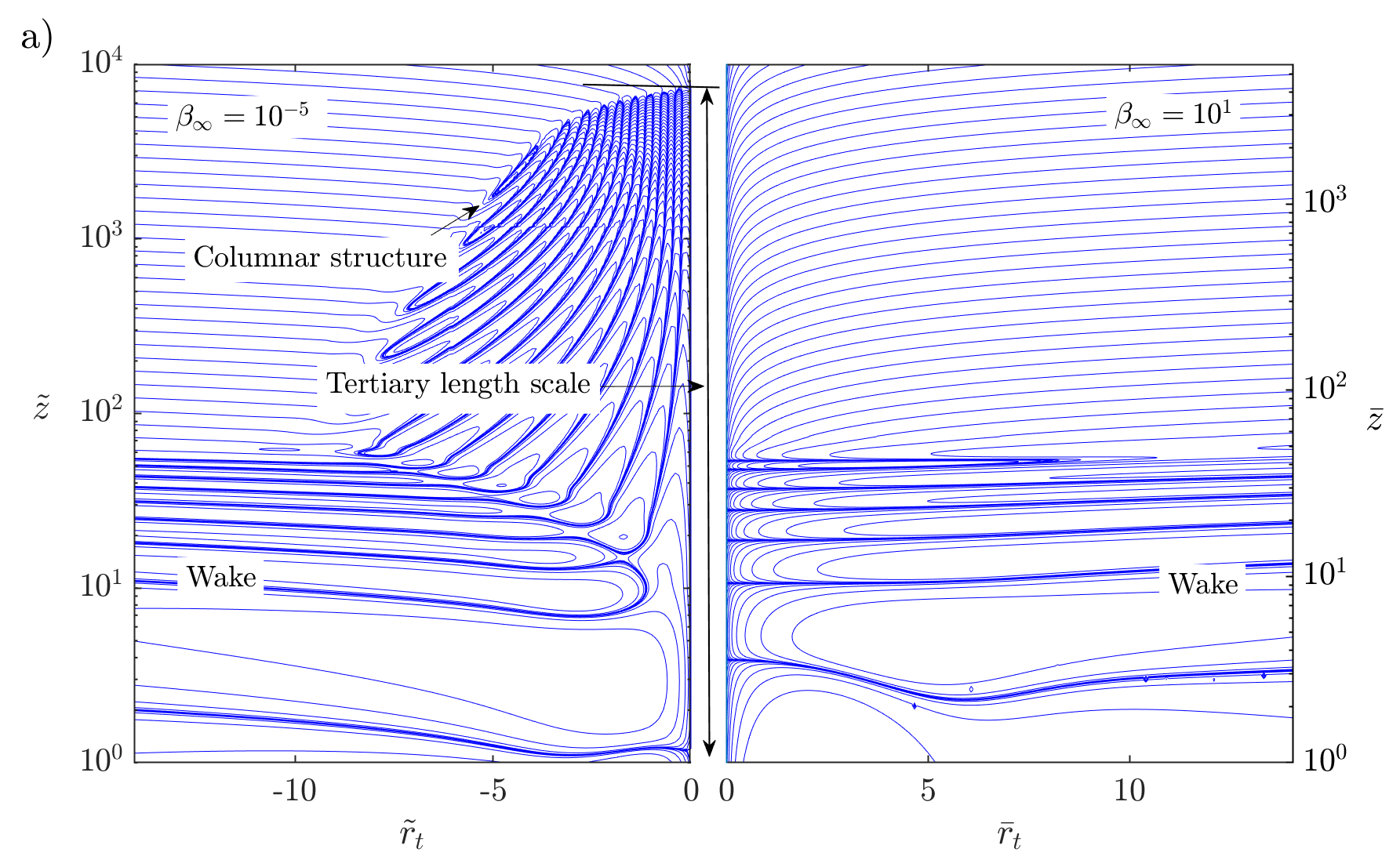}
    \end{minipage}

    \hspace{-0.4cm}
    \begin{minipage}[t]{\textwidth}
      \centering
      \includegraphics[width=\textwidth]{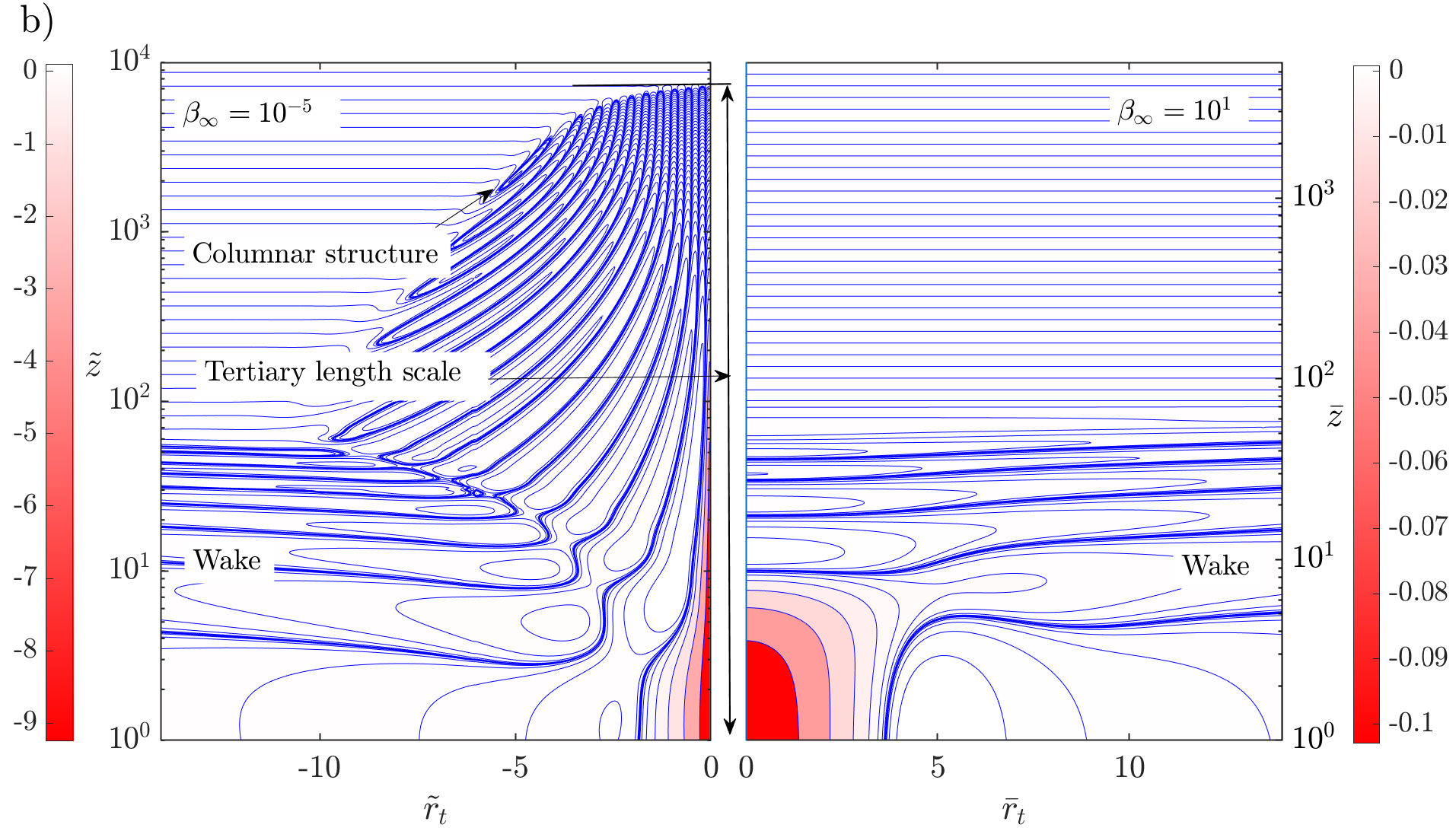}
    \end{minipage}
  \caption{a) Streamline, and b) isopycnal patterns in an axial (half) plane plane for $\beta_\infty = 10^{-5}$ (for $-14<\tilde{r}_t<0$) and $10$ (for $0<\bar{r}_t<14$); only the downstream region, corresponding to $\tilde{z} > 1$, is shown. Please note that the radial and axial coordinates for $\beta_\infty = 10^{-5}$ $(\tilde{r}_t, \tilde{z})$ are scaled by the large-$Pe$ stratification screening length of $\mathcal{O}[a(Ri_v)^{-1/3}]$, whereas those for $\beta_\infty = 10$ $(\bar{r}_t, \bar{z})$ are scaled by the small-$Pe$ screening length of $\mathcal{O}[a(Ri_vPe)^{-1/4}]$.}
  \label{fig:final}
\end{figure}

It is worth ending our investigation with figures \ref{fig:final}a and b - the first one containing streamline patterns and the second one containing iso-pycnals - that help contrast the structure of the disturbance fields in the diffusion-dominant and convection-dominant limits. This contrast is achieved by having each of these figures consist of two halves, one corresponding to $\beta_\infty \gg 1$, and the other to $\beta_\infty \ll 1$. It is worth mentioning that the diffusion-dominant halves in the said figures differ from the versions presented earlier\,(figures \ref{fig:4}a and \ref{fig:6}a) because, in the interest of a fair comparison, the length scale used in figures \ref{fig:final}a and b is now the small-$Pe$ screening length of $\mathcal{O}[a(Ri_vPe)^{-1/4}]$; the ratio of this length scale to $aRi_v^{-1/3}$ is $\beta_\infty^{1/4}$, and the large-$\beta_\infty$ halves in figures \ref{fig:final}a and b are rescaled versions of figures \ref{fig:4}a and \ref{fig:6}a, obtained by multiplying the axes with $\beta_\infty^{1/4}$.

\textbf{Declaration of interest.} The authors report no conflicts of interest
\backsection[Acknowledgements]{We acknowledge the use of the computing resources at HPCE, IIT Madras. R.P. acknowledges fruitful discussions with A. Varanasi during the early stages of this work. R.P. and A.R. acknowledge support from IIT Madras for its support of the ‘Geophysical Flows Lab’ research initiative under the Institute of Eminence framework}
\appendix
\section{Disturbance flow field Fourier integral simplification}
\label{sec:AppA}
\subsection{Stokes streamfunction}
The expression for the Stokes streamfunction, obtained via an inverse Fourier transform, is as given in \eqref{eq:psis}. For the sake of continuity, it is repeated here:
\begin{equation}
\label{eq:A1}
    \tilde{\psi}_s(\textbf{\~{r}}) = \dfrac{3\tilde{r}_t i}{4\pi^2}\displaystyle\int\dfrac{(ik_3+\beta_\infty k^2)k_2}{(ik_3+\beta_\infty k^2)k^4+k_t^2}e^{i\textbf{k}\cdot\textbf{\~{r}}}d\textbf{k}.
\end{equation}
For ease of numerical evaluation, the Stokeslet contribution can be separated out first in a manner similar to eq 3.4 of \citetalias{varanasi2022motion}, in which case \eqref{eq:A1} takes the form
\begin{equation}
\label{eq:A2}
    \tilde{\psi}_s(\textbf{\~{r}}) = \dfrac{3\tilde{r}_t i}{4\pi^2}\displaystyle\int\left(\dfrac{k_2}{k^4} -\dfrac{k_t^2k_2}{(ik_3+\beta_\infty k^2)k^4+k_t^2} \right) e^{i\textbf{k}\cdot\textbf{\~{r}}}d\textbf{k}.
\end{equation}
Here, the first term within brackets corresponds to the Stokeslet contribution and can be evaluated analytically. The 3D integral in the second term needs to be evaluated numerically; however, it can first be reduced to a 2D integral by a series of simplifications. Using spherical polar coordinates ($k,\theta,\phi$) in Fourier space for the second term, with the polar axis along $\boldsymbol{1}_3$, \eqref{eq:A2} can be written as 
\begin{equation}
\label{eq:A3}
    \tilde{\psi}_s(\tilde{r}_t,\tilde{z}) = \displaystyle -\dfrac{3\tilde{r}_t^2}{4\sqrt{\tilde{r}_t^2+\tilde{z}^2}} -\dfrac{3\tilde{r}_t i}{4\pi^2} \int_0^{\pi}d\theta\int_{-\pi}^{\pi}d\phi \int_0^{\infty}dk \dfrac{\sin^4\theta\sin\phi\exp(ik\bar{\delta}_1)}{k\left(i k^3 \cos\theta + \beta_{\infty} k^4 + \sin^2\theta\right)},
\end{equation}
where $k_t = \sqrt{k_1^2+k_2^2} = k\sin\theta$, $k_3 = k\cos\theta$, $k_2 = k\sin\theta\sin\phi$, and $\bar{\delta}_1 = \tilde{r}_t\sin\theta\sin\phi + \tilde{z}\cos\theta$. Here, the first term is the Stokeslet contribution in physical space. The second term can be simplified further based on angular symmetry, resulting in
\begin{equation}
\label{eq:A4}
    \tilde{\psi}_s = -\displaystyle \dfrac{3\tilde{r}_t^2}{4\sqrt{\tilde{r}_t^2+\tilde{z}^2}} - \dfrac{3\tilde{r}_t i}{2\pi^2} \int_0^{\pi/2}d\theta\int_{0}^{\pi/2}d\phi \int_{-\infty}^{\infty}dk \dfrac{\sin^4\theta\sin\phi\left(\exp(ik\bar{\delta}_1)-\exp(-ik\delta_1)\right)}{k\left(i k^3 \cos\theta + \beta_{\infty} k^4 + \sin^2\theta\right)},
\end{equation}
with $\delta_1 = \tilde{r}_t\sin\theta\sin\phi - \tilde{z}\cos\theta$. Note that $\delta_1$ and $\bar{\delta}_1$ in \eqref{eq:A3} and \eqref{eq:A4} have the same meaning as $\delta$ and $\bar{\delta}$ in \S~\ref{sec:Theory} but with $s = 1$ (hence the subscript) and with changed variables $\sin\theta = \sqrt{1-y^2}$ and $\sin\phi = \sqrt{1-x^2}$.

\begin{figure}
    \hspace{-0.5cm}
    \begin{minipage}[t]{1.05\textwidth}
      \centering
      \includegraphics[width=\textwidth]{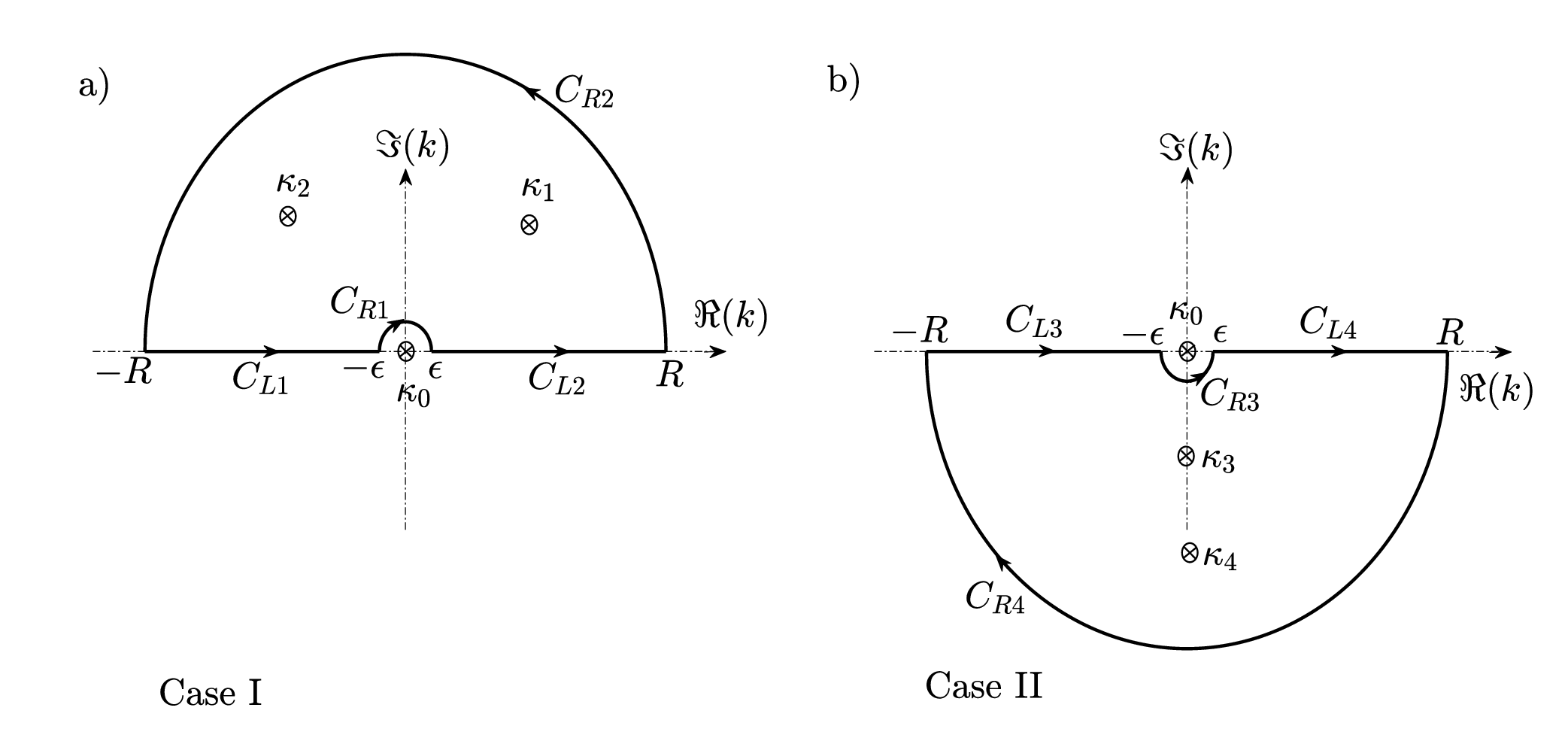}
    \end{minipage}
  \caption{Contours of integration used in solving the $k$-integral in \eqref{eq:A4}. Here, subfigure (a) shows a contour in the upper half of the complex$-k$ plane, used when the argument of the complex exponentials in \eqref{eq:A4} is positive (case I), whereas (b) shows a contour in the lower half of the complex$-k$ plane used when the argument of the complex exponentials in \eqref{eq:A4} is negative (case II). $C_{L1}-C_{L4}$ are integrals along the real line. $R$ is the radius of the curves $C_{R2}~\&~C_{R4}$ and $\epsilon$ is the radius of curves $C_{R1}~\&~C_{R3}$. $\kappa_0$ to $\kappa_4$ are zeros of the denominator in the integrand of \eqref{eq:A4}}
  \label{fig:contour}
\end{figure}

As mentioned in \S~\ref{sec:Theory}, we first evaluate the $k-$ integral in \eqref{eq:A4} using contour integration. However, the contour should be carefully chosen considering the signs of $\delta_1$ and $\bar{\delta}_1$ in the arguments of the exponentials, so as to satisfy Jordan's lemma. Here, we present the steps for positive $\tilde{z}$; the expressions for negative $\tilde{z}$ can be derived similarly. For positive $\tilde{z}$, $\bar{\delta}_1$ in \eqref{eq:A4} is always positive, whereas $\delta_1$ can be either positive for $\theta < \theta_c$ or negative for $\theta > \theta_c$; $\theta_c\left( = \arctan\left(\tilde{z}/\tilde{r}_t\sin\phi\right)\right)$ being the critical polar angle at which $\delta_1 = 0$. So, for the term involving the first exponential and for the term involving the second exponential when $\delta_1<0$, a contour in the upper half of the complex$-k$ plane, as shown in figure \ref{fig:contour}a (`case I'), should be chosen. When $\delta_1>0$, the contour for the term involving the second exponential should be chosen in the lower half of the complex$-k$ plane, as shown in figure \ref{fig:contour}b (`case II'). Therefore, one can write,
\begin{equation}
    \label{eq:A5}
    \int_{-\infty}^{\infty}dk I = \displaystyle \lim_{R\to\infty, \epsilon\to 0} \left(\int_{C_{L1}} + \int_{C_{L2}}\right)I = \lim_{R\to\infty, \epsilon\to 0}\left(\oint_{\mathrm{I}} - \int_{C_{R1}} - \cancelto{0}{\int_{C_{R2}}}\hspace{0.5cm}\right)I,
\end{equation}
where the LHS denotes the integral corresponding to case I in \eqref{eq:A4}. The integral along the curve $C_{R2}$ has been set to zero due to Jordan's lemma \citep[see Lemma 4.2.2 in][]{ablowitz2003complex}. For case II, $C_{L1}, C_{L2}, C_{R1}, C_{R2}$ and I in \eqref{eq:A5} should be replaced by $C_{L3}, C_{L4}, C_{R3}, C_{R4}$ and II, respectively. Also, note that the signs of the contour integrals\,($\oint_I$ or $\oint_{II}$), and the integrals along the curves $C_{R1}~\&~C_{R3}$, depend on sense\,(clockwise/anticlockwise) in which the contours/curves are traversed; see figures \ref{fig:contour}a,b. The contribution from the curves $C_{R1}$ and $C_{R3}$ can be evaluated \citep[using theorem 4.3.1 in][]{ablowitz2003complex} to be $-i\pi\sin^2\theta\sin\phi$ and $i\pi\sin^2\theta\sin\phi$, respectively. 

\begin{figure}
    \hspace{-0.5cm}
    \begin{minipage}[t]{0.55\textwidth}
      \centering
      \includegraphics[width=\textwidth]{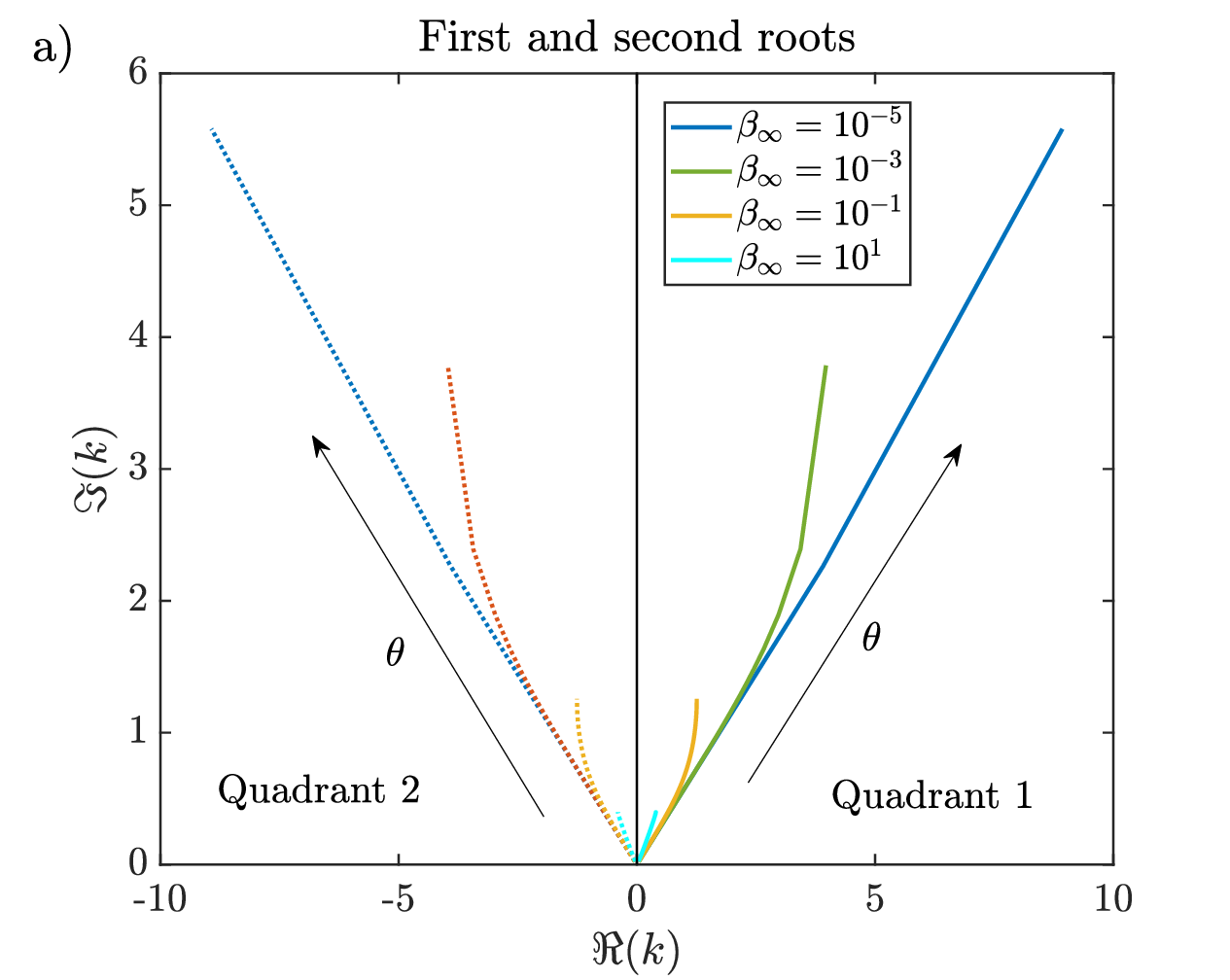}
    \end{minipage}
    \hspace{-0.4cm}
    \begin{minipage}[t]{0.55\textwidth}
      \centering
      \includegraphics[width=\textwidth]{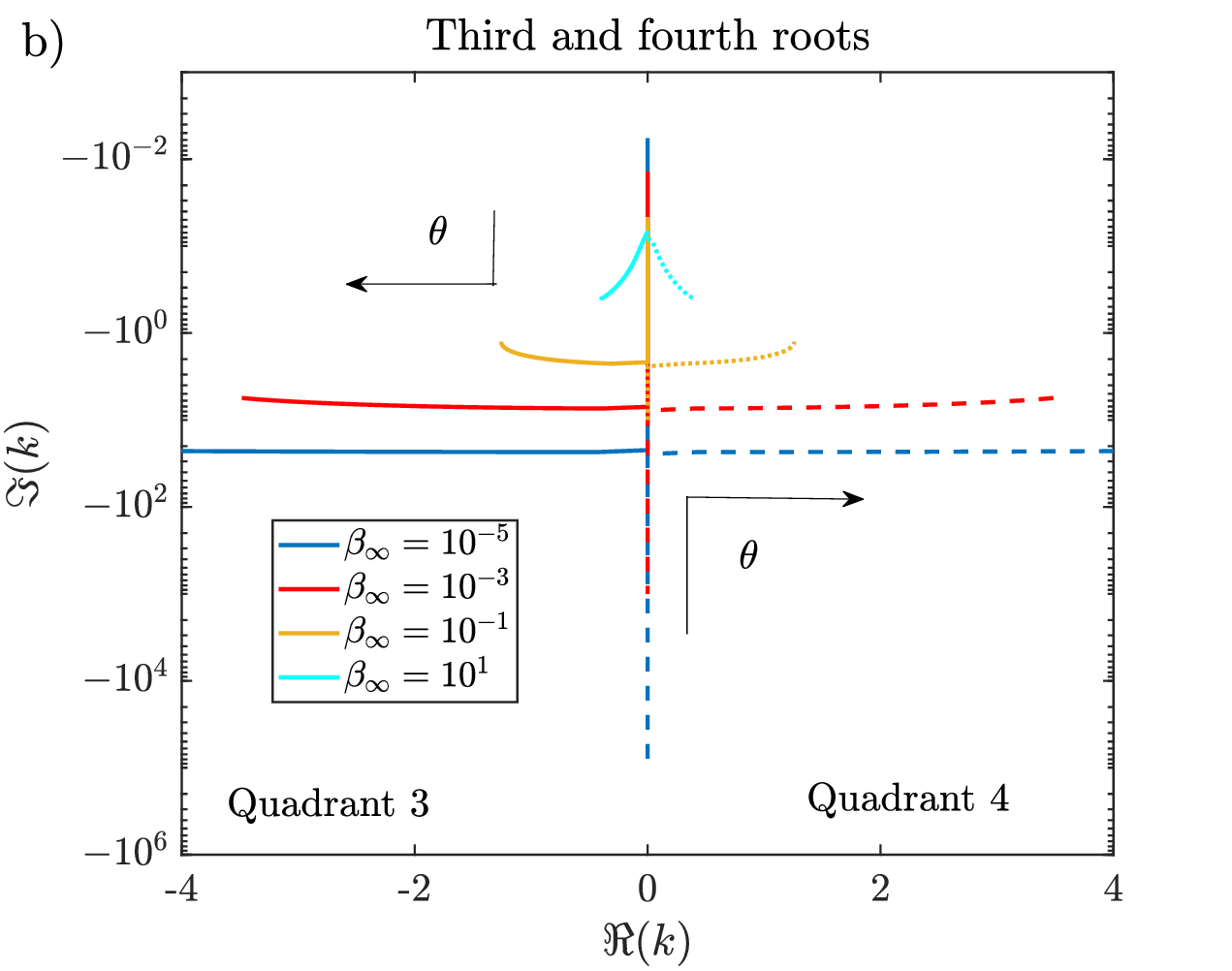}
    \end{minipage}
    \caption{The behavior of a) first and second roots, and b) third and fourth roots of the quartic polynomial ($\beta_\infty k^4+ik^3\cos\theta + \sin^2\theta$), as a function of $\theta$ and for various $\beta_\infty$, in the complex$-k$ plane. Here, $\theta$ varies from $0$ to $\pi/2$ and increases in the direction shown by the arrows. The continuous and dashed lines correspond to root 1 (3) and 2 (4) in subfigure a (b), respectively.}
  \label{fig:roots}
\end{figure}

To evaluate the contour integral that remains, the residues at the simple poles of the integrand need to be calculated \citep[see 4.1.10 in][]{ablowitz2003complex}. These correspond to the zeroes of the denominator of the integrand excluding the origin, and therefore, to the four roots of the quartic polynomial $\beta_\infty k^4 + ik^3\cos\theta +\sin^2\theta$. Figure \ref{fig:roots} shows the behavior of the roots in the complex$-k$ plane, as a function of $\theta$ ($0\leq \theta \leq \pi/2$), for various $\beta_\infty$. Two roots exist in the upper half of the complex$-k$ plane and are located symmetrically on either side of the positive imaginary axis. The remaining two roots exist in the lower half of the complex$-k$ plane; they can be purely imaginary or can lie symmetrically on either side of the negative imaginary axis, depending on $\theta$; note that one of the roots starts at $-\beta_\infty^{-1} i$ at $\theta = 0$, as shown in figure \ref{fig:roots}b. We number the roots $\kappa_1-\kappa_4$ in an anti-clockwise sense, so roots $\kappa_1$ and $\kappa_2$ lie in the upper half, while $\kappa_3$ and $\kappa_4$ lie in the lower half of the complex$-k$ plane.

After contour integration, the integral involving the fist exponential, in \eqref{eq:A4}, can be written as
\begin{equation}
\label{eq:A6}
   \displaystyle \dfrac{3\tilde{r}_t}{2 \pi} \int_0^{\pi/2}d\theta\int_{0}^{\pi/2}d\phi \left[2 \sum_{m=1}^{2}\dfrac{\sin^4\theta\sin\phi\exp(i \kappa_m\bar{\delta}_1)}{\kappa_m^3(3i\cos\theta+4\beta_\infty \kappa_m)} +  \sin^2\theta\sin\phi\right],
\end{equation}
where the first term in the brackets is due to the residues associated with $\kappa_1$ and $\kappa_2$, and the second term is due to the integral along $C_{R1}$. The integral in \eqref{eq:A4}, involving the second exponential, can be written as
\begin{align}
\label{eq:A7}
  \displaystyle -\dfrac{3\tilde{r}_t}{2\pi} \int_{0}^{\pi/2}d\phi &\left(\int_0^{\theta_c}d\theta \sum_{m=1}^{2} -  \int_{\theta_c}^{\pi/2}d\theta  \sum_{m=3}^{4} \right)\dfrac{2\sin^4\theta\sin\phi\exp(-i \kappa_m\delta_1)}{\kappa_m^3(3i\cos\theta+4\beta_\infty \kappa_m)} \\ \nonumber & -\dfrac{3\tilde{r}_t}{2\pi}  \int_{0}^{\pi/2}d\phi\left(\int_0^{\theta_c}d\theta -\int_{\theta_c}^{\pi/2}d\theta \right) \sin^2\theta\sin\phi,
\end{align}
where the first term is due to the roots of the quartic polynomial, and the second term is due to the integral along $C_{R1}$ and $C_{R3}$. The integrals corresponding to $C_{R1}$ and $C_{R3}$ have been given above, and taken together, will cancel the Stokeslet contribution in \eqref{eq:A4}. Finally, one can express the Stokes streamfunction as,
\begin{equation}
    \label{eq:A8}
    \tilde{\psi}_s = \dfrac{3\tilde{r}_t}{2 \pi} (T_{\psi 1}+T_{\psi 2}),
\end{equation}
where,
\begin{align}
\label{eq:A9}
   \displaystyle & T_{\psi 1}(\tilde{z}>0) =  \int_{0}^{\pi/2}d\phi \Bigg[\int_0^{\pi/2}d\theta \sum_{m=1}^{2}\dfrac{2\sin^4\theta\sin\phi\exp(i \kappa_m\bar{\delta}_1)}{\kappa_m^3(3i\cos\theta+4\beta_\infty \kappa_m)}\Bigg] \\ \nonumber & T_{\psi 2}(\tilde{z}>0) =- \int_{0}^{\pi/2}d\phi \Bigg[\left(\int_0^{\theta_c}d\theta \sum_{m=1}^{2} -  \int_{\theta_c}^{\pi/2}d\theta  \sum_{m=3}^{4} \right)\dfrac{2\sin^4\theta\sin\phi\exp(-i \kappa_m\delta_1)}{\kappa_m^3(3i\cos\theta+4\beta_\infty \kappa_m)} \Bigg].
\end{align}
Using similar arguments as above, the expressions for $\tilde{z}<0$ can also be written as
\begin{align}
\label{eq:A10}
   \displaystyle & T_{\psi 1}(\tilde{z}<0) =  \int_{0}^{\pi/2}d\phi \Bigg[ \int_0^{\pi/2}d\theta \sum_{m=3}^{4}\dfrac{2\sin^4\theta\sin\phi\exp(-i \kappa_m\delta_1)}{\kappa_m^3(3i\cos\theta+4\beta_\infty \kappa_m)} \Bigg] \\ \nonumber &  T_{\psi 2}(\tilde{z}<0) = -\int_{0}^{\pi/2}d\phi \Bigg[\left(\int_0^{\theta_c}d\theta \sum_{m=3}^{4} -  \int_{\theta_c}^{\pi/2}d\theta  \sum_{m=1}^{2} \right)\dfrac{2\sin^4\theta\sin\phi\exp(i \kappa_m\bar{\delta}_1)}{\kappa_m^3(3i\cos\theta+4\beta_\infty \kappa_m)}\Bigg].
\end{align}
The $\phi-$integral in the above expressions can be evaluated analytically only for those terms with the $\theta-$integral limits independent of $\phi$ i.e., for $T_{\psi 1}$ in \eqref{eq:A8} and in \eqref{eq:A9}, leading in these cases to a single $\theta-$integral in terms of Bessel and Struve functions as given in \eqref{eq:2.12}. Finally, by using the variables $(x,y) = (\sin\phi,\sin\theta)$, expressions \eqref{eq:2.11}-\eqref{eq:2.12} are recovered. 

As mentioned earlier, root $\kappa_4(\theta = 0) = -\beta_\infty^{-1} i$. Therefore, for $\beta_\infty \to 0$,  $\kappa_4$ approaches $-\infty i$, making the evaluation of the residue at $\kappa_4$ difficult. So, we set $\beta_\infty = 0$ in \eqref{eq:A4}, reducing the order of the polynomial in the denominator by one. The resulting expression can be simplified as
\begin{align}
\label{eq:A11}
    &\tilde{\psi}_s = -\displaystyle \dfrac{3\tilde{r}_t^2}{4\sqrt{\tilde{r}_t^2+\tilde{z}^2}} + \\ \nonumber & \dfrac{3\tilde{r}_t}{\pi^2} \int_0^{\pi/2}d\theta\int_{0}^{\pi/2}d\phi \int_{0}^{\infty}dk \dfrac{\sin^4\theta\sin\phi\left(\sin^2\theta(\sin k\bar{\delta}_1 + \sin k \delta_1) - k^3\cos\theta(\cos k\bar{\delta}_1 - \cos k\delta_1)\right)}{k(k^6\cos^2\theta + \sin^4\theta)},
\end{align}
In evaluating the $k-$integral for the terms including $\sin k\bar{\delta}_1$ and $\sin k\delta_1$, we use identity (3.738.1) from \citet{gradstein2007table}. For the terms including $\cos k\bar{\delta}_1$ and $\cos k\delta_1$, we use identity (3.738.2). Please note that the said identities are applicable only when the arguments of $\sin k\bar{\delta}_1$ (or $\sin k\delta_1$) and $\cos k\bar{\delta}_1$ (or $\cos k\delta_1$) are positive. Therefore, when $\delta_1$ ($\bar{\delta}_1$) is negative, for positive (negative) $\tilde{z}$, it should be replaced with $-|\delta_1|$ ($-|\bar{\delta}_1|$). So one obtains, for example, $\sin k\delta_1 = \sin (-k|\delta_1|) = -\sin k|\delta_1|$ and $\cos k\delta_1 = \cos (-k|\delta_1|) = \cos k|\delta_1|$. Writing the final expressions in a compact way, \eqref{eq:2.13} is recovered.

\subsection{Density}
For arbitrary $\beta_\infty$, the inverse Fourier transform integral for the density perturbation is given in \eqref{eq:2.10} 
\begin{equation}
\label{eq:A12}
    \tilde{\rho}_f(\textbf{\~{r}}) = \dfrac{-3}{4\pi^2}\displaystyle\int\dfrac{k_t^2}{(ik_3+\beta_\infty k^2)k^4+k_t^2}e^{i\textbf{k}\cdot\textbf{\~{r}}}d\textbf{k}.
\end{equation}
Again, use of spherical polar coordinates in Fourier space leads to 
\begin{equation}
\label{eq:A13}
    \tilde{\rho}_f = - \displaystyle \dfrac{3}{2\pi^2} \int_0^{\pi/2}d\theta\int_{0}^{\pi/2}d\phi \int_{-\infty}^{\infty}dk \dfrac{\sin^3\theta k^2\left(\exp(ik\bar{\delta}_1)+\exp(-ik\delta_1)\right)}{\left(i k^3 \cos\theta + \beta_{\infty} k^4 + \sin^2\theta\right)} = -\dfrac{3}{2\pi}(T_{\rho1} + T_{\rho2}).
\end{equation}
The $k$-integral here can be performed using contour integration similar to the earlier subsection. An important difference is that the contour need not be indented to exclude the origin, i.e., the small semi-circular curve ($C_{R1}$ and $C_{R3}$), as there is no pole at $k=0$. Therefore, the integration contour is a semi-circle either in the upper quadrants or the lower quadrants, depending on the sign of $\delta_1$ and $\bar{\delta}_1$. For $\tilde{z}>0$, after the $k-$integration, one obtains
\begin{equation}
\label{eq:A14}
    T_{\rho1} = 2i \displaystyle  \int_0^{\pi/2}d\theta\int_{0}^{\pi/2}d\phi  \sum_{m=1}^{2}\dfrac{\sin^3\theta \exp(i\kappa_m\bar{\delta}_1)}{\left(3i\cos\theta+4\beta_\infty \kappa_m\right)},
\end{equation}

\begin{equation}
\label{eq:A15}
    T_{\rho2} = 2i \displaystyle \int_{0}^{\pi/2}d\phi \left[\int_0^{\theta_c}d\theta \sum_{m=1}^{2}-\int_{\theta_c}^{\pi/2}d\theta \sum_{m=3}^{4}\right]\dfrac{\sin^3\theta \exp(-i\kappa_m\delta_1)}{\left(3i\cos\theta+4\beta_\infty \kappa_m\right)}.
\end{equation}
For $\tilde{z}<0$, 
\begin{equation}
\label{eq:A17}
    T_{\rho1} = -2i \displaystyle  \int_0^{\pi/2}d\theta\int_{0}^{\pi/2}d\phi  \sum_{m=3}^{4}\dfrac{\sin^3\theta \exp(-i\kappa_m\delta_1)}{\left(3i\cos\theta+4\beta_\infty \kappa_m\right)}.
\end{equation}

\begin{equation}
\label{eq:A16}
    T_{\rho2} = -2i \displaystyle \int_{0}^{\pi/2}d\phi \left[\int_0^{\theta_c}d\theta \sum_{m=3}^{4}-\int_{\theta_c}^{\pi/2}d\theta \sum_{m=1}^{2}\right]\dfrac{\sin^3\theta \exp(i\kappa_m\bar{\delta}_1)}{\left(3i\cos\theta+4\beta_\infty \kappa_m\right)},
\end{equation}
Similar to the case for Stokes streamfunction, the $\phi-$intergral in \eqref{eq:A14} and \eqref{eq:A17} can be evaluated in terms of Bessel and Struve functions. With a change of variables from $(\theta,\phi)$ to $(x,y) = (\sin\theta,\sin\phi)$, \eqref{eq:2.11} and \eqref{eq:2.12} can be recovered. For $\beta_\infty = 0$, we set $\beta_\infty$ to zero in \eqref{eq:A13} and write,
\begin{equation}
\label{eq:A18}
    \tilde{\rho}_f = - \displaystyle \dfrac{3}{\pi^2} \int_0^{\pi/2}d\theta\int_{0}^{\pi/2}d\phi \int_{0}^{\infty}dk k^2\sin^3\theta\dfrac{\sin^2\theta(\cos k\bar{\delta}_1 + \cos k\delta_1) + k^3 \cos \theta(\sin k\bar{\delta}_1 - \sin k\delta_1)}{k^6\cos^2\theta + \sin^4\theta},
\end{equation}
with the $k-$integral evaluated as described in the previous subsection to obtain the terms $T_{\rho1}$ and $T_{\rho2}$ in \eqref{eq:2.13}.
\section{Calculating the boundary of the downstream columnar structure}
\label{sec:AppB}
In order to find an expression for the boundary of the columnar structure, we first derive a closed-form expression for the Stokes streamfunction under the jet approximation, that is, the limit $k_3 \ll k_t$, corresponding to nearly vertical flow. In this limit, \eqref{eq:A1} simplifies to
\begin{equation}
    \label{eq:B1}
    \displaystyle \tilde{\psi}_s = \dfrac{3\tilde{r}_t i}{4\pi^2}\int \dfrac{(ik_3+\beta_\infty k_t^2)k_2}{(ik_3 + \beta_\infty k_t^2)k_t^4 + k_t^2}e^{i\boldsymbol{k}\cdot\tilde{\boldsymbol{r}}}d\boldsymbol{k}.
\end{equation}
Using cylindrical coordinates in Fourier space - ($k_t,\phi,k_3$) facilitates the evaluation of the $k_3-$integral using contour integration \citepalias{varanasi2022motion}. Subsequently, performing the $\phi-$integral results in the following 1D integral in terms of $k_t$:
\begin{equation}
    \label{eq:B2}
    \displaystyle \tilde{\psi}_s = 3\tilde{r}_t\int_0^\infty dk_t \dfrac{J_1(\tilde{r}_t k_t)e^{-z\left(\beta_\infty k_t^2 + 1/k_t^2\right)}}{k_t^4}.
\end{equation}
We now evaluate this integral in closed form in the non-diffusive limit ($\beta_\infty = 0$).
\subsection{Steepest-descent method for the far-field jet approximation integral}
\label{sec:AppB1}
For $\beta_\infty = 0$, using a change of variables $\lambda = \sqrt{z}/k_t$, the integral in \eqref{eq:B2} can be written as
\begin{equation}
\label{eq:B3}
    \displaystyle \frac{3\tilde{r}_t}{\tilde{z}^{3/2}}\int_0^\infty d\lambda~ \lambda^2 J_1\left(\frac{\tilde{r}_t\tilde{z}^{1/2}}{\lambda}\right)\exp(-\lambda^2).
\end{equation}
In the limit $\tilde{r}_t\tilde{z}^{1/2}  \gg 1$, the dominant contribution to the integral comes from $\lambda \sim \mathcal{O}(1)$, owing to the decaying exponential in the integrand. Therefore, $(\tilde{r}_t\tilde{z}^{1/2})/\lambda \gg 1$, and one can approximate the Bessel function in terms of its large-argument trigonometric representation $J_1(x) = \left(\sin x-\cos x\right)/\sqrt{\pi x}$, whence \eqref{eq:B3} simplifies to
\begin{equation}
\label{eq:B4}
    \displaystyle \frac{3(1+i)(\tilde{r}_t\tilde{z}^{1/2})^{1/2}}{2\sqrt{\pi}\tilde{z}^{2}}\int_{-\infty}^\infty d\lambda~ \lambda^{5/2} \exp\left(-\lambda^2+i\dfrac{\tilde{r}_t\tilde{z}^{1/2}}{\lambda}\right).
\end{equation}
A further change in variables, $\hat{\lambda} = \lambda (\tilde{r}_t\tilde{z}^{1/2})^{-1/3}$, leads to:
\begin{equation}
\label{eq:B5}
    \displaystyle \frac{3(1+i)\tilde{r}_t\tilde{z}^{1/2}}{2\sqrt{\pi}\tilde{z}^{2}}\int_{-\infty}^\infty d\hat{\lambda}~ \hat{\lambda}^{5/2} \exp\left((\tilde{r}_t\tilde{z}^{1/2})^{2/3}g(\hat{\lambda})\right), \textrm{where } g(\hat{\lambda}) = \left(-\hat{\lambda}^2+\dfrac{i}{\hat{\lambda}}\right).
\end{equation}
This integral can be determined using the steepest descent method, with the large parameter being $(\tilde{r}_t\tilde{z}^{1/2})^{2/3}$. Accordingly, the integration path along the real axis is deformed onto the appropriate constant phase contour ($\Im(g(\hat{\lambda}))$ is a constant) in the complex-$\hat{\lambda}$ plane. Next, the saddle points on this steepest descent contour are identified. These points correspond to $\Re(g(\hat{\lambda}))$ attaining a maximum, and for $\tilde{r}_t\tilde{z}^{1/2} \gg 1$, the dominant contribution to the integral arises from the neighborhood of the saddle point \citep{bender2013advanced}. Substituting $\hat{\lambda} = \hat{\lambda}_r + {i}\hat{\lambda}_i$ in $g(\hat{\lambda})$ above and writing the real and imaginary parts separately yields
\begin{equation}
\label{eq:B6}
    g(\hat{\lambda}_r, \hat{\lambda}_i) = \left(-\hat{\lambda}_r^2+\hat{\lambda}_i^2+\dfrac{\hat{\lambda}_i}{\hat{\lambda}_r^2+\hat{\lambda}_i^2}\right)+{i}\left(\dfrac{\hat{\lambda}_r}{\hat{\lambda}_r^2+\hat{\lambda}_i^2} - 2 \hat{\lambda}_r\hat{\lambda}_i\right),
\end{equation}
where the constant phase curves are given by 
\begin{equation}
\label{eq:B7}
    \Im(g(\hat{\lambda}_r,\hat{\lambda}_i)) = \left(\dfrac{\hat{\lambda}_r}{\hat{\lambda}_r^2+\hat{\lambda}_i^2} - 2 \hat{\lambda}_r\hat{\lambda}_i\right) = constant.
\end{equation}
The saddle points corresponding to $dg(\hat{\lambda})/d\hat{\lambda} = 0$ are given by
\begin{equation}
\label{eq:B8}
    \hat{\lambda}_n = \dfrac{1}{\sqrt[3]{2}}\exp\left(\dfrac{-i\pi}{3}\left(\dfrac{1}{2}+2n\right)\right), \textrm{  where } n \in [0,1,2].
\end{equation}
The constant phase curves that pass through the saddle points can now be drawn by replacing the $constant$ in \eqref{eq:B5} with $\Im(g(\hat{\lambda}_n))$. That is, 
\begin{equation}
\label{eq:B9}
    \left(\dfrac{\hat{\lambda}_r}{\hat{\lambda}_r^2+\hat{\lambda}_i^2} - 2 \hat{\lambda}_r\hat{\lambda}_i\right) = \Im(g(\hat{\lambda}_n)), \textrm{  where } n \in [0,1,2],
\end{equation}
\begin{figure}
\centering
    \begin{minipage}[t]{0.7\textwidth}
      \centering
      \includegraphics[width=\textwidth]{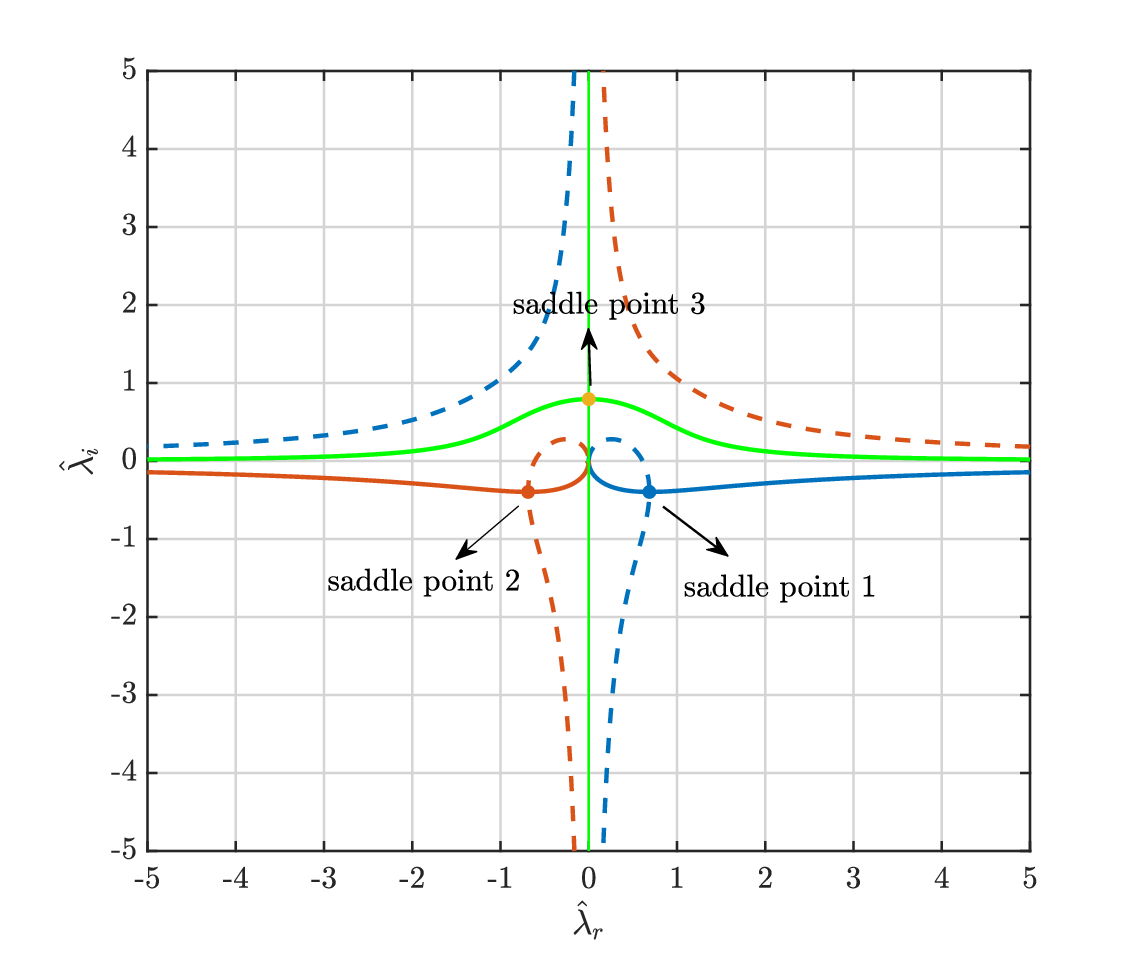}
    \end{minipage}
  \caption{Constant phase curves in the complex-$\hat{\lambda}$ plane from expression \eqref{eq:B7}. The blue, red, and green curves (both dashed and continuous) correspond to saddle points 1, 2, and 3, respectively; however, we consider only the continuous curves. The red and blue continuous curves are the steepest descent curves, whereas the green curve is the steepest ascent curve.}
  \label{fig:app1}
\end{figure}
which are depicted in figure \ref{fig:app1}. We only consider the red and blue curves (corresponding to saddle points 1 and 2)  to deform the original integration path (i.e. $\hat{\lambda} \in [-\infty, \infty]$) into; the green curve that passes through saddle point 3 is a steepest-ascent curve, and hence, cannot be used. In other words,
\begin{equation}
\label{eq:B10}
     \displaystyle \int_{-\infty}^\infty d\hat{\lambda}~ \hat{\lambda}^{5/2} \exp\left((\tilde{r}_t\tilde{z}^{1/2})^{2/3}g(\hat{\lambda})\right) = \left[\int_R + \int_B \right] d\hat{\lambda}~ \hat{\lambda}^{5/2} \exp\left((\tilde{r}_t\tilde{z}^{1/2})^{2/3}g(\hat{\lambda})\right),
\end{equation}
where $R$ and $B$ indicate the red and blue steepest descent contours, respectively. Next, the integrals are approximated close to saddle points 1 and 2, respectively; as already mentioned, these regions will have the largest contribution. This is is done by substituting $\hat{\lambda} = \hat{\lambda}_n + \Lambda$ and doing a Taylor series expansion of the integrand, for small $\Lambda$, whence one obtains
\begin{equation}
\label{eq:B11}
     \displaystyle   \left(e^{(\tilde{r}_t\tilde{z}^{1/2})^{2/3}g(\hat{\lambda}_1)}\hat{\lambda}_1^{5/2}+e^{(\tilde{r}_t\tilde{z}^{1/2})^{2/3}g(\hat{\lambda}_2)}\hat{\lambda}_2^{5/2}\right)\int_{-\infty} ^{\infty} d\Lambda~  \exp\left(-3(\tilde{r}_t\tilde{z}^{1/2})^{2/3}\Lambda^2\right),
\end{equation}
at the leading order in $\Lambda$. Evaluating the Gaussian integral above yields
\begin{equation}
     \displaystyle   \left(e^{(\tilde{r}_t\tilde{z}^{1/2})^{2/3}g(\hat{\lambda}_1)}\hat{\lambda}_1^{5/2}+e^{(\tilde{r}_t\tilde{z}^{1/2})^{2/3}g(\hat{\lambda}_2)}\hat{\lambda}_2^{5/2}\right)\sqrt{\dfrac{\pi}{3\left(\tilde{r}_t\tilde{z}^{1/2}\right)^{2/3}}}.
\end{equation}
Substituting $\hat{\lambda}_1$ and $\hat{\lambda}_2$ in the expression above, one obtains the Stokes streamfunction in the limit of $\tilde{r}_t\tilde{z}^{1/2}  \gg 1$, as given in \eqref{eq:3.6}. 
\subsection{Boundary of the columnar structure}
\label{sec:AppB2}
As described in \S~\ref{sec:3.2}, the boundary separating the columnar structure, and the far-field region described by the algebraic wake asymptote, is given by (see \eqref{eq:3.7})
\begin{equation}
\label{eq:B12}
    \dfrac{1620\tilde{r}_t^2}{\tilde{z}^7} = \dfrac{\sqrt{3}}{\sqrt[3]{2}}\dfrac{(\tilde{r}_J\tilde{z}^{1/2})^{4/3}}{\tilde{z}^2}\exp\left(-\dfrac{3}{2\sqrt[3]{4}} (\tilde{r}_J\tilde{z}^{1/2})^{2/3}\right).
\end{equation}
This can be written as
\begin{equation}
\label{eq:B13}
    \dfrac{B}{\tilde{z}^6} = \dfrac{\exp\left(-A \xi\right)}{\xi},
\end{equation}
where $A =3/2\sqrt[3]{4}$, $B = 1620\sqrt[3]{2}/\sqrt{3}$ are constants and $\xi = (\tilde{r}_J\tilde{z}^{1/2})^{2/3}$. Taking a logarithm on both sides gives
\begin{equation}
\label{eq:B14}
    \ln\left(\tilde{z}^6/B\right) = A\xi + \ln\xi.
\end{equation}
For the columnar structure $\tilde{z} \gg 1$, and the LHS in \eqref{eq:B14} is a logarithmically large quantity. Assuming $\xi \gg 1$ in anticipation, the first term on the RHS is dominant over the second. Therefore, at leading (logarithmic) order, one obtains 
\begin{equation}
\label{eq:B15}
    \xi \sim L_1 \vcentcolon = \dfrac{1}{A}\left[\ln\left(\dfrac{\tilde{z}^6}{B}\right)\right].
\end{equation}
In order to obtain a better approximation, one writes $\xi = L_1 + L_2$, where $L_1$ is given in \eqref{eq:B15} and the correction $L_2 \ll L_1$. Substituting $\xi = L_1 + L_2$ in \eqref{eq:B14} gives
\begin{equation}
\label{eq:B16}
    \ln\left(\dfrac{\tilde{z}^6}{B}\right) = A L_1+ A L_2 + \ln(L_1+L_2).
\end{equation}
The first term in the RHS cancels the LHS due to \eqref{eq:B15}. Next, the third term in \eqref{eq:B16} can be written as $\ln(L_1+L_2) = \ln L_1 + \ln(1+L_2/L_1) \approx \ln L_1 + L_2/L_1$. Therefore, \eqref{eq:B16} simplifies to
\begin{equation}
    \label{eq:B17}
    A L_2 = - \ln L_1 - \dfrac{L_2}{L_1}.
\end{equation}
Since $L_2/L_1 \ll L_2$, one obtains
\begin{equation}
\label{eq:B18}
    L_2 = -\dfrac{1}{A}\ln\left(L_1\right).
\end{equation}
Using $L_1$ and $L_2$ from \eqref{eq:B15} and \eqref{eq:B18}, one obtains
\begin{equation}
\label{eq:B19}
    \xi = L_1+L_2 = \dfrac{1}{A}\left[\ln\left(\dfrac{\tilde{z}^6}{B}\right) - \ln\left(\dfrac{1}{A}\ln\left(\dfrac{\tilde{z}^6}{B}\right)\right)\right],
\end{equation}
to second order. Substituting $A, B$ and $\xi$ in (\ref{eq:B19}) results in an explicit expression for the radial extent of the columnar structure boundary ($\tilde{r}_J$) in terms of $\tilde{z}$, given in \eqref{eq:3.7}. Figure \ref{fig:14}a shows a comparison of the boundary calculated from $L_1$ alone\,(red curve), and using $L_1+L_2$\,(black dashed curve), with that obtained from the streamline plot\,(blue curve). The two-term approximation is better than the single-term one, and shows an exact match with the numerically obtained boundary. 

\section{Tertiary screening length}
\label{sec:AppC}
\begin{figure}
    \hspace{-0.5cm}
    \begin{minipage}[t]{0.55\textwidth}
      \centering
      \includegraphics[width=\textwidth]{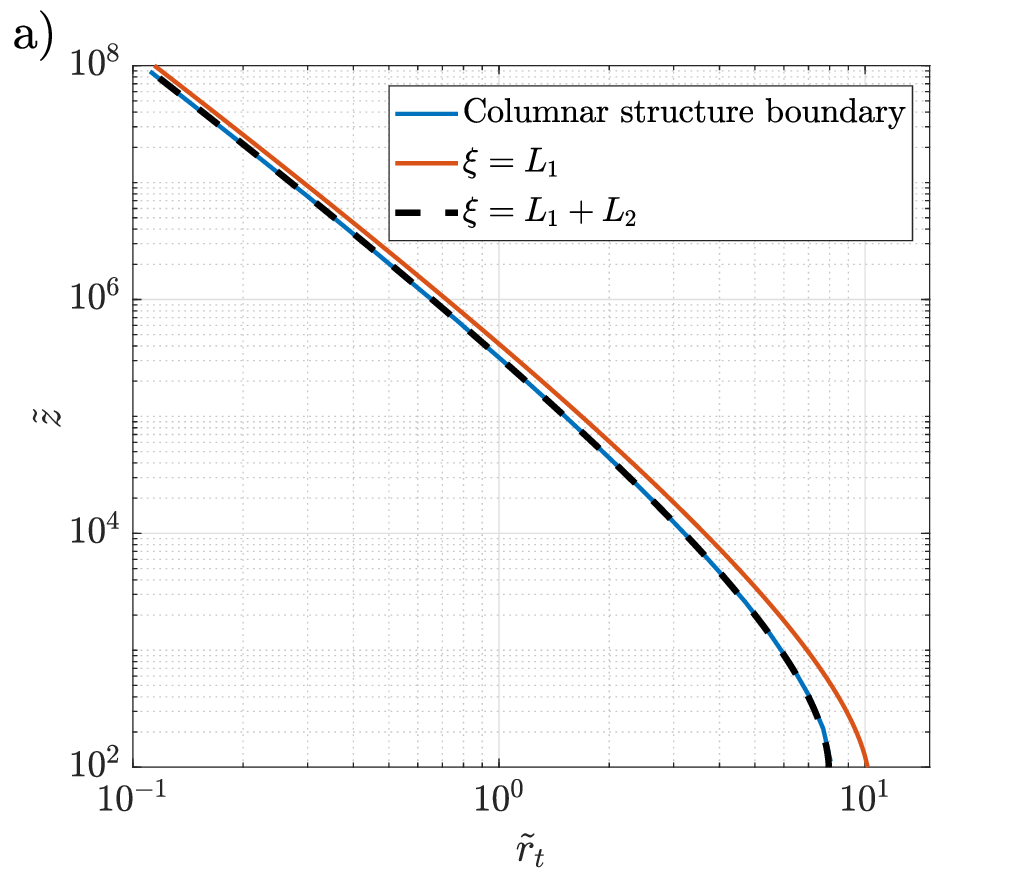}
    \end{minipage}
    \hspace{-0.4cm}
    \begin{minipage}[t]{0.55\textwidth}
      \centering
      \includegraphics[width=\textwidth]{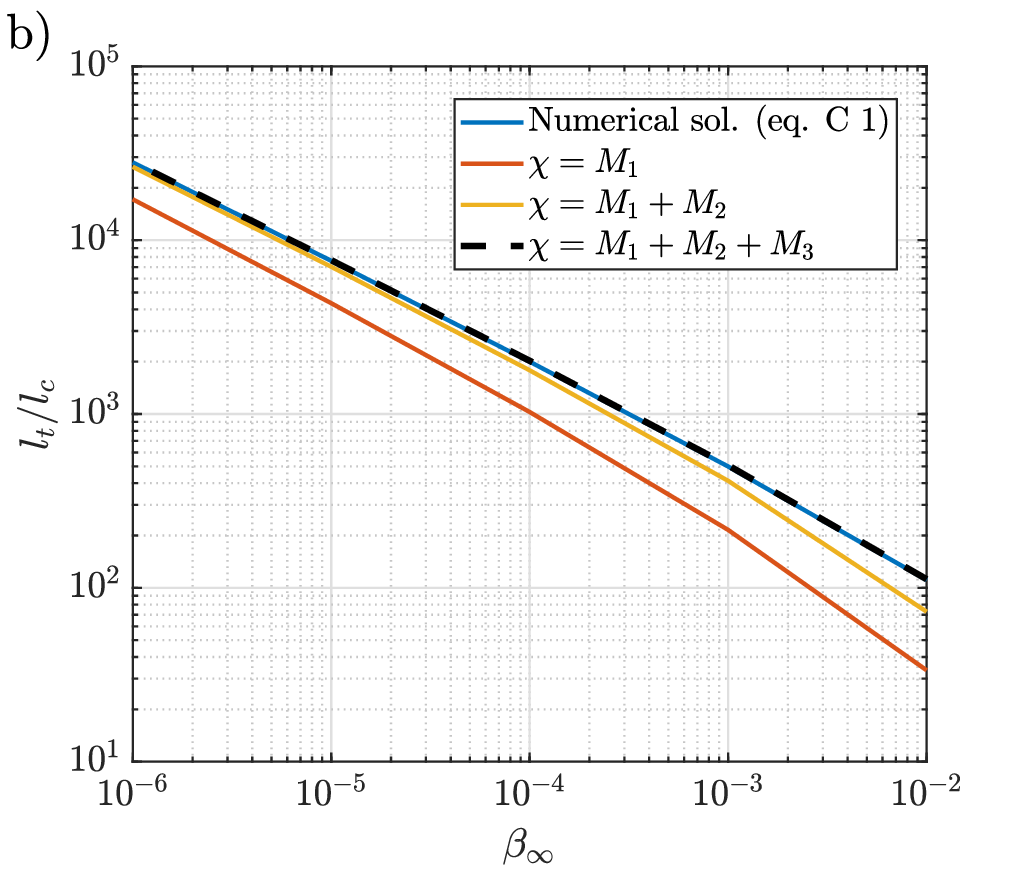}
    \end{minipage}
    \caption{a) Comparison of the columnar structure boundary obtained in Appendix \ref{sec:AppB2} with the boundary drawn from ends of different recirculating cells in the streamline plot of $\beta_\infty = 0$ (blue curve). The red and black dashed curves are from the single and two term asymptotic solution from  \eqref{eq:B15} and \eqref{eq:B19}, respectively. b) Comparison of the tertiary screening length obtained in Appendix \ref{sec:AppC} and that from numerically solving \eqref{eq:C1} (blue curve). The red, green and black curves are from the single, two and three term asymptotic solutions given in \eqref{eq:C4}, \eqref{eq:C6}, and \eqref{eq:C7}, respectively.}
  \label{fig:14}
\end{figure}
As described in \S~\ref{sec:3.2}, we equate the large-argument form of the modified Bessel function expression of the axial velocity (along the rear-stagnation streamline) with the expected far-field decay asymptote to obtain 
\begin{equation}
\label{eq:C1}
    3\beta_\infty^{1/2}\sqrt{\frac{\pi}{2}}\dfrac{\exp\left(-2\beta_\infty^{1/2}\tilde{z}_t\right)}{\left(-2\beta_\infty^{1/2}\tilde{z}_t\right)^{1/2}} = \dfrac{3240}{\tilde{z}_t^7}.
\end{equation}
This equation can be simplified by setting $\chi = \beta_\infty^{1/2}\tilde{z}$ and $C = 2160\beta_\infty^{3}/\sqrt{\pi}$, so \eqref{eq:C1} takes the form
\begin{equation}
\label{eq:C2}
    \exp(-2 \chi) = \dfrac{C}{\chi^{13/2}}.
\end{equation}
Taking a logarithm on both sides of \eqref{eq:C2} gives,
\begin{equation}
\label{eq:C3}
    \ln \left(\dfrac{1}{C}\right) = 2\chi - \dfrac{13}{2}\ln(\chi).
\end{equation}
In the convection dominant limit ($\beta_\infty \ll 1$), the LHS of \eqref{eq:C3} is a logarithmically large quantity. Assuming $\chi \gg 1$ therefore, one finds that the first term in the RHS is dominant over the second; a reasonable assumption the secondary screening length is given by $l_s \sim \mathcal{O}(al_c\beta_\infty^{-1/2})$, and $\chi \gg 1$ implies that the tertiary screening length is asymptotically larger in relation. Therefore, one obtains, to the leading (logarithmic) order,
\begin{equation}
    \label{eq:C4}
    \chi \sim M_1 \vcentcolon= \dfrac{1}{2}\ln\left(\dfrac{1}{C}\right).
\end{equation}
Similar to Appendix \ref{sec:AppB2}, a correction to $\chi$ can be found by substituting $\chi = M_1 + M_2$, where $M_1$ is given by \eqref{eq:C4}, in \eqref{eq:C3}. Grouping terms of the same order yields
\begin{equation}
    \label{eq:C5}
    \left(\ln\left(\dfrac{1}{C}\right) - 2M_1\right) = \left(2M_2 - \dfrac{13}{2}\ln M_1\right) - \dfrac{13}{2}\dfrac{M_2}{M_1}.
\end{equation}
In obtaining the expression above, $\ln(M_1+M_2)$ due to the second term in \eqref{eq:C3} is written as $\ln M_1 + \ln(1+M_2/M_1) \approx \ln M_1 + M_2/M_1$. 
One then finds
\begin{equation}
    \label{eq:C6}
    M_2 = \dfrac{13}{4}\ln M_1.
\end{equation}
Figure \ref{fig:14}b shows a comparison of the tertiary screening length calculated from $M_1$ alone\,(red curve),  and using $M_1+M_2$\,(green curve), with that obtained from numerically solving \eqref{eq:C1}\,(blue curve). In contrast to $\tilde{r}_J$ above, there is a small deviation from the numerical results even relative to the two-term approximation. A better approximation is therefore obtained by considering $\chi = M_1 + M_2 + M_3$ and substituting in \eqref{eq:C2}. This results in
\begin{equation}
    \label{eq:C7}
    \left(\ln\left(\dfrac{1}{C}\right)-2M_1\right) =  \left(2M_2-\dfrac{13}{2}\ln M_1\right) + 2M_3 - \dfrac{13}{2}\ln\left(1+\dfrac{M_2+M_3}{M_1}\right).
\end{equation}
Again, the LHS and the first two terms in the RHS cancel due to \eqref{eq:C4} and \eqref{eq:C6}, respectively. Further, noticing that $M_3/M_1 \ll M_3$, one finds
\begin{equation}
    \label{eq:C8}
    M_3 = \dfrac{13}{4}\dfrac{M_2}{M_1}.
\end{equation}
Finally, substituting $M_1$, $M_2$ and $M_3$, yields
\begin{equation}
\label{eq:C9}
    \chi = M_1 + M_2 + M_3 = \left(\dfrac{1}{2}\ln\left(\dfrac{1}{C}\right)+\dfrac{13}{4}\ln\left(\dfrac{1}{2}\ln\left(\dfrac{1}{C}\right)\right) + \dfrac{169}{16}\dfrac{\ln\left(\dfrac{1}{2}\ln\left(\dfrac{1}{C}\right)\right)}{\dfrac{1}{2}\ln\left(\dfrac{1}{C}\right)} \right)
\end{equation}
The tertiary screening length calculated from the expression above (black dashed curve) matches well with that from numerically solving \eqref{eq:C1} (blue curve), as shown in figure \ref{fig:14}b.
\bibliography{main}
\bibliographystyle{jfm}
\end{document}